\documentclass[onecolumn]{emulateapj}
\usepackage{natbib,amsmath}

\newcommand{\beq}{\begin{equation}}
\newcommand{\eeq}{\end{equation}}

\def\be{\begin{equation}}
\def\ee{\end{equation}}
\def\bea{\begin{eqnarray}}
\def\eea{\end{eqnarray}}



\def\myputfigure#1#2#3#4#5%
{\vskip#5pt\makebox[0pt]{\hskip#2in
\includegraphics[width=#3\textwidth]{#1}}\vskip#4pt\hfill}

\newcommand{\nh}{$N_{\rm H}$}
\newcommand{\mnh}{N_{\rm H}}

\def\sci#1{{\; \times \; 10^{#1}}}
\def\msol{M_\odot}
\def\delv{\Delta v}
\newcommand{\lyb}{Ly$\beta$}
\newcommand{\kms}{km~s$^{-1}$}
\newcommand{\mkms}{{\rm \; km\;s^{-1}}}
\newcommand{\sdssj}{SDSSJ1204+0221}
\newcommand{\fgqso}{SDSSJ1204+0221FG}
\newcommand{\bgqso}{SDSSJ1204+0221BG}
\newcommand{\slls}{SLLS/SDSSJ1204+0221BG}

\newcommand{\mnhi}{N_{\rm HI}}
\newcommand{\nhi}{$N_{\rm HI}$}
\newcommand{\lya}{Ly$\alpha$}
\newcommand{\cm}[1]{\, {\rm cm^{#1}}}
\newcommand{\N}[1]{{N({\rm #1})}}
\newcommand{\hkpc}{$h^{-1}$\,kpc}
\newcommand{\hMpc}{$h^{-1}$\,Mpc}
\newcommand{\rperp}{R_\perp}

\bibpunct{(}{)}{;}{a}{}{,}

\begin{document}

\lefthead{QUASARS PROBING QUASARS III}
\righthead{PROCHASKA \& HENNAWI}

\submitted{Accepted to ApJ, September 10 2008}

\title{Quasars Probing Quasars III: New Clues to Feedback, Quenching, and the 
Physics of Massive Galaxy Formation}

\author{J. Xavier Prochaska\altaffilmark{1,2} \& Joseph F. Hennawi\altaffilmark{3,4}}
 
\altaffiltext{1}{Department of Astronomy and Astrophysics, 
  UCO/Lick Observatory; University of California, 1156 High Street, Santa Cruz, 
  CA 95064; xavier@ucolick.org}
\altaffiltext{2}{Visiting Astronomer, W.M. Keck Telescope.
The Keck Observatory is a joint facility of the University
of California and the California Institute of Technology.}
\altaffiltext{3}{Department of Astronomy, University of California
  Berkeley, Berkeley, CA 94720; joeh@berkeley.edu}
\altaffiltext{4}{NSF Astronomy and Astrophysics Postdoctoral
Fellow}

\begin{abstract}
  Galaxies hosting $z\sim 2$ quasars are the high-$z$ progenitors of
  today's massive `red-and-dead' galaxies. With close pairs of quasars
  at different redshifts, a background quasar can be used to study a
  foreground quasar's halo gas in \emph{absorption}, providing a wealth of
  information about feedback, quenching, and
  the physics of massive galaxy formation.  We present a Keck/HIRES
  spectrum of the bright background quasar in a projected pair with
  angular separation $\theta = 13.3''$ corresponding to $\rperp
  =108$\,kpc at the redshift of the foreground quasar $z_{\rm fg}=
  2.4360 \pm 0.0005$, precisely determined from Gemini/GNIRS near-IR
  spectroscopy. Our echelle spectrum reveals optically thick gas
  ($\mnhi \approx 10^{19.7} \cm{-2}$) coincident with the foreground
  quasar redshift. The ionic transitions of associated metal-lines
  reveal the following properties of the foreground quasar's halo: (1)
  the kinematics are extreme with absorption extending to $+780 \mkms$
  relative to $z_{\rm fg}$; (2) the metallicity is nearly solar; (3)
  the temperature of the predominantly ionized gas is $T \lesssim
  20,000$K; (4) the electron density is $n_{\rm e} \sim 1 \cm{-3}$
  indicating a characteristic size $\sim 10 - 100$\,pc for the
  absorbing `clouds'; (7) there is little (if any) warm gas
  $10^{5}\,{\rm K} \lesssim T \lesssim 10^{6}\,{\rm K}$; (8) the gas
  is unlikely illuminated by the foreground quasar, implying
  anisotropic or intermittent emission. The mass of cold $T\sim
  10^{4}{\rm K}$ gas implied by our observations is significant,
  amounting to a few percent of the total expected baryonic mass density
  of the foreground quasar's dark halo at $r\sim 100\,{\rm kpc}$. The
  origin of this material is still unclear, and we discuss several
  possibilities in the context of current models of feedback and
  massive galaxy formation.
\end{abstract}

\keywords{quasars: general -- intergalactic medium -- quasars: absorption lines -- cosmology: general -- surveys: observations}

\section{Introduction}
\label{sec:intro}

Over the course of a quasar's lifetime, the accretion of material onto
its $\sim 10^9\,\msol$ supermassive black hole will liberate an
enormous energy $E=\epsilon M_{\rm BH} c^2 \simeq 2\sci{62}\,{\rm
  ergs}$\footnote{This energy is appropriate for a typical quasar near
  the peak of activity at $z\sim 2.5$, a radiative efficiency
  $\epsilon\sim 0.1$, and an Eddington ratio of $0.1$.} affecting its
environment from pc to Gpc scales. On cosmological scales, the
collective emission from quasars dominates the ultraviolet background
radiation field \citep{hm96} that maintains roughly $90\%$ of the
universe as a highly ionized plasma \citep{gp65}, while the harder
photons produce the cosmic X-ray background
\citep{fi99,erz02,Ueda03,Cao05}. On Mpc scales, the ultraviolet flux
photoionizes the nearby IGM resulting in the proximity effect, or the
lower optical depth for \lya\ absorption at the quasar redshift
\citep{bdo88,sbd+00}. On the smallest scales of 1-100\,pc, the
interplay between black hole accretion, radiation, and the surrounding
gas, produce the characteristic emission from the well-studied broad
and narrow emission line regions \citep[see][]{ferost06}.

But the degree to which a quasar can influence its host galaxy on
scales $\sim\,{\rm kpc-Mpc}$ is much less clear. Indeed, just a few
percent of the energy $\sim 10^{62}\,{\rm ergs}$ emitted by a quasar
is comparable to the binding energy of a massive galaxy $\sim
10^{60}$\,erg, and thus capable of ejecting its interstellar medium (ISM)
or shock-heating it to temperatures $T\sim 10^{7}$\,K.  The bulges of
all local galaxies harbor supermassive black holes \citep{KormRich95},
the masses of which are tightly correlated with the properties of
their host spheroids, measured on scales $\sim$ kpc
\citep{Magorrian98,FerrMerr00,Gebhardt00}.  This has led many to
speculate that some ``feedback'' mechanism couples the quasar phase of
rapid supermassive black hole growth with the evolution of its host
galaxy
\citep[e.g.][]{SilkRees98,kh00,WyitheLoeb03,gds+04,ScannOh04,Springel05,Scann05,Kawata05,Menci06,Cox06,LuMo07,Sijacki07,Hopkins08}.

Besides explaining the correlation between black holes and bulges,
feedback from an AGN has also been invoked on even larger scales, as
the energy source which ``quenches'' star-formation in massive
galaxies, leaving them ``red and dead''.  The observed bimodality in
the galaxy color-magnitude diagram
\citep{Strateva01,baldry04,bell04,blanton05,faber07} and the sharp
cutoff at the bright end of the galaxy luminosity function
\citep{bbf+03} both point to some physical mechanism which shuts off
star-formation in massive galaxies. Otherwise, large quantities of gas
should have accreted onto the progenitors of ellipticals
resulting in an unseen
population of very massive blue galaxies. It is generally believed
that the coupling between the quasar and the galaxy is ``kinetic'' in
nature, involving galactic-scale outflows which suppress gas accretion
and inject heat, ultimately shutting off star-formation
\citep[e.g.][]{kh00,ScannOh04,Springel05,Cox06,hhc+07}.


An important laboratory for studying the astrophysics of quasar
feedback and quenching are luminous high redshift radio galaxies
\citep[HzRGs; for a review see][]{Mccarthy93,mileyd08}.  In these
sources outflowing collimated relativistic plasma, or ``radio jets'',
can extend to several tens of kpc from the optical host galaxies,
providing the most compelling evidence that an AGN can impact its
galactic scale surroundings. It is unclear, however, whether this jet
energy $\gtrsim 10^{60}$\,erg \citep[e.g.][]{Miley80} couples strongly
to the ambient ISM \citep{bc89}.  In addition to the tremendous radio
power, $z\gtrsim 2$ HzrGs are often associated with giant Ly$\alpha$
recombination nebulae, with sizes of up to $\sim 200$\,kpc, sometimes
extending well beyond the radio emission \citep[see][for a
review]{villar07}.  Ionizing radiation from the central AGN, thought
to be obscured from our vantage point, is likely to play a significant
role in exciting this emission. But the extreme kinematics 
of these nebulae 
\citep[${\rm FWHM}\gtrsim 1000\mkms$][]{nle+06,reuland07,villar07},
their complex irregular morphology, the tendency for the line emission
to be aligned with the axis of the radio jets, and their near solar
metallicities \citep{vernet01,humphrey08}, strongly suggest that
jet-gas interactions or feedback from the AGN are giving rise to a
large-scale outflow, and shocks from these
motions could also be triggering the emission. But the comoving number
density of HzRGs and giant \lya\ nebulae, 
$n \sim 10^{-8}\ , {\rm Mpc}^{-3}$ 
\citep{mileyd08}, is about three orders of magnitude
smaller than the number density of luminous quasars. Thus only about
one in a thousand of the supermassive black holes in the Universe
exhibit such dramatic evidence for feedback during their active
phase\footnote{Because the implied lifetime of the jets and nebulae
  are $\sim 10^8$ years, they cannot represent a short lived phase of
  evolution in every quasar.}, which would do little to quench
star-formation in the entire population of local massive galaxies. Is
feedback occurring in the typical quasar in the Universe?

While it has been argued that heating from a quasar is responsible
for quenching massive galaxies \citep{sdh05,hopkins07}, 
an alternative scenario is related to the physics of structure formation
in a hierarchical Universe.  This idea goes back to a classic argument
first made by \citet{ro77} that the ability to cool on a
dynamical timescale is what sets the upper bound for the mass of
luminous galaxies. At low halo masses the cooling time of gas is
shorter than the dynamical time resulting in a ``cold accretion''
mode, limited only by the supply of infalling material.  For halos
with $M \gtrsim 10^{12}\msol$, the cooling time exceeds the dynamical
time, which is the criterion for the formation of a stable accretion
shock \citep{db06}.  In these more massive halos, subsequent
infalling gas is shock heated and pressure supported by a hot halo in
quasi-hydrostatic equilibrium. However, cosmological hydrodynamical
simulations have shown that although the majority of gas is
shock-heated to the virial temperature, some level of ``cold mode''
accretion still takes place in even the most massive halos
\citep{kkw+05,ocvirk08}. It is unknown whether the
formation of a hot-halo alone can quench future star-formation and
explain the red and dead galaxy population. 
\citet{db08} recently proposed a quenching scenario whereby the
gravitational energy of cosmological accretion can serve as a 
heat source to quenching star formation, provided that a
cold ($T\sim 10^4$\,K) phase of pressure-confined (and thus long lived)
gas clumps ($M\sim 10^5-10^8\msol$) form, which can penetrate the halo and 
deliver the necessary heat. 

Understanding the role that quasar feedback, hot halos, cold
accretion, and pressure confined clumps play in quenching
star-formation, requires studying the physical state of the gas on
scales 10\,kpc-1\,Mpc in the high redshift progenitors of local massive
red galaxies.  \citet[][see also Wechsler et al.\ 1998]{cst+07} model
the clustering and number density
of $z\sim 2-3$ star-forming galaxies and deduced host
dark halo masses $M \lesssim 10^{12}~\msol$. These low masses allowed
them to convincingly argue that $z\sim 2-3$ star-forming galaxies
\emph{do not} evolve into the quenched red and dead galaxies that we
see today, but rather evolve into blue $\sim L_{\ast}$ galaxies
similar to the Milky Way. But the strong clustering of luminous
quasars at $z\sim 2$ implies larger dark halo masses $\gtrsim
10^{13}~\msol$ \citep{croom01,pmn04,croom05}, indicating that they
\emph{are indeed the progenitors of local red and dead galaxies}. Of
course, we are only observing a small fraction $\sim 10^{-2}-10^{-3}$
of these progenitors as quasars at any given time because the quasar
lifetime is much shorter than the age of the Universe.

However quasar spectra have thus far provided little insight into the
state of gas in their galactic environments for two reasons. First,
the large ionizing flux from the quasar typically photoionizes the
hydrogen in and around the galactic host \citep{hp07,phh08,cmb+08},
which would otherwise be detected as strong \ion{H}{1} absorption at
the quasar redshift. Second, the interpretation of material that is
detected in absorption along the line-of-sight to a quasar is limited
by the unknown distance between the material and the quasar. A
systematic search for the signatures of photoionized gas near quasars
(e.g.\ via \ion{N}{5}, \ion{O}{6} absorption) is only now being 
carried out on statistical datasets \citep{tripp08,fbp08}.  
Observers do frequently identify narrow
associated absorption lines (NAALs) or so-called ``associated absorbers'' 
which are believed to arise in the quasar environment. Some of these are
attributed to gas on sub-pc scales \citep[e.g.][]{elvis00}, but other
examples exhibit ionization states that suggest the gas lies at
several tens of kpc \citep{dcr+04,rps+07}. Note that these associated
absorbers are distinct from the broad absorption line (BAL) features 
which arise from dense clumps of material that have been ionized and
accelerated by the quasar, and are generally believed to be confined
to sub-kpc scales \citep[but see][]{dab+01}.

In a series of papers, we have introduced a novel technique to study
the physical state of gas in the ISM and halo of luminous quasars,
which has the potential to provide powerful constraints on feedback,
quenching and the physics of massive galaxy formation.  Namely, we use
a background quasar (b/g quasar) sightline to probe the state of gas
in absorption in the vicinity of a foreground quasar \citep[f/g
quasar; see also][]{bhm+06}.  Although such projected quasar pair
sightlines are extremely rare, \citet[][see also Hennawi 2004]{hso+06}
showed that it is straightforward to select $z\gtrsim 2$ projected
quasar pairs from the imaging and spectroscopy provided by the Sloan
Digital Sky Survey \citep[SDSS;][]{yaa+00}.  To date about 90 pairs of
quasars have been uncovered with impact parameter $R < 300~{\rm kpc}$
and $z_{\rm fg} > 1.6$\footnote{The lower limit on redshift is
  motivated by the ability to detect redshifted \lya\ absorption above
  the atmospheric cutoff $\lambda > 3200$~\AA}. Spectroscopic
observations of the b/g quasar in each pair reveals the nature of the
IGM transverse to the f/g quasar on scales of a few 10\,kpc to several
Mpc. This approach has the advantage of tracing diffuse gas over a
wide range of density and temperature, ranging from cold neutral
material $T\sim 100\,{\rm K}$ to collisionally ionized plasma $T
\approx 10^6$\,K, and with column densities in the range $N\sim
10^{12}-10^{22}\,{\rm cm^{-2}}$.  In Paper~I \citep{hpb+06}, we
searched 149 b/g quasar spectra for optically thick absorption in the
vicinity of $1.8 < z_{\rm fg} < 4.0$ luminous f/g quasars, and
uncovered a sample of 27 new quasar-absorber pairs with impact
parameters ranging from $30\,{\rm kpc}-2.5\,{\rm Mpc}$.  In Paper~II
\citep{hp07}, we analyzed the clustering of these transverse absorbers
with the foreground quasar and measured a large clustering signal on
galactic scales.  We also refer the reader to the manuscript
by \cite{tytler08} who consider several applications of quasar
pair spectroscopy.

In this paper, we present the first high resolution spectrum of the
b/g quasar in the close projected quasar pair \sdssj.  By mining the
sky for very rare close associations of quasars
\citep{thesis,hso+06,hpb+06}, we previously discovered this rare
system with angular separation $\theta = 13.3''$ corresponding to
impact parameter $\rperp =108$\,kpc at the redshift of the foreground
quasar $z_{\rm fg}= 2.4360 \pm 0.0005$, precisely determined from
Gemini/GNIRS near-IR spectroscopy.  The spectral and photometric
properties of \fgqso\ make it an unremarkable quasar at $z \sim
2.4$. We estimate a bolometric luminosity of $L_{\rm QSO} \simeq
1.4\sci{46}{\rm erg\, s^{-1}}$ placing it near the `knee' of the
$z\sim 2.5$ quasar luminosity function \citep{croom04,richards06}, and
corresponding to a supermassive black hole $M_{\rm BH} \simeq
1.1\sci{9}\left(\frac{f_{\rm Edd}}{0.1}\right)^{-1}\,\msol$, if it
accretes at one tenth of the Eddington limit.  The impact parameter of
our background sightline $\rperp =108$\,kpc, easily resolves the
expected virial radius $r_{vir} = 250\,{\rm kpc} (M/10^{13.3}
\msol)^{1/3}$ of the f/g quasar host, and pierces its halo at about
the `cooling radius', where gas shock-heated to the virial temperature
should take about a Hubble time to cool.  The only remarkable thing
about \sdssj\ is that it has a $z_{\rm bg} = 2.53$ b/g quasar
in close projection
which is bright enough ($r=19.0$) for high resolution spectroscopy.
Our Keck HIRES Echelle spectrum of \bgqso, the first ever to probe the
halo gas of a f/g quasar, resolves the velocity fields of the
absorbing gas and allows us to measure precise column densities for
\ion{H}{1} and the ionic transitions of metals like Si, C, N, O, and
Fe. These measurements allow us to place constraints on the physical
state of the gas near the f/g quasar, such as its kinematics,
temperature, ionization structure, chemical enrichment patterns,
volume density, the size of the absorbers, the intensity of the
impingent radiation field, as well as test for the presence of hot
collisionally ionized gas.  In $\S$~\ref{sec:obs} we present the
observations and provide column density measurements.  We constrain
the ionization state, estimate relative chemical abundances, and
constrain the electron density of the gas in $\S$~\ref{sec:ion}.  In
$\S$~\ref{sec:discuss} we further discuss our results and how they
relate to other observations of quasars, absorption line systems, and
high redshift galaxies. The implications of our results for models of
feedback, quenching, and massive galaxy formation are presented in
$\S$~\ref{sec:model}, and we conclude with a summary in
$\S$~\ref{sec:summ}.  The reader who is not concerned with the details
of the observations and absorption line modeling can read
Table~\ref{tab:summ} summarizing those results and then skip to
$\S$~\ref{sec:discuss}.  Throughout the manuscript, we use the
cosmological parameters $\Omega_m = 0.30$, $\Omega_\Lambda =0.70$,
$h=0.70$, consistent to within $1-\sigma$ with the parameters measured
by the WMAP experiment \citep{wmap05} and we adopt the solar chemical
composition compiled by \cite{gas07}.

\begin{figure}
\begin{center}
\includegraphics[width=6.8in]{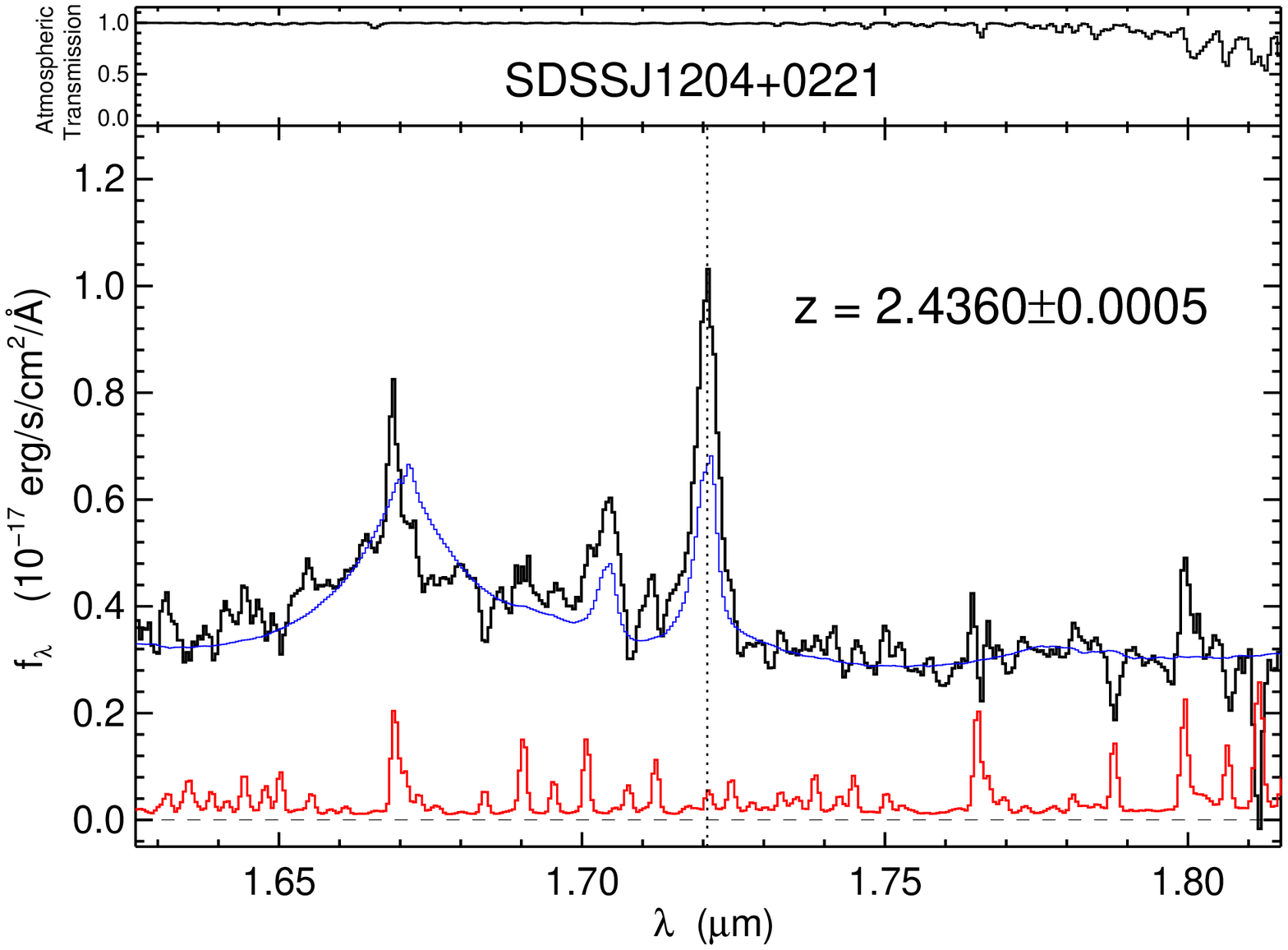}
\end{center}
\caption{GNIRS spectrum of \fgqso\ centered on the
  H$\beta$-[\ion{O}{3}] emission line complex (H$\beta$~$\lambda
  4861$, [\ion{O}{3}]~$\lambda 4959$, and [\ion{O}{3}]~$\lambda 5007$)
  which is redshifted into the $H$-band. This spectrum was used to
  determine the systemic redshift of the quasar, $z_{\rm fg} = 2.4360 \pm
  0.0005$, from the strong [\ion{O}{3}]~$\lambda 5007$ emission
  line. 
  The lower red histogram
  is the 1-$\sigma$ noise and the smooth blue curve overplotted on the
  data is the composite quasar template of \citet{vanden01} redshifted to
  $z_{\rm fg}$. The upper panel shows the atmospheric
  transmission for the corresponding wavelengths.}
\label{fig:GNIRS}
\end{figure}

\section{Observations and Analysis}
\label{sec:obs}

The principal goal of our analysis is to search for and characterize
gas associated with the foreground quasar \fgqso\ in the high
resolution spectrum of a background quasar \bgqso\ identified in close
projection.  To establish a physical association, we demand that the
absorption and \fgqso\ have nearly identical velocity.  To this end,
we must obtain a precise measurements for the quasar systemic
redshift.  But it is well known that the primary rest-frame ultraviolet
emission lines which are redshifted into the optical for $z\gtrsim 2$
quasars can differ by up to $\sim 3000\mkms$ from systemic, due to
outflowing/inflowing material in the broad line regions of quasars
\citep{Gaskell82,TF92,vanden01,Richards02}, with a more typical error
being $\sim 1000\mkms$. An accurate redshift can be determined
from narrow ($\sigma \lesssim 200\mkms$) forbidden emission lines,
such as [\ion{O}{2}]~$\lambda 3727$ or [\ion{O}{3}]~$\lambda 5007$
which arise from the narrow line region, but at $z \gtrsim 2$,
measurements of these lines require spectra covering the near
infrared. 

To this end we observed \fgqso\ using the the Gemini Near Infra-Red
Spectrograph \citep[GNIRS;][]{GNIRS} on the Gemini-South telescope on
March 27, 2006. We used the $0.15\arcsec$/pixel camera and the 32 lines
mm$^{-1}$ grating in cross dispersed mode, giving complete coverage
over the wavelength range 0.9-2.4~$\mu{\rm m}$. The slit width was
$0.45\arcsec$ ($\sim$ 3 pixels) giving a resolving power of $R\simeq
1100$ or a FWHM~$\simeq 270\mkms$.  The total exposure time
was 5440s, which was broken up into 16$\times$340s exposures to prevent
the brightest sky lines from saturating the detector. The GNIRS
spectra were reduced using standard techniques with a custom data
reduction pipeline written in the Interactive Data Language (IDL) and
described in Hennawi \& Prochaska (in prep.). 
Wavelength solutions were determined by
comparing extracted spectra of the night sky to an atlas of OH sky
emission lines, and heliocentric corrections were applied to the
spectra. The RMS deviation in our wavelength fits were typically
1-1.5~\AA~(or about 0.2-0.3 pixels). In the $H$-band, this corresponds
to a velocity uncertainty of $20-30\mkms$, but since our fits
typically use 50 lines, we believe that our wavelength solutions are
accurate to better than $\lesssim 5~\mkms$.

We computed a systemic redshift from the strong [\ion{O}{3}]~$\lambda
5007$ emission line\footnote{The [\ion{O}{3}] 4959 line is plagued
by sky emission lines and the H$\beta$ lines ia not as reliable
a diagnostic as [\ion{O}{3}] for QSO redshifts.} 
which is redshifted to $\sim 1.7~\mu{\rm m}$ in
the $H$-band for $z\sim 2.4$ (see Figure~\ref{fig:GNIRS}).  We found
that the most effective line-centering algorithm was to iterate a
flux-weighted line-centering scheme until the line-center converged to
within a specified tolerance. We achieved more stable results when the
pixel values were weighted by a Gaussian kernel (true flux-weighting
would correspond to a box-car kernel) with dispersion set to
$\sigma_{\rm [O~III]} = 6.04$~\AA, which is the average dispersion of
the [\ion{O}{3}] emission line measured by \citet{vanden01}.  We do
not estimate formal errors for the line centering, as they are smaller
than the intrinsic error incurred by using [\ion{O}{3}] as a proxy for
the systemic frame.  \citet{boroson05} measured the distribution of
velocity shifts of the [\ion{O}{3}] line center about the systemic
frame defined by low-ionization forbidden lines. He found that
[\ion{O}{3}] has an average blueshift of $\Delta v = 27~\mkms$ from
systemic and a dispersion of $\sigma = 44~\mkms$ about this value. To
account for the average shift, we add $\Delta v = 27~\mkms$ to the
vacuum rest wavelength of $5008.24$~\AA~when computing the redshift of
the line. For the $1\sigma$ error on our [\ion{O}{3}] near-IR
redshifts we adopt $\sigma = 44~\mkms$. We thus determine the systemic
redshift of \fgqso\ to be $z_{\rm fg} = 2.4360 \pm 0.0005$. All
velocities are reported relative to this redshift value for the
remainder of this paper.  Figure~\ref{fig:GNIRS} shows part of the
$H$-band region of our GNIRS spectrum of \fgqso\ centered on the
H$\beta$-[\ion{O}{3}] emission line complex (H$\beta$~$\lambda 4861$,
[\ion{O}{3}]~$\lambda 4959$, and [\ion{O}{3}]~$\lambda 5007$) and our
determination of the systemic redshift.

\begin{figure}
\begin{center}
\includegraphics[width=6.6in]{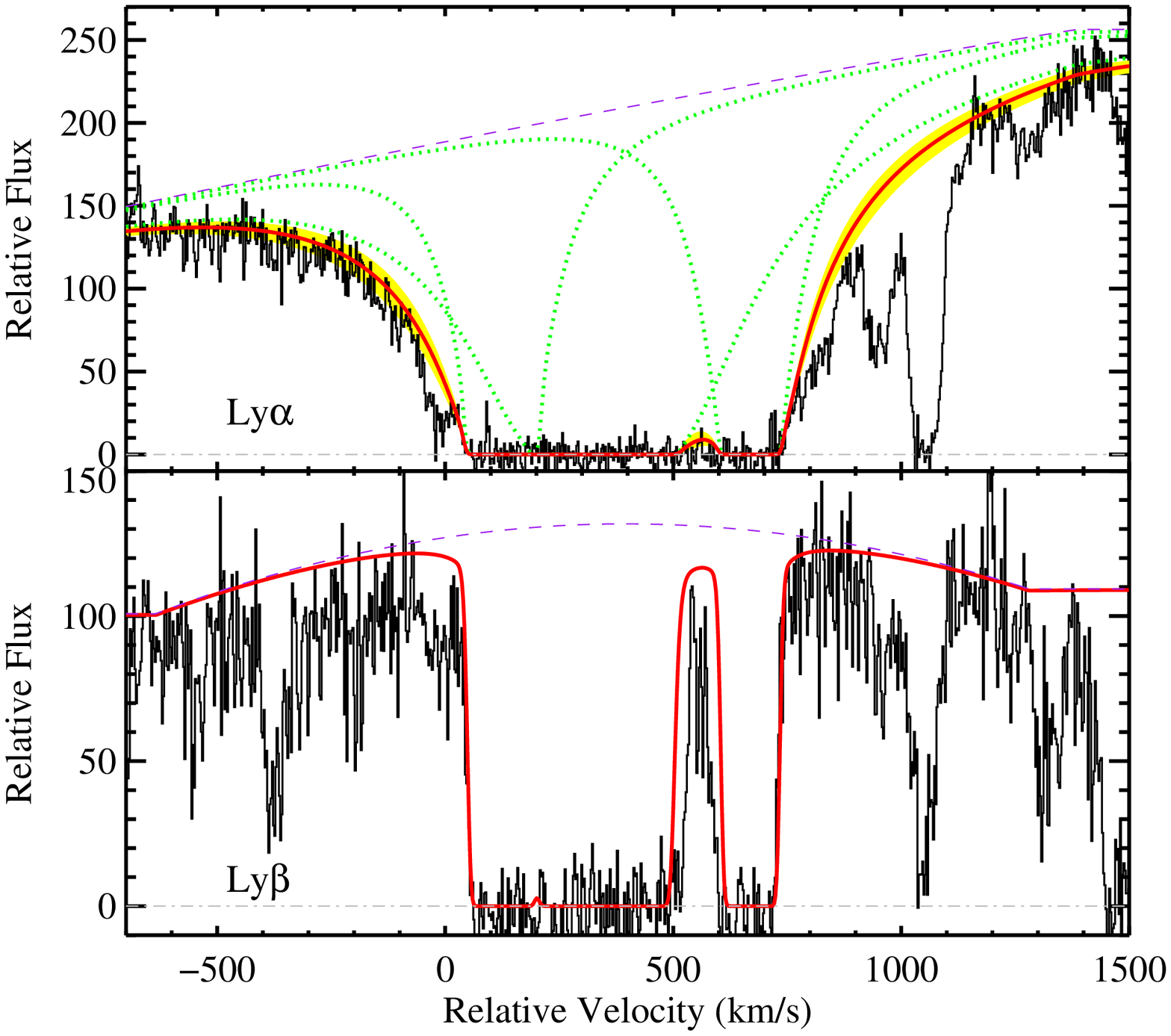}
\end{center}
\caption{\lya\ and \lyb\ profiles for the super Lyman limit
system identified 
in the spectrum of the background quasar \bgqso\
at a velocity consistent with the foreground quasar \fgqso. 
The relative velocity $v=0\mkms$ corresponds to the best-fit
value for the redshift of \fgqso, $z_{\rm fg} = 2.4360$. 
The purple dashed line indicates our estimate of the quasar
continuum in these relative flux units.
We have fitted these line-profiles with three \ion{H}{1} components,
each centered at the peak optical depth of low-ion absorption
associated with this \ion{H}{1} gas.
The green dotted lines show the \lya\ profiles for each component,
and the solid red curve is the convolved fit.  The \lyb\ profile
constrains the Doppler parameters of the outer two components to have
$b < 25 \mkms$, while the \lya\ profile requires these components
have $\mnhi < 10^{19} \cm{-2}$.
With these constraints on the outer components, we derive
$\mnhi = 10^{19.60 \pm 0.15} \cm{-2}$ for the central component
and note a mild degeneracy between this value and that for 
the outer components which is reflected in our error estimate.
}
\label{fig:HI}
\end{figure}

High resolution optical spectroscopy of the b/g quasar \bgqso\
was obtained on the nights of 
UT~13 April 2005 and 3 May 2005 (one 5400s exposure per night), 
using the HIRESb spectrometer \citep{vogt94} 
on the Keck\,I telescope.
The data were acquired through the C5 decker affording a 
FWHM~$\approx 8 \mkms$ resolution and processed with the 
HIRedux\footnote{http://www.ucolick.org/$\sim$xavier/HIRedux/index.html}
pipeline \citep{bbp08}.  
The signal-to-noise (S/N) ratio
per 2.6\kms\ pixel is approximately 15 at 4000\AA\ and the
combined spectrum offers nearly continuous wavelength coverage
from $\lambda \approx 3450$\AA\ to 6400\AA.   
A search for gas in the spectrum of \bgqso\ at $z \approx z_{\rm fg}$ 
revealed a strong \lya\ profile and a series of 
metal-line transitions.
In Figure~\ref{fig:HI}, we present the \lya\ and \lyb\
profiles for this absorption system. 
One readily observes the damping
wings of the \lya\ profile which require a total \ion{H}{1}
column density $\mnhi \gtrsim 10^{19} \cm{-2}$ marking this absorber  
as a `super' Lyman limit system \citep[SLLS; also referred to as
sub-damped \lya\ systems][]{peroux02}. 
For the remainder of the paper, we refer to this absorption system
as \slls.

We have modeled
the \lya\ and \lyb\ profiles by introducing three 
\ion{H}{1} components at the velocities corresponding to the
peak optical depths of three sets of metal-line complexes 
(defined below as subsystems A, B, and C).  Having fixed the redshifts of the 
Lyman series, we
then varied the \nhi\ values and Doppler parameters ($b$-values)
of these three components.
For subsystems A and C, the \lya\ profile restricts the
\ion{H}{1} column densities to be $\mnhi^{A,C} < 10^{19} \cm{-2}$;
larger values give a damped \lya\ profile that contradicts the observed
fluxes at $\lambda \approx 4177.5$\AA\ and 4187.5\AA\ 
(i.e.\ $\delta v \approx -100\mkms$ and $+750\mkms$ in Figure~\ref{fig:HI}).  
Similarly, the \lyb\ profile
demands effective\footnote{The metal-line profiles show these 
subsystems are the blend of several narrow components but we only consider
a single \ion{H}{1} cloud for each.  Therefore, the $b$-values reported for
the \ion{H}{1} gas likely
overestimate the values of the individual components.}
Doppler parameters $b_{A,C} < 25 \mkms$.  Allowing for these
constraints, we derive a value of $\mnhi^B = 10^{19.60 \pm 0.15} \cm{-2}$
for subsystem B that is independent of its nearly unconstrained
Doppler parameter.  The error in this \nhi\ estimate includes
uncertainty due to the continuum placement and 
line-blending with subsystems A and C.  In the figure, we have
overplotted a best-fit solution which assumes
$\mnhi = 10^{18.6} \cm{-2}$ for the subsystems A and C.

\begin{figure}
\begin{center}
\includegraphics[height=6.8in,angle=90]{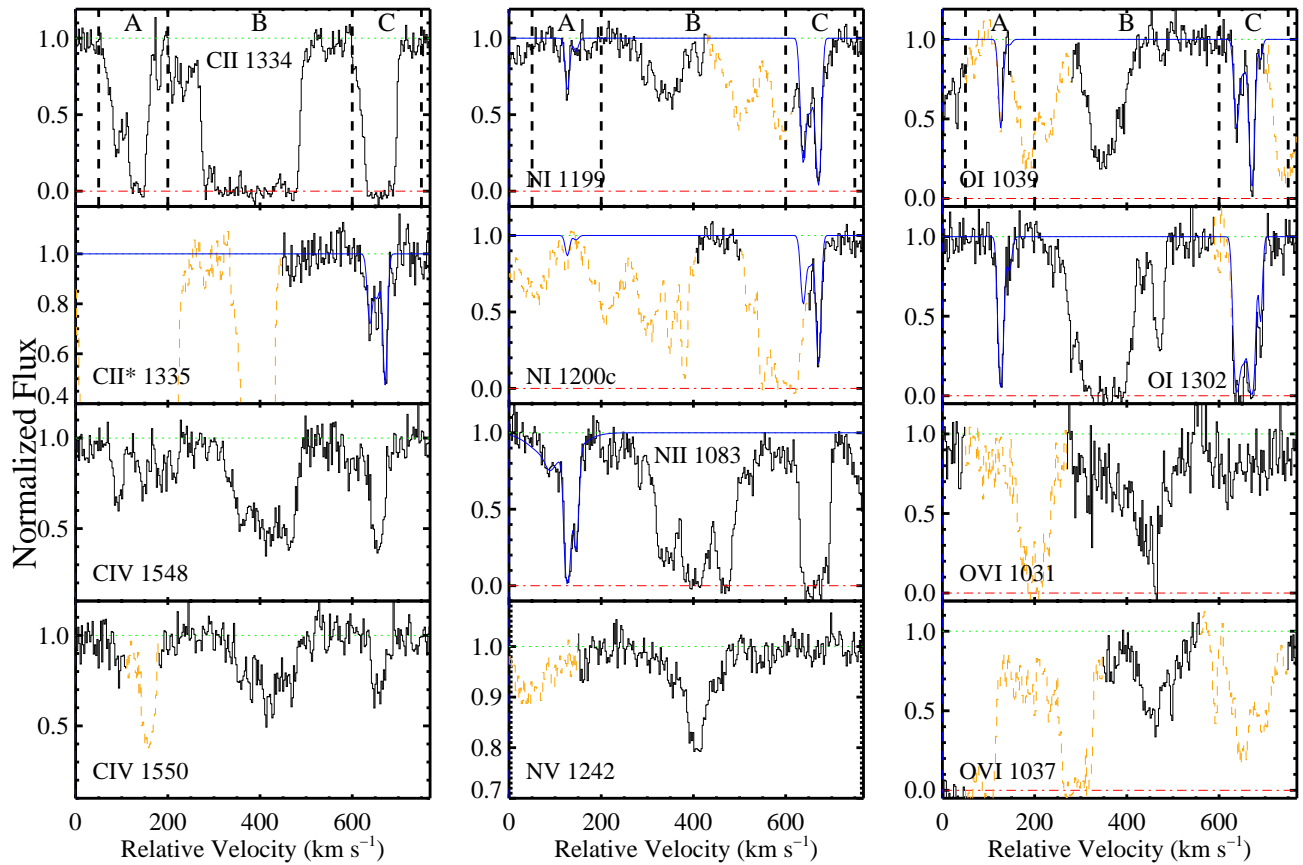}
\end{center}
\caption{
Metal-line transitions from the super Lyman limit system identified
at the redshift of \fgqso.  The velocities are relative to the 
measured redshift of \fgqso, $z_{\rm fg} = 2.4360$.  The absorption
occurs in three velocity intervals which we designate as subsystems
A,B,C as denoted by the vertical dashed lines in the upper panel
of each subplot.  
Note the strong low-ion and \ion{N}{2} absorption,
the absence of \ion{Si}{2}$^*$~1264, the relatively weak 
\ion{C}{4} and \ion{Si}{4} profiles, and the likely absence of
\ion{N}{5} and \ion{O}{6} absorption.
For a subset of the transitions from subsystems A and C we overplot
the line-profile fits derived using the VPFIT software package
and mark the components included (dark blue) and not
included (light red) in the fit as listed in Table~\ref{tab:vpfit}.
Absorption that is presumed unrelated to \slls\
(e.g.\ \lya\ features from unrelated redshifts) 
is presented as a dashed, orange line.
}
\label{fig:metals}
\end{figure}

\begin{figure}
\begin{center}
\includegraphics[height=6.8in,angle=90]{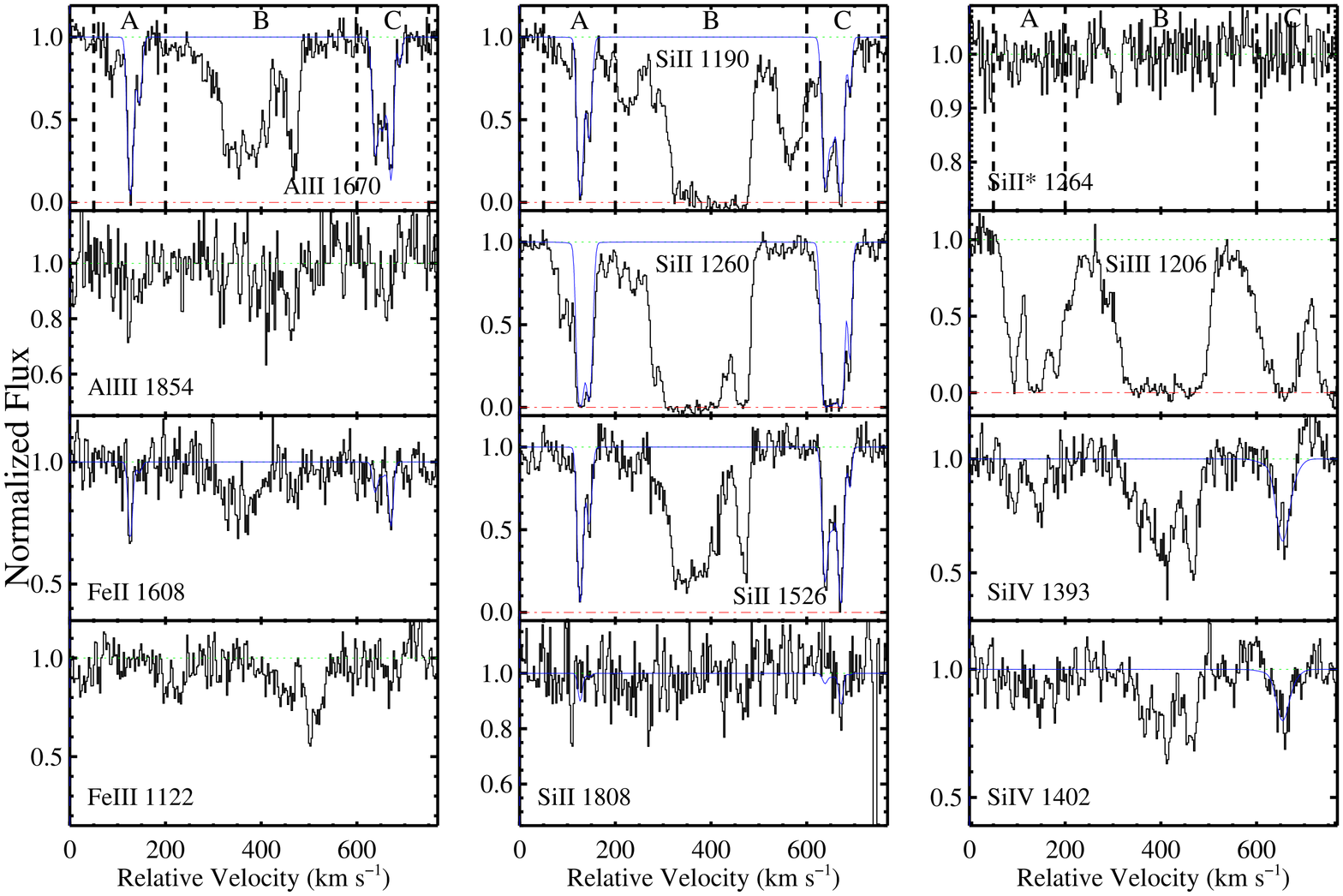}
\end{center}
\end{figure}

\begin{deluxetable}{lccccccccc}
\tablewidth{0pc}
\tablecaption{VOIGT PROFILE SOLUTIONS \label{tab:vpfit}}
\tabletypesize{\footnotesize}
\tablehead{\colhead{Comp} & \colhead{$z$} & \colhead{$\sigma(z)$} 
& \colhead{$v^a$} & \colhead{$b$} & \colhead{$\sigma(b)$}
& \colhead{Ion} & \colhead{$\log N$} & \colhead{$\sigma(N)$} \\
& & ($10^{-5}$)& (\kms) & (\kms) & (\kms) & & ($\cm{-2}$)  }
\startdata
\cutinhead{\ion{H}{1}}
A&2.43744&&&&&&\ion{H}{1}&18.60&0.40\\
B&2.44000&&&&&&\ion{H}{1}&19.60&0.15\\
C&2.44367&&&&&&\ion{H}{1}&18.60&0.40\\
\cutinhead{Metals}
1&2.437010&1.0&$+  88$&12.22& 7.77&\ion{N}{2}&12.84& 0.80\\
2&2.437150&1.0&$+ 100$&60.71&15.80&\ion{N}{2}&13.87& 0.14\\
3&2.437451&0.2&$+ 127$& 4.49& 0.31&\ion{N}{1}&13.09& 0.06\\
&&&&&&\ion{N}{2}&14.84& 0.22\\
&&&&&&\ion{O}{1}&14.60& 0.06\\
&&&&&&\ion{Al}{2}&12.97& 0.13\\
&&&&&&\ion{Si}{2}&14.07& 0.11\\
&&&&&&\ion{Fe}{2}&13.26& 0.08\\
4&2.437664&0.5&$+ 145$& 5.82& 0.86&\ion{N}{1}&12.46& 0.27\\
&&&&&&\ion{N}{2}&13.82& 0.05\\
&&&&&&\ion{O}{1}&13.28& 0.12\\
&&&&&&\ion{Al}{2}&11.92& 0.06\\
&&&&&&\ion{Si}{2}&13.23& 0.05\\
&&&&&&\ion{Fe}{2}&12.40& 0.50\\
5&2.443318&0.6&$+ 639$& 5.48& 0.60&\ion{C}{2}*&13.00& 0.09\\
&&&&&&\ion{N}{1}&13.75& 0.05\\
&&&&&&\ion{O}{1}&14.62& 0.06\\
&&&&&&\ion{Al}{2}&12.29& 0.07\\
&&&&&&\ion{Si}{2}&13.65& 0.05\\
&&&&&&\ion{Si}{4}&12.05& 1.40\\
&&&&&&\ion{Fe}{2}&12.82& 0.19\\
6&2.443497&1.3&$+ 655$& 9.05& 2.83&\ion{C}{2}*&12.94& 0.13\\
&&&&&&\ion{N}{1}&13.46& 0.11\\
&&&&&&\ion{O}{1}&14.22& 0.16\\
&&&&&&\ion{Al}{2}&12.26& 0.09\\
&&&&&&\ion{Si}{2}&13.38& 0.11\\
&&&&&&\ion{Si}{4}&12.76& 0.35\\
&&&&&&\ion{Fe}{2}&12.66& 0.31\\
7&2.443702&0.2&$+ 672$& 4.35& 0.27&\ion{C}{2}*&13.35& 0.04\\
&&&&&&\ion{N}{1}&14.33& 0.05\\
&&&&&&\ion{O}{1}&15.35& 0.12\\
&&&&&&\ion{Al}{2}&12.57& 0.09\\
&&&&&&\ion{Si}{2}&14.12& 0.08\\
&&&&&&\ion{Si}{4}&12.20& 1.32\\
&&&&&&\ion{Fe}{2}&13.16& 0.09\\
8&2.443915&1.0&$+ 691$& 3.00& 2.20&\ion{O}{1}&13.74& 0.21\\
&&&&&&\ion{Al}{2}&11.36& 0.14\\
&&&&&&\ion{Si}{2}&12.76& 0.09\\
\enddata
\tablenotetext{a}{Velocity relative to the redshift of the foreground quasar \fgqso, $z_{fg}=2.4360$.}
\end{deluxetable}
 
\clearpage
\begin{deluxetable}{lcccccccc}
\tablewidth{0pc}
\tablecaption{IONIC COLUMN DENSITIES \label{tab:colm}}
\tabletypesize{\footnotesize}
\tablehead{\colhead{Ion} & \colhead{$\lambda_{\rm rest}$} & \colhead{$\log f$}
& \colhead{$v_{int}^a$} 
& \colhead{$\log N_{\rm AODM}$} & \colhead{$\log N_{\rm VPFIT}^b$} \\
& (\AA) & & (\kms) &  }
\startdata
\cutinhead{A}
\ion{C}{2}&&&&$> 14.63$\\
&1036.3367 &$ -0.9097$&$[  37, 187]$&$> 14.63$&\\
\ion{C}{4}&&&&$ 13.60 \pm 0.03$\\
&1548.1950 &$ -0.7194$&$[  57, 227]$&$ 13.60 \pm 0.03$&\\
\ion{N}{1}&&&&&$ 13.18 \pm 0.07$\\
\ion{N}{2}&&&&&$ 14.93 \pm 0.18$\\
\ion{N}{5}&&&&$< 13.10$\\
&1242.8040 &$ -1.1066$&$[  87, 167]$&$< 13.10$&\\
\ion{O}{1}&&&&&$ 14.62 \pm 0.06$\\
\ion{O}{6}&&&&$< 13.67$\\
&1031.9261 &$ -0.8765$&$[  67, 147]$&$< 13.67$&\\
\ion{Al}{2}&&&&&$ 13.01 \pm 0.12$\\
\ion{Al}{3}&&&&$< 12.41$\\
&1854.7164 &$ -0.2684$&$[  67, 187]$&$< 12.41$&\\
\ion{Si}{2}&&&&&$ 14.13 \pm 0.09$\\
\ion{Si}{2}*&&&&$< 11.93$\\
&1264.7377 &$ -0.0441$&$[  37, 197]$&$< 11.93$&\\
\ion{Si}{3}&&&&$> 13.66$\\
&1206.5000 &$  0.2201$&$[  37, 217]$&$> 13.66$&\\
\ion{Si}{4}&&&&$ 12.93 \pm 0.03$\\
&1393.7550 &$ -0.2774$&$[  57, 187]$&$ 12.90 \pm 0.03$&\\
&1402.7700 &$ -0.5817$&$[  57, 187]$&$ 13.03 \pm 0.05$&\\
\ion{Si}{6}&&&&&$< 11.93$\\
\ion{Fe}{2}&&&&&$ 13.32 \pm 0.09$\\
\ion{Fe}{3}&&&&$< 13.37$\\
&1122.5260 &$ -1.2684$&$[  67, 167]$&$< 13.37$&\\
\cutinhead{B}
\ion{C}{2}&&&&$> 15.24$\\
&1036.3367 &$ -0.9097$&$[ 249, 499]$&$> 15.24$&\\
\ion{C}{4}&&&&$ 14.09 \pm 0.03$\\
&1548.1950 &$ -0.7194$&$[ 289, 499]$&$ 14.11 \pm 0.03$&\\
&1550.7700 &$ -1.0213$&$[ 289, 499]$&$ 14.06 \pm 0.03$&\\
\ion{N}{1}&&&&$ 14.00 \pm 0.03$\\
&1199.5496 &$ -0.8794$&$[ 249, 429]$&$ 14.00 \pm 0.03$&\\
\ion{N}{2}&&&&$> 15.10$\\
&1083.9900 &$ -0.9867$&$[ 249, 499]$&$> 15.10$&\\
\ion{N}{5}&&&&$< 13.81$\\
&1242.8040 &$ -1.1066$&$[ 289, 499]$&$< 13.81$&\\
\ion{O}{1}&&&&$ 15.61 \pm 0.03$\\
&1039.2304 &$ -2.0364$&$[ 289, 499]$&$ 15.61 \pm 0.03$&\\
&1302.1685 &$ -1.3110$&$[ 349, 449]$&$> 15.03$&\\
\ion{O}{6}&&&&$< 14.58$\\
&1031.9261 &$ -0.8765$&$[ 289, 499]$&$< 14.58$&\\
\ion{Al}{2}&&&&$ 13.32 \pm 0.03$\\
&1670.7874 &$  0.2742$&$[ 249, 499]$&$ 13.32 \pm 0.03$&\\
\ion{Al}{3}&&&&$ 12.68 \pm 0.08$\\
&1854.7164 &$ -0.2684$&$[ 349, 499]$&$ 12.68 \pm 0.08$&\\
\ion{Si}{2}&&&&$ 14.60 \pm 0.03$\\
&1264.7377 &$ -0.0441$&$[ 249, 509]$&$< 12.04$&\\
&1304.3702 &$ -1.0269$&$[ 249, 499]$&$ 14.64 \pm 0.03$&\\
&1526.7066 &$ -0.8962$&$[ 249, 499]$&$ 14.58 \pm 0.03$&\\
&1808.0130 &$ -2.6603$&$[ 299, 399]$&$< 14.65$&\\
\ion{Si}{2}*&&&&$< 12.04$\\
&1264.7377 &$ -0.0441$&$[ 249, 509]$&$< 12.04$&\\
\ion{Si}{3}&&&&$> 14.02$\\
&1206.5000 &$  0.2201$&$[ 249, 509]$&$> 14.02$&\\
\ion{Si}{4}&&&&$ 13.50 \pm 0.03$\\
&1393.7550 &$ -0.2774$&$[ 289, 499]$&$ 13.50 \pm 0.03$&\\
&1402.7700 &$ -0.5817$&$[ 289, 499]$&$ 13.50 \pm 0.03$&\\
\ion{Si}{6}&&&&&$< 12.04$\\
\ion{Fe}{2}&&&&$ 13.87 \pm 0.04$\\
&1608.4511 &$ -1.2366$&$[ 299, 449]$&$ 13.87 \pm 0.04$&\\
\ion{Fe}{3}&&&&$< 13.35$\\
&1122.5260 &$ -1.2684$&$[ 299, 399]$&$< 13.35$&\\
\enddata
\tablenotetext{a}{Velocity interval for the AODM
column density measurement.
Velocities are relative to the redshift of the foreground quasar \fgqso, $z_{fg}=2.4360$.}
\tablenotetext{b}{Sum of all fitted components.}
\end{deluxetable}

\begin{deluxetable}{lcccccccc}
\tablenum{2}
\tablewidth{0pc}
\tabletypesize{\scriptsize}
\tablehead{\colhead{Ion} & \colhead{$\lambda_{\rm rest}$} & \colhead{$\log f$}
& \colhead{$v_{int}^a$} 
& \colhead{$\log N_{\rm AODM}$} & \colhead{$\log N_{\rm VPFIT}^b$} \\
& (\AA) & & (\kms) &  }
\startdata
\cutinhead{C}
\ion{C}{2}&&&&$> 14.79$\\
&1036.3367 &$ -0.9097$&$[ 598, 698]$&$> 14.79$&\\
\ion{C}{2}*&&&&&$ 13.61 \pm 0.04$\\
\ion{C}{4}&&&&$ 13.66 \pm 0.03$\\
&1548.1950 &$ -0.7194$&$[ 568, 728]$&$ 13.71 \pm 0.03$&\\
&1550.7700 &$ -1.0213$&$[ 568, 728]$&$ 13.48 \pm 0.06$&\\
\ion{N}{1}&&&&&$ 14.47 \pm 0.04$\\
\ion{N}{2}&&&&$> 14.80$\\
&1083.9900 &$ -0.9867$&$[ 598, 728]$&$> 14.80$&\\
\ion{N}{5}&&&&$< 12.58$\\
&1242.8040 &$ -1.1066$&$[ 618, 718]$&$< 12.58$&\\
\ion{O}{1}&&&&&$ 15.46 \pm 0.09$\\
\ion{O}{6}&&&&$< 13.70$\\
&1031.9261 &$ -0.8765$&$[ 678, 818]$&$< 13.70$&\\
\ion{Al}{2}&&&&&$ 12.89 \pm 0.05$\\
\ion{Al}{3}&&&&$< 12.28$\\
&1854.7164 &$ -0.2684$&$[ 648, 718]$&$< 12.28$&\\
\ion{Si}{2}&&&&&$ 14.31 \pm 0.06$\\
\ion{Si}{2}*&&&&$< 11.82$\\
&1264.7377 &$ -0.0441$&$[ 628, 728]$&$< 11.82$&\\
\ion{Si}{3}&&&&$> 13.59$\\
&1206.5000 &$  0.2201$&$[ 628, 728]$&$> 13.59$&\\
\ion{Si}{4}&&&&&$ 12.93 \pm 0.39$\\
\ion{Si}{6}&&&&&$< 11.82$\\
\ion{Fe}{2}&&&&&$ 13.41 \pm 0.09$\\
\ion{Fe}{3}&&&&$< 13.37$\\
&1122.5260 &$ -1.2684$&$[ 678, 778]$&$< 13.37$&\\
\enddata
\tablenotetext{a}{Velocity interval for the AODM
column density measurement.
Velocities are relative to the redshift of the foreground quasar \fgqso, $z_{fg}=2.4360$.}
\tablenotetext{b}{Sum of all fitted components.}
\end{deluxetable}

In Figure~\ref{fig:metals}
we present a sub-set of the metal-line transitions observed along the
sightline near the redshift of the foreground quasar.  All velocities
are relative to the redshift of \fgqso, $z_{\rm fg} = 2.4360$. 
The various ions observed near this redshift show absorption 
in roughly three distinct velocity intervals 
spanning a total interval $\Delta v \approx 650 \mkms$.  We define
these three velocity intervals as `subsystems',
A: $+50\mkms < \delta v <+200\mkms$; 
B: $+200\mkms<\delta v<+600\mkms$; and 
C: $+600\mkms<\delta v<+750\mkms$.
We have measured column densities for the ions observed 
in these subsystems with two methods:
(i) by fitting Voigt-profiles to the data using the VPFIT software
package (kindly provided by R. Carswell); and
(ii) by integrating the apparent optical depth profile
\citep[AODM;][]{savage91}.
The Voigt-profile fitting is most reliably performed on unsaturated
transitions of ions which show similar velocity structure.
In this case, one can `tie' the redshifts of the components
which comprise the velocity profile
for all of the ions analyzed and only
allow the column densities and Doppler parameters to vary.
The VPFIT software package minimizes the $\chi^2$ of the profile
fits and reports a best value of the redshift, $b$-value,
and column density for each component introduced by the user.

We were able to employ the Voigt-profile procedure on the majority of low and
intermediate ions for subsystems A and C.
With the exception of \ion{Si}{4} in subsystem C\footnote{ 
Even in this case, a reasonable fit required a substantially different
Doppler parameter.},  however, 
we could not recover a consistent solution assuming the 
same component
structure for the low and high-ions.   Therefore, we calculate
the high-ion column densities using the AODM method integrated
across the entire velocity interval of each subsystem.
We conclude that the high-ion gas arises in a distinct
phase with a unique velocity field.  In the ionization modeling
that follows ($\S$~\ref{sec:ion}), one should keep in mind that
the subsystems are likely multi-phase absorbers.  
This means the resulting high-ion to low-ion ratios 
should be considered upper limits with respect to the nature
of the lowest ionization phase. 
The results of the line-profile fits are overplotted in 
Figure~\ref{fig:metals} and tabulated in 
Table~\ref{tab:vpfit}.

The velocity profiles of subsystem B are more complicated
than subsystems A and C.
This complex absorption precludes
a well constrained line-profile solution based
on individual Voigt profiles.
Furthermore, it is apparent from a visual inspection of the \ion{Si}{2} and
\ion{C}{4} transitions that the ratios of the low-ion to
high-ions columns vary significantly from $\delta v=+300$ to +500\kms.
Therefore, subsystem B is comprised of gas in at least two
different ionization phases.  This includes the most highly
ionized gas on the sightline (at $\delta v\approx +430\mkms$), where one
observes strong \ion{C}{4} and \ion{Si}{4} absorption but very
little \ion{O}{1} or \ion{Si}{2} gas.  For subsystem B, 
therefore, we report total column densities based on the AODM
and caution that an analysis of the gas properties must consider
multi-phase material.

\section{Ionization Modeling and Physical Conditions}
\label{sec:ion}

In this section we constrain the ionization state of the
gas in the three subsystems identified with the super-LLS
identified at $z \approx 2.4360$ toward \bgqso, named \slls.
One goal is to test the hypothesis that 
the nearby quasar \fgqso\
is shining on the gas.  A test of this hypothesis is to compare
the intensity of the ionizing radiation field
that reproduces the observed ionic ratios with the predicted
flux of the quasar at the impact parameter of the sightline.
We constrain the intensity by comparing observed ionic ratios
with photoionization models.  
Another goal is to constrain the chemical abundances of the gas.  
These measurements reflect the 
integrated star formation history of the galaxy that hosts
(or hosted)
the absorber and, in turn, star-formation in this quasar environment.
Standard practice is to gauge
the contributions of Type\,II nucleosynthesis relative to 
Type\,I abundances from estimates of the $\alpha$/Fe ratio
where $\alpha$-elements include O, Mg, Si, and S. 
The data also constrain the N abundance
which is expected to trace nucleosynthesis by 
intermediate mass stars and therefore offers unique constraints
on the timescales of star formation \citep{hek00,henrypro07}.
This requires an assessment of the ionization
state to estimate corrections to the observed ionic ratios.

The analysis focuses on
multiple ionization states
of individual elements (e.g.\ N$^0$/N$^+$, Si$^+$/Si$^{+3}$)
to avoid uncertainties related to intrinsic abundances.
For an optically thick absorber, one expects absorption
from ions with ionization potentials (IPs) of 
one to a few Ryd.
Ions with lower ionization potentials should be 
absent because \ion{H}{1} gas is essentially
transparent to photons with energies less than 1\,Ryd.
Photons with energies above a few Ryd will be significantly
attenuated by the \ion{H}{1} gas and for $h\nu > 4$\,Ryd
by helium.
Our analysis is most sensitive to the shape and
intensity of the radiation field at 1 to 4\,Ryd, although
the nature
of the radiation field at higher energies is constrained
by observations of \ion{N}{5} and \ion{O}{6} transitions:
IP(N$^{+4}$) = 77.4\,eV and IP(O$^{+5}$) = 113.9\,eV.

Finally, we will constrain the electron density $n_{\rm e}$ of the
gas by comparing the relative populations of the $J=3/2$ and
$J=1/2$ fine-structure levels of C$^+$ and Si$^+$ ions
\citep[e.g.][]{pro99,silva02}.
We will argue the gas is partially ionized and
therefore assume collisions by electrons\footnote{The
impact parameter of \bgqso\ from \fgqso\ is sufficiently
large that indirect UV pumping by \fgqso\ does
not contribute to the excitation of these lines.}
dominate excitation to the $J=3/2$ upper level.
With an estimate of the gas temperature one estimates $n_{\rm e}$
by measuring the relative populations of the ground and
excited states.

In the following sub-sections,
we start with the simpler
subsystems A and C before considering the
more complex subsystem B.

\subsection{Subsystem A: $+50\mkms<\delta v<+200\mkms$}

Before delving into detailed photoionization models, one
can build intuition by making qualitative inspection of
the ions detected.
Examining subsystem A in Figure~\ref{fig:metals}, one
notes strong absorption from a series of low-ions, e.g.\
O$^0$, Si$^+$, N$^0$, Fe$^+$.
This is characteristic of Lyman limit systems where the large
\ion{H}{1} opacity self-shields gas from local or background
UV sources.  
Elements therefore occupy the first ionization state (termed 
the low-ion) with an ionization potential greater than 1\,Ryd.
One also identifies
strong \ion{N}{2} absorption that traces the observed \ion{N}{1} profile
in subsystem~A.  The measured ratio of these ionization states
is large, $\log[\N{N^+}/\N{N^0}] = +1.8$\,dex, revealing the
gas is partially ionized.  On the other hand,
there is only weak \ion{Si}{4} and \ion{C}{4} absorption at
these velocities indicating the ionization state of the gas 
is not extreme.
Furthermore, the absence of strong \ion{N}{5} and \ion{O}{6} absorption
at these velocities limits the flux of photons with 
energies $h\nu \gtrsim 4$\,Ryd and also rules out
a collisionally ionized gas with $T \approx 10^5$\,K (see Figure~\ref{fig:cie}).
Qualitatively, the data suggest a partially ionized gas with
$T < 10^5$\,K.

There are two main mechanisms that produce a partially ionized gas: 
collisional ionization and photoionization.
Under the assumption of collisional ionization equilibrium 
\citep[CIE;][]{sd93}, the detection of 
low-ions like O$^0$ and N$^0$
requires the electron temperature to be $T_e < 30,000$K. 
This constraint is supported by the small $b$-values measured for 
the N$^0$ gas at $v\approx +130\mkms$: $b({\rm N^0}) = 4.5 \pm 0.3\mkms$
(Table~\ref{tab:vpfit}). If we assume the gas is broadened by purely 
thermal motions,  we can set a 2-$\sigma$ upper limit to 
the temperature $T_e<25,000$K.
We achieve a similar limit by fitting the N gas and Si gas independently
at the same redshift
and use the difference in Doppler parameters to estimate $T$.
Nevertheless, a CIE model with $T_e \approx 25,000$K does
roughly reproduce 
the observed ionic ratios of Si$^+$/Si$^{+3}$, 
Al$^+$/Al$^{++}$, and Fe$^+$/Fe$^{++}$ with N$^+$/N$^0$ discrepant 
(the model underpredicts this ratio).
Therefore, one cannot rule out a model where collisional 
ionization is the primary mechanism producing the observed ionic ratios,
although to invoke this scenario one must introduce a heat source
to maintain the gas at this temperature because the cooling time
is short ($t_{\rm cool} \lesssim 10^4$\,yr) for any reasonable density. 
In the following photoionization modelling, we will however assume 
that photoionization is dominant and thereby set an 
upper limit to the intensity of the ionizing radiation field.
Finally, because the gas is partially ionized one presumes that
$T_{\rm e} > 10,000$K.  In the following and throughout the remainder of the
paper we will adopt an electron temperature $T_{\rm e} = 20,000$K.

To explore photoionization,
we have used the Cloudy software package \citep{ferlandetal98}
to calculate the ionization state of plane-parallel slabs
with total column density $\log \mnhi = 18.6$, density $n_{\rm H} = 0.1\cm{-3}$,
and metallicity [M/H]$=-0.5$, while varying the ionization parameter
$U \equiv \Phi/n_{\rm H} c$ where $\Phi$ is the flux of ionizing photons 
having $h\nu \ge 1$\,Ryd.  
The results are largely insensitive to this choice of volume density
(they are nearly homologous with $U$),
but they do vary with metallicity because this affects the
cooling rate of the gas and the electron density.  
We will demonstrate that the assumed metallicity is consistent 
with the [O/H] value inferred from the observed O$^0$/H$^0$ ratio.
For the calculations, 
we assume a power-law spectrum $f_\nu \propto \nu^{-1.57}$ which is 
representative of $z \sim 2$ quasars \citep{telfer02}.  
The results are nearly identical if we instead
use the extragalactic UV background (EUVB) 
field computed by \cite{hm96} because the radiation fields have
very similar slopes at the energies that  
span the ionization potentials of the observed ions ($h\nu = 1- 3$\,Ryd).
When the input $U$ parameter is varied,
the algorithm varies the number of gas slabs 
to maintain a constant \nhi\ value.
Because subsystem A is optically thick, the results are
sensitive to the assumed \nhi\ value; 
unfortunately, this quantity is not well
constrained by the observations ($\S$~\ref{sec:obs}). 
Larger \nhi\ values imply more self-shielding of the inner
regions which, in turn, demand a more intense radiation field
to explain the observed ionization states.
The various ions respond differently to the shielding 
effects and pose an independent constraint on the \nhi\ value
under the assumptions of this photoionization model.
For example, we cannot reproduce the 
observed Fe$^+$/Fe$^{++}$, Si$^+$/Si$^{+3}$, and N$^0$/N$^+$ ratios
when $\mnhi > 10^{19} \cm{-2}$; 
the N$^0$/N$^+$ ratios require $\log U = -2$
while the other two ratios set an upper limit $\log U < -3$.
Under the assumption of a single-phase photoionization
model with $f_\nu \propto \nu^{-1.57}$, the data requires
$\mnhi < 10^{18.75} \cm{-2}$ with a preferred value of
$\mnhi \approx 10^{18.6} \cm{-2}$.  
This central value is consistent with the line-profile analysis of
\lya\ and \lyb\ for subsystem~A (Figure~\ref{fig:HI}).

\begin{figure}
\begin{center}
\includegraphics[height=6.8in,angle=90]{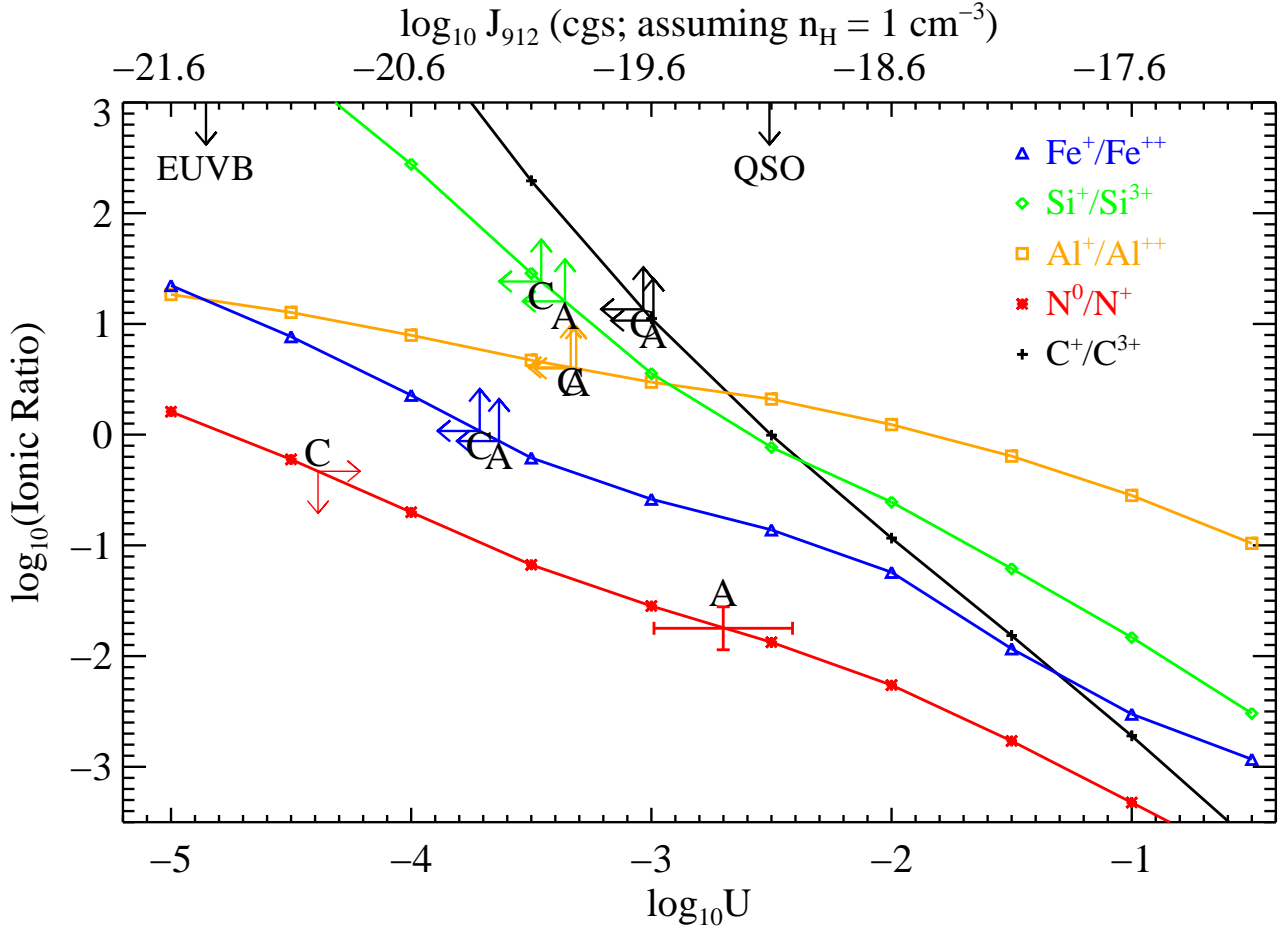}
\end{center}
\caption{
The solid curves show predicted ionic ratios as a function of the
ionization parameter $U$ for a series of ion pairs.  These curves
were calculated using the Cloudy software package assuming a plane-parallel
slab of constant density material, a $f_\nu \propto \nu^{-1.57}$ power-law
spectrum, and solar relative abundances with an absolute metallicity
[M/H]~$=-0.5$\,dex.  Overplotted on the curves are observational
constraints for subsystems A and C.  With the exception of
N$^0$/N$^+$ in subsystem~A, 
the observations indicate $\log U < -3$ (see the text for a full
discussion).  
Note that
the constraints for C and Si are reported as upper limits to $U$
because the high-ion absorption (C$^{+3}$, Si$^{+3}$) is likely
associated with a different phase.
The intensities for the extragalactic UV background (EUVB) and
the foreground quasar at a distance of 100\,kpc are shwon on the 
top axis.
}
\label{fig:cldyA}
\end{figure}

Figure~\ref{fig:cldyA} presents the results of the
Cloudy calculations for a series of ionic ratios
for a wide range of ionization parameters
assuming $\mnhi = 10^{18.6} \cm{-2}$.
As noted above, we examine multiple ionization states of the same
element (e.g.\ N$^0$/N$^+$) to provide constraints that are
independent of intrinsic abundances.
The observational constraints (Table~\ref{tab:colm}) on the ionic
ratios are indicated by vertical error intervals (or lower and upper limits)
and the corresponding horiontal error intervals (or lower and upper limits) on 
 $\log U$ are indicated. 
One notes that even with this lower \nhi\ value,
the constraints are not fully consistent with a single $U$ value;
in particular the observed N$^0$/N$^+$ ratio indicates
$\log U \gtrsim -3$ whereas the other ratios imply 
$\log U < -3$.
Although this inconsistency may suggest the low and
intermediate ions arise in a multi-phase or non-equilibrium
medium, we caution that the systematic uncertainties
inherent to photoionization modelling of optically thick absorbers 
are large
(e.g.\ the assumed spectral shape, cloud geometry, and uncertain atomic data).
In the following we will adopt an ionization
parameter $\log U = -3.0$\,dex with an uncertainty
of 0.3\,dex.
This $U$ value is similar to the results derived for other LLS at
these redshifts \citep[e.g.][]{pro99}.  It implies
an ionization fraction $x \equiv {\rm H^+/H} = 0.96$ and
a total hydrogen column density 
$N_H \equiv \mnhi/(1-x) =  10^{20.0} \cm{-2} (\mnhi/10^{18.6} \cm{-2})$.
Note that the error in $1-x$ is roughly linear with the
uncertainty in $U$ for $\log U > -4$\,dex.

\begin{figure}
\begin{center}
\includegraphics[height=6.8in,angle=90]{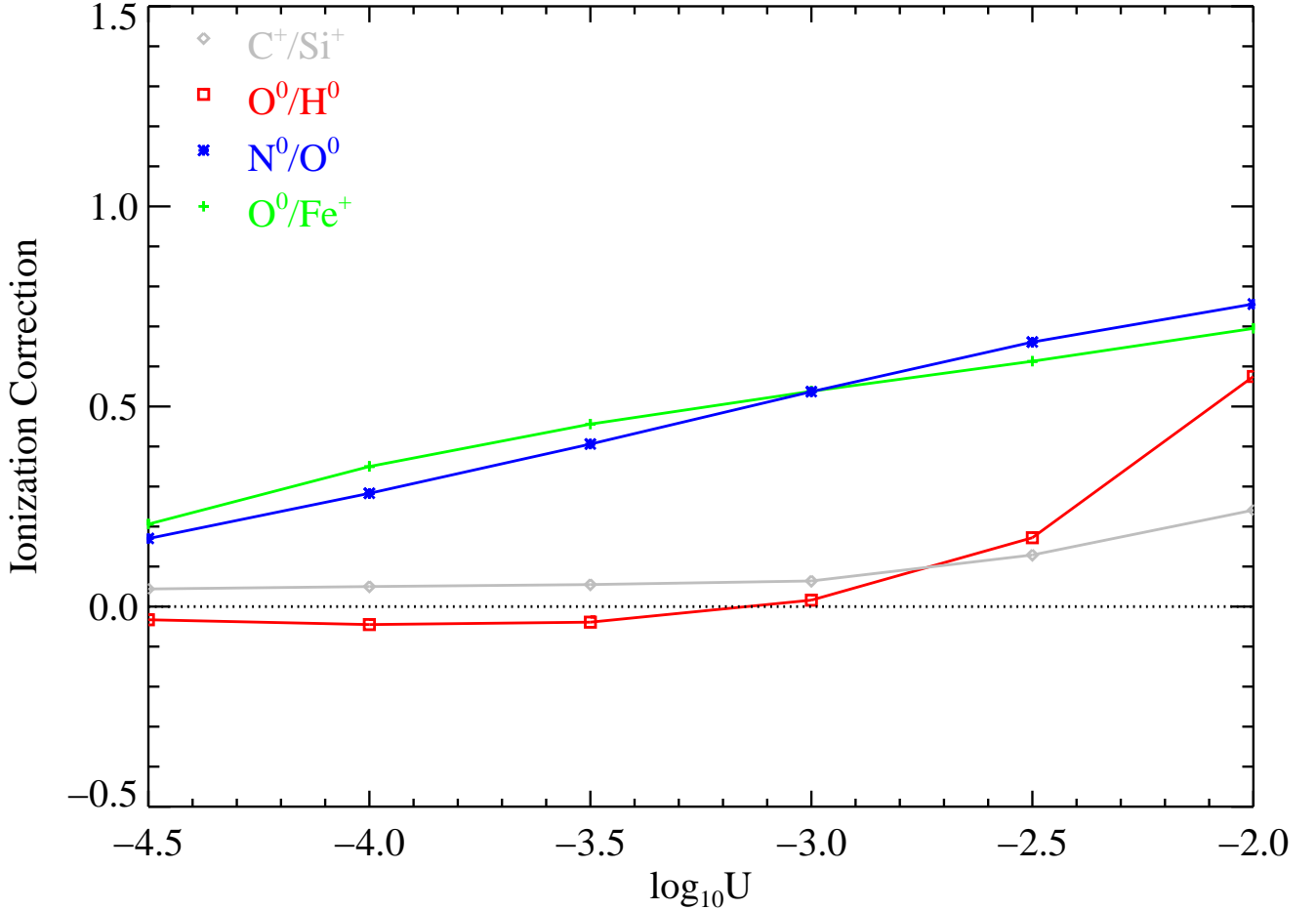}
\end{center}
\caption{
Estimated ionization corrections (I.C.) for several ionic
ratios as a function of ionization parameter $U$.  These values
represent the offset that must be applied to an observed ionic
ratio to give the relative elemental abundances, 
e.g.\ log(N/O) = log(N$^0$/H$^0$) + I.C.$_{(N/O)}$.
The calculations were performed with the Cloudy software
package and assume $\mnhi = 10^{19} \cm{-2}$, 
a plane-parallel constant density slab with $n_{\rm H} = 10^{-1} \cm{-3}$,
a gas metallicity [M/H]~$=-0.5$\,dex with solar relative abundances,
and a power-law ionizing spectrum with $f_\nu \propto \nu^{-1.57}$.
Note that the corrections for O$^0$/H$^0$ are small for 
$\log U < -2.5$\,dex whereas the N$^+$/O$^0$ and O$^0$/Fe$^+$ ratios
require modest corrections for $\log U > -4.5$\,dex.
}
\label{fig:ioncorr}
\end{figure}

At low ionization parameters ($\log U < -2.5$; Figure~\ref{fig:ioncorr}), 
the O$^0$/H$^0$
ratio is a good estimator of the oxygen 
abundance\footnote{Note that the same is true for any 
CIE model with $T_e < 10^{4.5}$\,K.} because the two atoms have
very similar ionization potential and also because charge-exchange
reactions mediate their ionization fractions.
We calculate [O/H]~$\equiv \log({\rm O/H}) - \log({\rm O/H})_\odot
= -0.65 - \log_{10}(\mnhi/10^{18.6} \cm{-2})$
assuming a solar oxygen abundance $\log({\rm O/H})_\odot = -3.34$\,dex.
The uncertainty in this measurement (and any other chemical 
abundance measurements of subsystem~A) is 
dominated by the poor constraint on the \nhi\ value.
The \lya\ and \lyb\ profiles indicate $\mnhi < 10^{19} \cm{-2}$ which
sets a lower limit to the metallicity of 1/10 solar abundance.

\begin{deluxetable}{lccc}
\tablewidth{0pc}
\tablecaption{ELEMENTAL ABUNDANCES\label{tab:xh}}
\tabletypesize{\footnotesize}
\tablehead{\colhead{Ion} &
\colhead{[X/H]} & \colhead{[X/O$^{0}$]}}
\startdata
\cutinhead{Subsystem A$^a$}
C$^{+}$ & $>-1.27$ & $>-0.65$ \\
N$^{0}$ & $-0.65\pm0.14$ & $-0.03\pm0.08$ \\
O$^{0}$ & $-0.62\pm0.06$ & $$ \\
Al$^{+}$ & $-1.02\pm0.19$ & $-0.40\pm0.25$ \\
Si$^{+}$ & $-0.95\pm0.11$ & $-0.33\pm0.17$ \\
Fe$^{+}$ & $-1.26\pm0.03$ & $-0.64\pm0.05$ \\
\cutinhead{Subsystem C$^a$}
C$^{+}$ & $>-1.10$ & $>-1.32$ \\
N$^{0}$ & $ 0.65\pm0.14$ & $ 0.43\pm0.08$ \\
O$^{0}$ & $ 0.22\pm0.06$ & $$ \\
Al$^{+}$ & $-1.14\pm0.19$ & $-1.36\pm0.25$ \\
Si$^{+}$ & $-0.77\pm0.11$ & $-0.98\pm0.17$ \\
Fe$^{+}$ & $-1.16\pm0.03$ & $-1.38\pm0.05$ \\
\enddata
\tablenotetext{a}{Assumes $\mnhi = 10^{18.6} \cm{-2}$ and log $U = -3.0\pm 0.3$\,dex}
\end{deluxetable}

Relative abundance ratios (e.g.\ O/Fe)
are less sensitive to the \ion{H}{1} column density
but are not strictly independent of \nhi\ because it dominates
the opacity in our ionization models.  Adopting $\log U = -3$
and $\log \mnhi = 10^{18.6} \cm{-2}$, we find the following
relative abundance values: [O/Fe] =$+0.6 \pm 0.1$, [O/C]~$<+0.75$, 
and [N/O]=$-0.1 \pm 0.2$\,dex (Table~\ref{tab:xh}).
The reported errors are dominated by uncertainty in the $U$ parameter.
We interpret these chemical abundance measurements in 
$\S$~\ref{sec:chem}.

Independent of photoionization modeling, the data yield
a measurement of the electron density from the observed 
populations of the fine-structure levels of Si$^+$ and/or C$^+$.  
The $J=1/2$ ground-state of these ions has a corresponding
$J=3/2$ fine-structure level which can be populated by collisions
with ions and atoms, via indirect UV pumping, and also via
a magnetic-dipole transition.
For a partially ionized gas (demanded by the observed N$^+$/N$^0$ ratio),
collisions with free electrons should dominate excitation
to the $J=3/2$ upper level.   Indirect UV pumping could also
contribute, but the quasars (foreground and background)
are too distant and we assume that there are
no important local sources (e.g.\ OB stars).  
In any case, contributions from UV pumping would only
tighten the following upper limit on $n_{\rm e}$.
Unfortunately, 
the transitions from the $J=3/2$ excited level of C$^+$
are blended with the resonant \ion{C}{2} 
transitions of subsystem~B.
Therefore, only the Si$^+$ levels (which have a wider energy separation)
are available for analysis.  We set an upper limit to the relative populations
based on the non-detection of the \ion{Si}{2}*~1264 transition, 
$\log[\N{Si^+_{J=3/2}}]/[\N{Si^+_{J=1/2}}] < -2.2$\,dex.
For a gas with $T_{\rm e} \gg 413$\,K, the level populations for
excitation dominated by electron collisions is given by:

\begin{equation}
\frac{\N{Si^+_{J=1/2}}}{\N{Si^+_{J=3/2}}} = 
\frac{1}{2} + \frac{640 T_4^{1/2}}{n_{\rm e}} \;\;\; (T \gg 413{\rm K})
\;\;\; .
\label{eqn:siii}
\end{equation}
Adopting $T_{\rm e}=20,000$K, 
the observations place an upper limit on the electron density,
$n_{\rm e} < 6 \cm{-3}$.

One can use this constraint on the electron density
to constrain the hydrogen volume density $n_{\rm H}$ by
assuming the ionization fraction from our
photoionization model, i.e., 
$x = 0.96$.  We calculate 

\begin{equation}
n_{\rm H} = n_{\rm e}/x < 6.2 \cm{-3} \;\;\; .
\label{eqn:nH}
\end{equation}
One can also estimate a lower limit for the
characteristic size of the system,

\begin{equation}
\ell \equiv \frac{x}{1-x} \frac{\mnhi}{n_{\rm e}} \;\;\; .
\label{eqn:ell}
\end{equation}
We derive $\ell > 5.2 (\mnhi/10^{18.6} \cm{-2})$\,pc.

\subsection{Subsystem C: $+600\mkms<\delta v<+780\mkms$}

In comparison with subsystem A, we find similar results
for the ionization state of subsystem C.
These include strong \ion{N}{2} absorption
relative to \ion{N}{1} which indicates a partially ionized gas.
Again, a CIE model with $T_e \approx 20,000$\,K
yields ionic ratios consistent with the majority of
observed ionic ratios,
but we will focus on photoionization models of the gas.
Like subsystem A, we also find that models with $\mnhi < 10^{19} \cm{-2}$
are preferred and the observed ionic ratios 
(Fe$^+$/Fe$^{++}$, N$^0$/N$^+$, Si$^+$/Si$^{+3}$) imply
$\mnhi \approx 10^{18.6}\cm{-2}$ and $\log U \lesssim -3$.
Using the same Cloudy model assumed for subsystem~A,
we derive the chemical abundances listed in Table~\ref{tab:xh}.

In contrast to subsystem A, we measure larger metal
column densities and a correspondingly higher oxygen
abundance, [O/H]~$= +0.2 - \log_{10}(\mnhi/10^{18.6}\cm{-2})$,
if the subsystems have comparable \nhi\ values.
Even if we have underestimated the \ion{H}{1} column density
of subsystem~C
by a factor of 0.5\,dex (the maximum value allowed by the 
observed \lya\ profile; Figure~\ref{fig:HI}),
the gas metallicity would be roughly solar.  This lower limit
([O/H]~$>-0.25$) matches the highest values observed
for damped \lya\ systems \citep{pwh+07}
and all but a few super-LLS studied to date \citep{pkm06,poh+06}. 
The data also suggest variations between the oxygen abundances
of subsystems~A and C (and also B, see below).  
Abundance variations like these are rarely studied 
in optically thick absorption line systems because it is 
difficult to
resolve the \ion{H}{1} Lyman series \citep{ppb+08}.

The relative chemical abundances of O/Fe and N/O are also
significantly higher in subsystem~C than subsystem~A:
[O/Fe]~$= +1.3 \pm 0.1$\,dex, [N/O]=~$+0.3 \pm 0.2$\,dex.
Although these values include ionization corrections
(Figure~\ref{fig:ioncorr}), the gas is indisputably
$\alpha$-enhanced and the N/O ratio must be greater than one half solar.
This N/O value is unique for optically thick absorbers and
the large $\alpha$-enhancement also 
exceeds most measurements of O/Fe
in extragalactic environments.
We interpret these results further in $\S$~\ref{sec:chem}.

We derive a constraint on the electron density of
subsystem C from observations of fine-structure levels 
of C$^+$ and Si$^+$.
Assuming that excitation of the upper level is
dominated by electron collisions, 
the populations of the excited state to the ground state
for C$^+$ are related to the gas temperature and electron density, e.g.,

\begin{equation}
\frac{\N{C^+_{J=1/2}}}{\N{C^+_{J=3/2}}} = 
\frac{1}{2} + \frac{37 T_4^{1/2}}{n_{\rm e}} \;\;\; (T \gg 92{\rm K}) \;\; .
\label{eqn:cii}
\end{equation}
As for subsystem~A, we assume $T_{\rm e}=20,000$K, and obtain

\begin{equation}
n_{\rm e} = \frac{106}{2 \frac{\N{C^+_{J=1/2}}}{\N{C^+_{J=3/2}}} - 1} \;\;\; .
\end{equation}
Although the column density of the C$^+_{J=3/2}$ excited state is well
constrained by the spectra, the resonance \ion{C}{2}~1036 and \ion{C}{2}~1334
transitions are heavily saturated.  A conservative lower limit to 
$\N{C^+_{J=1/2}}$ is derived by integrating the optical depth
profile assuming a normalized flux equal to the $1 \sigma$ error estimate,
$\N{C^+_{J=1/2}} > 10^{14.8} \cm{-2}$.  
Combined with our measurement of $\N{C^+_{J=3/2}}$, we set
an upper limit to the electron density $n_{\rm e} < 3.5 \cm{-3}$.
Similar to subsystem~A, we also derive an upper limit to $n_{\rm e}$ from the
non-detection of the Si$^+$ excited level,
$n_{\rm e} < 2.5 \cm{-3}$ for $T_{\rm e} = 20,000$K (Equation~\ref{eqn:siii}).
Assuming the same photoionization model ($\log U = -3, x=0.96$)
as for subsystem~A,
we infer a hydrogen volume density $n_{\rm H} < 2.6 \cm{-3}$
and a lower limit to the characteristic size, 
$\ell > 15.5 (\mnhi/10^{18.6} \cm{-2})$\,pc.

Because we have a positive detection for the \ion{C}{2}$^*$~1335
transition, we may further constrain the volume density by
estimating the total C$^+$ column density from a proxy ion.
A good choice is Si$^+$ because our Cloudy calculations indicate
Si$^+$/Si~$\approx \; \rm C^+/C$ for $\log U \approx -3$\,dex.
This requires, however, that one adopt a value for the intrinsic Si/C
ratio.  If we assume solar relative abundances [Si/C]=0, which is
consistent with the C/Si ratio deduced from the ratio of \ion{C}{4}
to \ion{Si}{4} assuming the ionization correction
of our best fit photoioinzation model,
 we estimate
$\log \N{C^+} = 15.2$\,dex.  From equation~\ref{eqn:cii} we
estimate $n_{\rm e} = 1.7 \cm{-3}$ which nearly matches the upper limit
we have derived above.  A more conservative estimate is to assume
[Si/C]~$>-1$ and report a lower limit to the electron density,
$n_{\rm e} > 0.2 \cm{-3}$.

\begin{deluxetable}{cccc}
\tablewidth{0pc}
\tablecaption{SUMMARY OF PROPERTIES FOR \slls\label{tab:summ}}
\tabletypesize{\footnotesize}
\tablehead{\colhead{Property} &\colhead{A} & \colhead{B} & \colhead{C} }
\startdata
log (\nhi/$\cm{-2}$) &$18.6\pm 0.4$&$19.60\pm0.15$&$18.6\pm 0.4$\\
log U &$-3.0\pm 0.3$&$-3.0\pm 0.3$&$-3.0\pm 0.3$\\
$\log T_e$ (K)&$4.3\pm 0.15$&$4.3\pm 0.15$&$4.3\pm 0.15$\\
$\lbrack$O/H]$^a$&$-0.6$&$-0.6$&$+ 0.2$\\
log(O$^0$/Fe$^+$) - log(O/Fe)$_\odot$&$+ 0.1$&$+ 0.5$&$+ 0.8$\\
$\lbrack$O/Fe]$^b$&$+ 0.6$&$+ 0.7$&$+ 1.4$\\
log(N$^0$/O$^0$) - log(N/O)$_\odot$&$-0.6$&$-0.7$&$-0.1$\\
$\lbrack$N/O]$^b$&$-0.0$&$-0.5$&$+ 0.4$\\
$1-x$ &$0.04\pm0.01$&$0.18\pm0.05$&$0.04\pm0.01$\\
$\log (N_{\rm H}/\cm{-2})^c$ &20.00&20.34&20.00\\
$n_e \; (\cm{-3})$ &$<6.0$&$<2.5$&$1.7^{+0.3}_{-1}$\\
$n_H \; (\cm{-3})$ &$<6.2$&$<3.1$&$1.8^{+0.3}_{-1}$\\
$\ell \; ({\rm pc})^c$ &$>  5.2$&$> 23.3$&$ 18.2$\\
\enddata
\tablenotetext{a}{Oxygen metallicity estimated from 
the observed O$^0$/H$^0$ ratio assuming no ionization correction.
Because ionization corrections would increase these values, one may
consider them lower limits to the oxygen abundances subject to
the large uncertainies in $\log \mnhi$ for subsystems A and C.}
\tablenotetext{b}{Abundance derived from low-ion ratios and the ionization corrections calculated assuming $\log U = -3.0 \pm 0.3$\,dex,
$\mnhi$ as listed, [O/H]=~$-0.5$, and $f_\nu = \nu^{-1.57}$.}
\tablenotetext{c}{We estimate a log uncertainty of 0.8\,dex for 
Subsystems A and C where both \nhi\ and $x$ are uncertain and 
0.4\,dex for Subsytem B where the \nhi\ value is better constrained.
A similar consideration holds for $\ell$.}
 
\end{deluxetable}

\subsection{Subsystem B: $+200\mkms<\delta v<+600\mkms$}

As noted in $\S$~\ref{sec:obs}, the velocity profiles of
subsystem~B are complex and preclude a
well-constrained solution using line-profile fitting
techniques.  Our expectation is that the velocity profiles
are comprised of a series of overlapping components each
with a line-profile similar to the components observed
in subsystems~A and C.  With higher S/N data, one might
be able to discern such structure. 
For now, we use the AODM to measure ionic column densities
for this subsystem.  
This approach is problematic, however, for constraining
the ionization state because the ionic ratios vary across
the velocity profiles (e.g.\ N$^+$/N$^0$ is significantly
larger at $\delta v=+470\mkms$ than $\delta v=+350\mkms$).
Therefore, we cautiously report constraints on the
ionization state and chemical abundances.
We examine the integrated ionic column densities
(and their ratios) which give optical depth weighted
averages.

The integrated ionic column density ratios 
(e.g.\ N$^0$/N$^+$, Fe$^+$/Fe$^{++}$) in subsystem~B
are consistent with those observed for subsystems~A and C.
This suggests that the average ionization state of subsystem~B
is similar to that of subsystems~A and C.
Empirically, the low observed Fe$^+$/Fe$^{++}$ and Al$^+$/Al$^{++}$ 
ratios imply a modestly ionized gas with $\log U < -2.5$\,dex,
and the oxygen abundance should follow 
the O$^0$/H$^0$ ratio:
[O/H]~$= -0.64 \pm 0.15$\,dex.  
This value is lower than the O/H abundance derived
for subsystem~C and suggests modest abundance
variations exist between the subsystems.
We also note that the observed O$^0$/Fe$^+$ ratio for
subsystem~B requires at least a modest $\alpha$-enhancement,
[O/Fe]~$>+0.4$\,dex and that the observed N$^0$/O$^0$
ratio implies [N/O]~$>-0.7$\,dex.
These values are similar to the results derived for subsystem~A.
For the remainder of the paper, we will adopt $\log U = -3.0$
for this subsystem.  Because this subsystem has a significantly
higher \nhi\ value, this implies a higher neutral fraction
($1-x = 0.2$) compared to subsystems A and C.

Finally, the non-detection of \ion{Si}{2}*~1264 absorption sets
an upper limit to the electron density $n_{\rm e} < 2.5 \cm{-3}$ under
the assumption that electron collisions dominate the excitation
rate and that $T_{\rm e} = 20,000$\,K (Equation~\ref{eqn:siii}).
If we assume the gas has an ionization fraction near unity, we
set a similar upper limit for the hydrogen volume density and
a size estimate $\ell > 21$\,pc.

\section{Discussion of the Observations}
\label{sec:discuss}

In this section we discuss the observed properties of \slls\ in terms
of the galactic environment around high $z$ quasars.  We examine the
velocity field of this gas, its chemical abundances, and then test
whether \fgqso\ may be shining on the absorbing material.  As our
discussion focuses on a single system, the degree to which we can
generalize these results is questionable. It is nevertheless fruitful
to explore various scenarios for interpreting the observations.
Table~\ref{tab:summ} summarizes the physical characteristics of \slls\
derived in the previous section.

\subsection{Kinematic Constraints}
\label{sec:kinematic}

The most robust measurements for \slls\ from our high resolution
spectrum of \bgqso\ are on the velocity field of the gas.  Both the
absolute (i.e.\ redshift) and relative velocities constrain the origin
of the gas and its relation to the foreground quasar \fgqso.  The
redshift of \fgqso\ was measured from our observations of the
rest-frame optical [\ion{O}{3}]~$\lambda 5007$ emission line using the
GNIRS spectrometer on the Gemini-S telescope ($\S$~\ref{sec:obs},
Figure~\ref{fig:GNIRS}).
We achieved a velocity precision of $\approx 40$\kms\ giving $z_{\rm fg} =
2.4360 \pm 0.0005$.  Figures~\ref{fig:HI} and \ref{fig:metals}, which
present the velocity profiles of \slls\ relative to $z_{\rm fg}$ show
strong \ion{H}{1} and metal-line absorption from $\delta v \approx
+70\mkms$ to +700\,\kms.  These absorption profiles have two notable
characteristics.  First, the gas along the \bgqso\ sightline is
asymmetrically distributed relative to $z_{\rm fg}$, i.e.\ within a
2000\kms\ window centered on $z = z_{\rm fg}$, there is significant
absorption only at $z > z_{\rm fg}$.  The nearest `cloud' with
$\tau_{Ly\alpha} > 1$ and $\delta v < 0 \mkms$ lies off of
Figure~\ref{fig:HI} at $\delta v \approx -1500\mkms$, which
corresponds to $\approx 6 h^{-1}$ Mpc (proper) under the assumption of
Hubble expansion.  Second, the total velocity width is very large,
$\delv \approx 650 \mkms$.  One may compare this value against the
velocity widths measured for systems with comparable \ion{H}{1}
surface density, i.e.\ the damped \lya\ (DLA) and super Lyman limit
systems.  Regarding the former, the velocity width of \slls\ exceeds
greater than $99\%$ of the $\delv$ values for randomly selected DLAs
at $z>2$ \citep{pw97,pw01}.  Regarding the SLLS population
\citep{peroux_slls03}, none have as large of a velocity width.  The
equivalent width of the metal-line transitions (a proxy for velocity
width) is also large compared to systems selected on the basis of
metal-line absorption \citep[e.g.\ the \ion{Mg}{2} systems;][]{ppb06}.
In these respects, {\it the velocity field of the gas near \fgqso\ is
extreme}.

Several factors lead us to conclude that the distance between the gas
in \newline \slls\ and \fgqso\ is comparable to the impact parameter.  Recall
that \slls\ exhibits a multi-component kinematic structure with
unusually large velocity width between the components. If the relative
motions of the clouds and the quasar are to be interpreted as Hubble
flow, their implied line-of-sight separation would be $\approx 2.8$
\,Mpc (for $\Delta v \approx 700\mkms$). The chance alignment of several
galaxies across this large distance is unlikely however, which we can
quantify with clustering arguments.  Quasars arise in highly biased
regions of the Universe. They are observed to cluster strongly on
$\sim 10$\,\hMpc\ scales \citep{croom01,pmn04,croom05,myers07a,shen07}
and the correlation function steepens considerably on smaller scales
$\sim 100$\,\hkpc\ \citep{thesis,hso+06,myers07b,myers08}.  High
redshift quasars also exhibit a large cross-correlation amplitude with
star-forming galaxies \citep{as05,coil06}, and it is possible that
galaxies clustered around the quasar are giving rise to the \lya\ and
metal-line absorption seen in \slls.  Another possibility is that the
absorption in \slls\ is not due to a nearby galaxy, but is rather halo
gas distributed around the quasar with some density profile. Either
scenario results in a clustering pattern around the quasar, which we
measured in \cite{hp07}. The probability $P(<r|\rperp, v_{\rm max})$
that the total distance between the quasar and the absorber is less
than $r$, given a fixed impact parameter $\rperp$ and the knowledge
that absorption occurs within a velocity interval $\pm v_{\rm max}$,
is simply an integral over this quasar-absorber correlation
function.

\begin{figure}
\begin{center}
\includegraphics[width=6.8in]{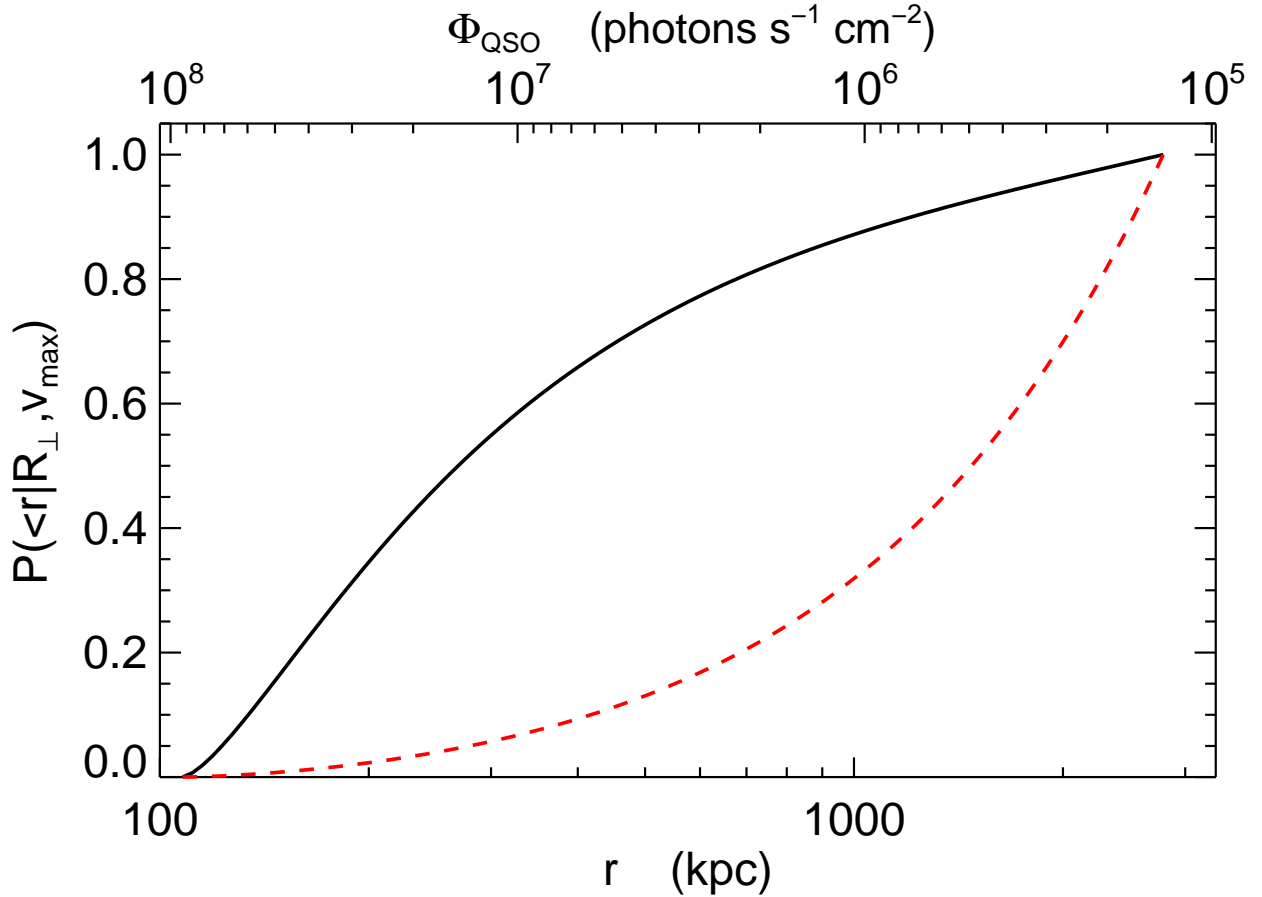}
\end{center}
  \caption{Probability distribution of the distance from the
    foreground quasar. The function $P(<r|R_{\perp}, v_{\rm max})$
    represents the probability that the distance from the foreground
    quasar is less than $r$, given a fixed impact parameter
    $R_{\perp}$ and knowledge that the absorber lies in the velocity
    interval $\pm v_{\rm max}$. The impact parameter for \sdssj\ is
    used and a velocity interval of $v_{\rm max} = 700\mkms$ was
    assumed.  This probability is computed from the quasar absorber
    correlation function $\xi_{\rm QA}=(r\slash r_0)^{-\gamma}$
    measured by \citet{hp07}, with $\gamma=2$ and $r_0 =
    5.8$\,\hMpc. The solid (black) curve illustrates the probability
    distribution accounting for clustering with $\xi_{\rm QA}$. The
  dashed (red) curve shows the probability if there is no
  clustering. The upper x-axis shows the ionizing photon flux
  $\Phi_{\rm QSO}$ implied by the SDSS magnitude of \fgqso\ at the
  given distance.}
  \label{fig:cluster}
\end{figure}

Figure~\ref{fig:cluster} shows this cumulative probability
for $v_{\rm max} = 700\mkms$ and the impact parameter of $\rperp =
108$\,kpc, from which we deduce that the probability for the gas in
\slls\ lying within 500\,kpc of \fgqso\ is $\approx 70\%$. Perhaps an even
stronger argument that the gas in \slls\ lies at a distance comparable
to the impact parameter is that the physical
characteristics (ionization state, \ion{H}{1} surface density,
chemical abundances) of the subsystems comprising \slls\ show only
modest variations. Whereas, several of these same properties ---
specifically the metallicity, relative abundance pattern, and the
velocity field --- are highly anomalous compared to the population of
intervening optically thick absorbers. In particular, the only place
that such large N/O ratios are observed is in quasar
environments. Finally, it is worth emphasizing that the very
small sizes deduced for the subsystems of \slls\ ($d \lesssim
100$\,pc) imply that their relative separations are also comparable to
the impact parameter. The alternative, which would be for the
subsystems to be part of a single structure with size $d \lesssim
100$\,pc, would imply an unphysically large velocity shear across this
small distance.

\begin{figure}
\begin{center}
\includegraphics[width=6.0in]{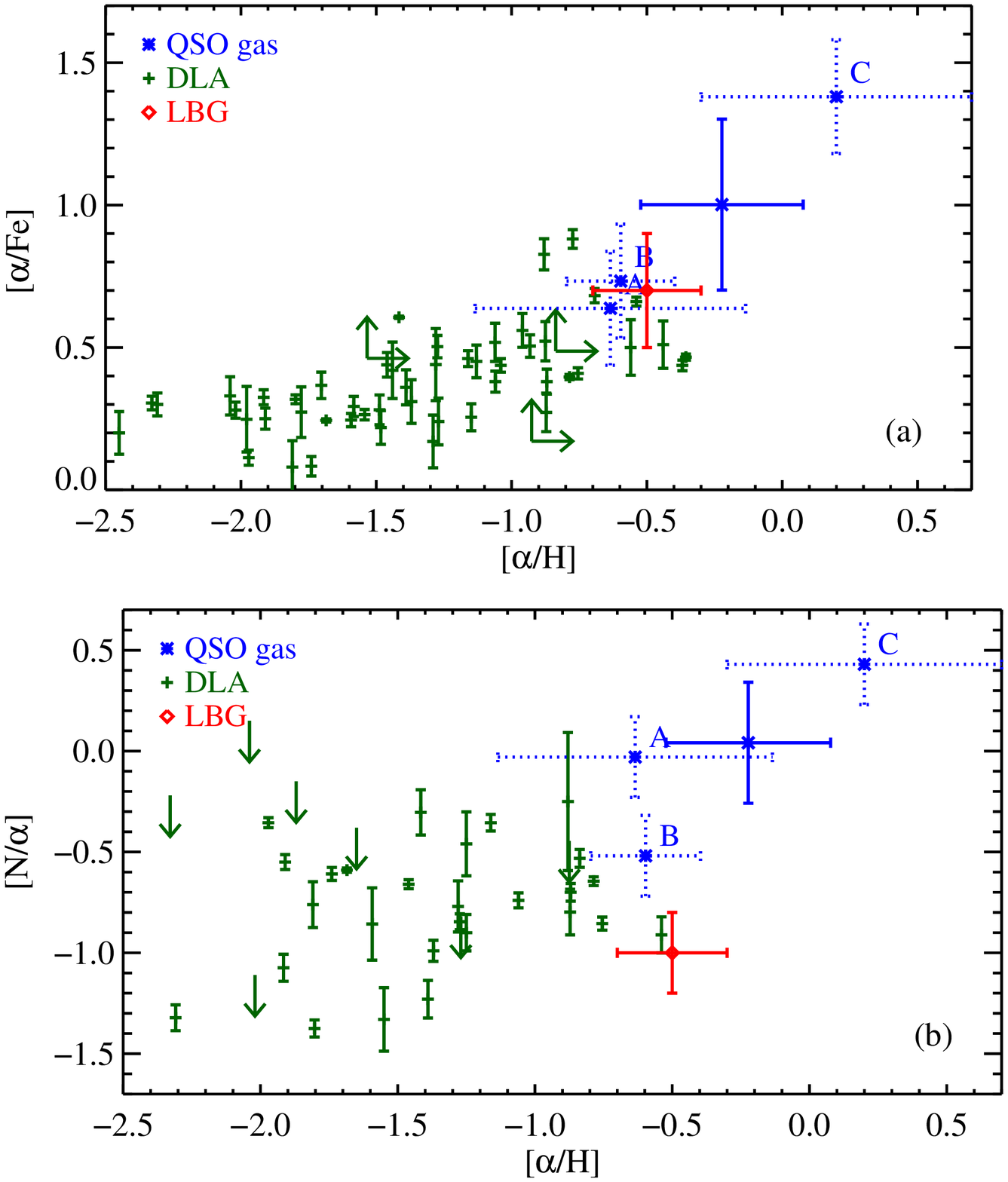}
\end{center}
\caption{
Observed gas-phase (a) $\alpha$/Fe and (b) N/$\alpha$ 
abundances as a function of 
alpha abundance [$\alpha$/H] for galaxies intervening
quasar sightlines \citep[DLAs;][]{pw01,pwh+07}, the three subsystems 
associated with \fgqso\ (blue dotted), and the \nh-weighted average values
(blue solid).  
We note the gas associated with \fgqso\ has
systematically higher metallicity and higher $\alpha$/Fe, N/O ratios 
than the majority (or all) DLAs.  This indicates gas that
was enriched in a `starburst' fashion, i.e.\ a system with a very
high star formation efficiency \citep{cmv03}.
}
\label{fig:relabnd}
\end{figure}

\subsection{Clues from Chemical Enrichment}
\label{sec:chem}

Clues to the origin of \slls\ are encoded in the enrichment pattern of
its gas.  In particular, the relative abundances constrain the
timescales and intensity of star-formation of the galaxy (or galaxies)
that produced the observed metals.  Because the gas is partially
ionized and we do not observe transitions from all ionization states,
we proceed cautiously to interpret the observed ionic ratios.  One
robust conclusion from our abundance analysis ($\S$~\ref{sec:ion};
Table~\ref{tab:summ}) is that the gas is highly enriched.  The
observed O$^0$/H$^0$ ratio indicates [O/H]~$>-1$ across the $\approx
650 \mkms$ velocity interval spanning the entire system and we derive
an integrated average abundance: [O/H]$_{\rm TOT} > -0.5$\,dex.  This
metallicity is expressed as a lower limit for several reasons.
First, ionization corrections to the observed O$^0$/H$^0$ ratios
will give larger [O/H] values
(Figure~\ref{fig:ioncorr}).  Second, the lower limit assumes the
nearly maximal \nhi\ values permitted by the data for the three
subsystems.  
If we adopt a modest ionization correction and lower \nhi\ values, the oxygen
metallicity for \slls\ approaches the solar abundance.  
Finally, a proper derivation of the oxygen metallicity for the
SLLS is to take an \nh-weighted average of each subsystem.
This is especially important for predominantly ionized gas.
Using our fidcuial values, we find $\rm <[O/H]> = -0.23$\,dex assuming
no ionization corrections.
We emphasize that even a 1/3 solar metallicity ([O/H]~$= -0.5$\,dex) 
represents a very high enrichment level.  
It exceeds the metallicity of nearly all
optically thick absorbers at these redshifts: 98\% of sightlines
randomly intersecting galaxies \citep[DLAs;][]{pgw+03} and 95\% of the
super-LLS population \citep{pdd+07}.  Such high enrichment levels are
generally observed only in gas within and around star-burst galaxies
\citep{prs+02,simcoe06} and in quasar environments
\citep{dhs03,dcr+04,agk+07}.

The observations also constrain the relative abundances of O, Fe, and N,
e.g.\ the O/Fe and N/O ratios.  These diagnostics are valuable
for constraining the star formation history for 
\slls\ because each element is associated
with a unique nucleosynthetic site and has a unique production timescale.
Oxygen is made in massive stars and one predicts minimal delay
($t < 100$\,Myr) 
for its nucleosynthesis and release to the ISM
following the onset of star formation. 
The production of nitrogen and iron, however, is expected to be
delayed by several 100\,Myr to $\approx 1$\,Gyr
because these elements are made in 
intermediate mass stars and Type\,Ia SN respectively.
By examining the relative abundances of O, N, and Fe,
therefore, one may constrain the duration of the star-formation
episode that produced the metals. 

Unfortunately, the relative abundance ratios of O, N, and Fe are more
sensitive to the ionization state of the gas.  The most conservative
treatment is to report lower limits to O/Fe and N/O from the
observed ionic column density ratios of O$^0$/Fe$^+$ and
N$^0$/O$^0$.  As with O$^0$/H$^0$,
corrections to these ionic ratios are positive 
definite (Figure~\ref{fig:ioncorr}).
We measure lower limits to [O/Fe] of $+0.1, +0.5, +0.8$\,dex for 
subsystems~A,B,C respectively and conclude
the gas has a super-solar $\alpha$/Fe ratio.  For even a modest
ionization correction, the integrated average
$\alpha$/Fe ratio approaches ten times the solar ratio and
exceeds the value measured for nearly all intervening
absorption line systems. In Figure~\ref{fig:relabnd},
we present an estimate for these values for the
various subsystems and also for the \nh-weighted averages against
measurements for a set of high-$z$ galactic observations.

In the local universe, large $\alpha$/Fe ratios are observed
in metal-poor stars \citep[e.g.][]{mcw97}, massive early-type
galaxies \citep{tfw+00}, and highly depleted ISM gas \citep[e.g.][]{ss96}.
The first two examples reflect nucleosynthetic enhancements and
are attributed to the Type~II supernovae
of massive stars.  The latter example, however, results from the highly
refractory nature of Fe relative to O, i.e.\ one observes enhanced O/Fe
gas-phase abundances in the ISM
because Fe is preferentially depleted into dust grains.  
The observed $\alpha$/Fe enhancement in \slls\
could be due to one or both of these effects,
but our expectation is that differential depletion
is not dominant.
A high depletion level would be unusual
for gas with modest surface and volume densities that
is predominantly ionized with a temperature of $T \approx 20,000$K.   
It is more likely that the super-solar $\alpha$/Fe ratios reflect
an intrinsic, nucleosynthetic enhancement.  Granted the
high metallicity of \slls, the $\alpha$/Fe enhancement
suggests a star formation history representive of bright
spheroids \cite[e.g.\ bulges, elliptical galaxies][]{gfs+07}.  
Chemical evolution modeling suggests these stellar systems
underwent a short period of intense star formation where
the gas was rapidly enriched by supernovae from massive
stars before Type~Ia SN could contribute
\citep[e.g.][]{matteucci94}.   We conclude that the metals
for \slls\ were produced in a starburst lasting less than $\sim 1$\,Gyr.  

The N/O ratio of the gas offers additional constraints on
the star-formation history.  Similar to O/Fe, ionization corrections
may be important and
the conservative approach is to adopt the N$^0$/O$^0$ ratio as
a lower limit to N/O.  We find [N/O]~$> -0.7$\,dex for 
subsystems~A and B and [N/O]~$> -0.2$\,dex for 
subsystem~C (Table~\ref{tab:summ}).
The observations, specifically the large N$^+$/N$^0$ ratio,
indicate non-zero ionization corrections. 
We estimate the correction from photoionization
modeling to be $+0.25$ to $+0.5$\,dex for $\log U = -4$ to $-3$\,dex and
conclude that N/O is at least 1/3 the solar ratio in each
subsystem and may be
super-solar in subsystem~C.  These values are 
larger than N/O values observed for quasar absorption line systems
\citep[Figure~\ref{fig:relabnd}, e.g.][]{henrypro07}.  
Indeed, solar N/O ratios 
are commonly observed only in quasar environments
\citep{hamannf99,dhs03,dcr+04,agk+07}.
This further reflects the high 
metallicity of \slls\ because N behaves as a 
`secondary element' with its yield scaling as
the square of the enrichment level (M/H)$^2$.  
Therefore, one expects near-solar or even super-solar
N/O abundances in high metallicity environments.
The exception to this expectation is if the system is too young
for many intermediate mass stars to have
cycled through the AGB phase.
The observation of nearly solar
N/O in \slls, therefore, implies the gas was enriched over
an episode of at least a few 100\,Myr \citep{mp93,rsm+02}.

As noted above, the high metallicity and solar or super-solar N/O and
$\alpha$/Fe ratios for \slls\ describe a chemical abundance pattern
that matches the abundances commonly measured for gas near quasars.
This includes the high metallicities inferred from emission lines from
the quasar broad line region and the relative abundances derived for
narrow associated absorption lines \citep{dhs03,dcr+04}.  We are inclined,
therefore, to causally connect the enrichment pattern of \slls\ with
\fgqso.  A key question that follows is whether the metals were
produced by the quasar's host galaxy or whether they originated in a
neighboring galaxy or galaxies.  The latter hypothesis would imply
that galaxies near quasars have enrichment histories similar to that
inferred for the quasar environment, which could be tenable
considering that early-type galaxies near the centers of modern
clusters have very uniform colors and spectra indicating a common age
of formation and star-formation history \citep[e.g.][]{ble92}.
Alternatively, as we discuss further in the next section, the metals
may have been generated during an early episode of star formation in
the host galaxy of \fgqso, and then transported to large radii $\rperp
\approx 100$\,kpc via a large scale outflow in $\approx 100$\,Myr
time.  To achieve a high N/O ratio while maintaining a super-solar
$\alpha$/Fe ratio, the data suggest a starburst lasting several
100\,Myr but less than $\approx 1$\,Gyr.  It is intriguing that the
implied sequence of early, intense star-formation followed by an
optically bright quasar phase and possibly a large scale outflow,
follows the general prescription for quasar activity proposed by
Hopkins and collaborators \citep{Springel05,Cox06,hhc+07}.

\begin{figure}
\begin{center}
\includegraphics[width=6.8in]{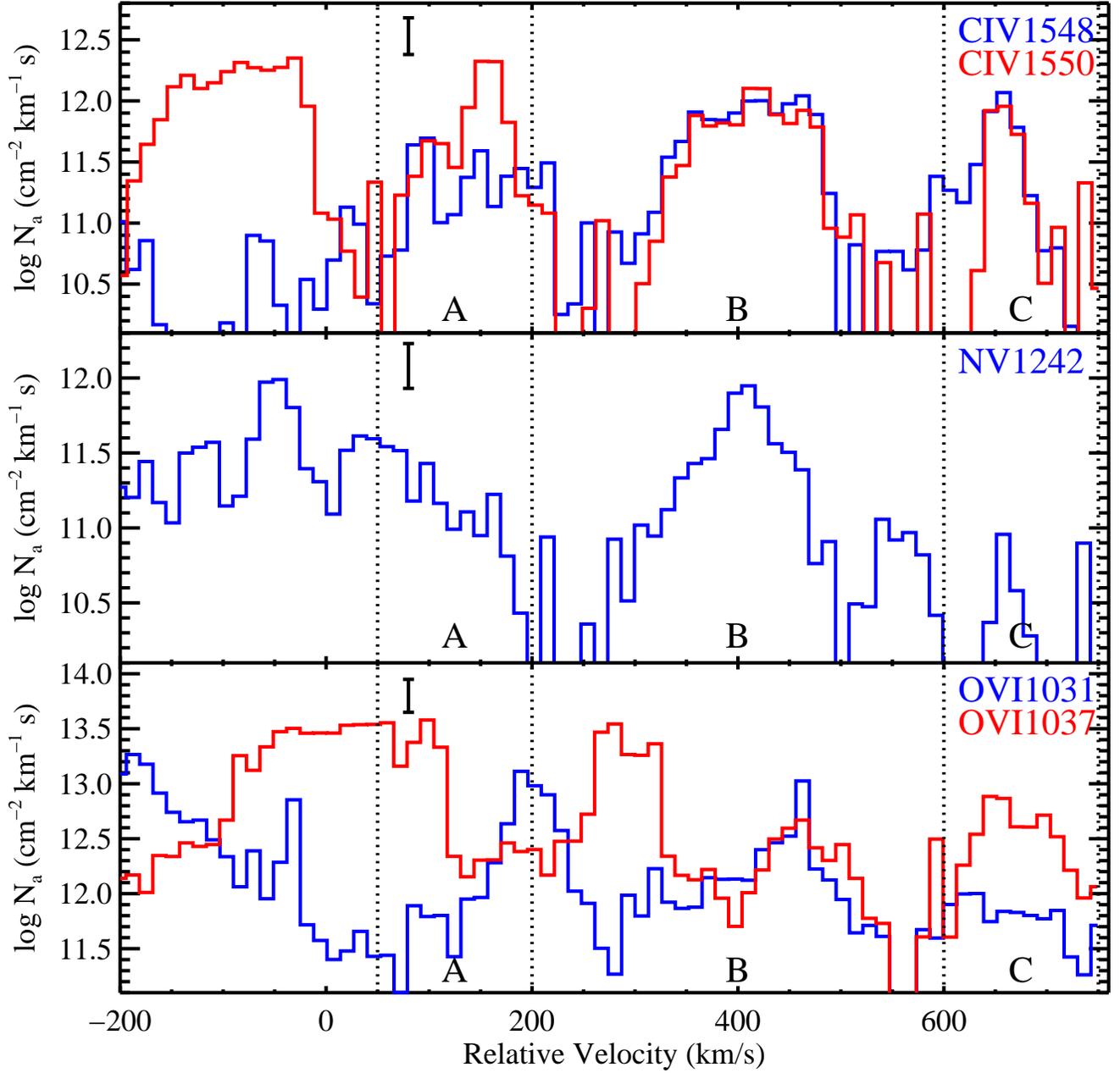}
\end{center}
\caption{
Apparent column density $N_a$ profiles for 
the NV~1242 transition, and the \ion{C}{4} and \ion{O}{6} doublets
for velocities relative to $z_{\rm fg} = 2.4360$. 
The profiles are significantly contaminated by coincident line-blending
which is best revealed by comparing the $N_a$ profiles for the pair
of transitions in each doublet.  After accounting for this line-blending,
we conclude that there is weak (or negligible) \ion{N}{5} and \ion{O}{6}
absorption in subsystem~C and at $\delta v < 0 \mkms$.
Values and upper limits to the integrated ionic column densities
are given in Table~\ref{tab:high}.
}
\label{fig:highion}
\end{figure}

\begin{deluxetable}{lcccccccc}
\tablewidth{0pc}
\tablecaption{HIGH-ION COLUMN DENSITIES \label{tab:high}}
\tabletypesize{\footnotesize}
\tablehead{\colhead{Ion} & 
\colhead{\fgqso$^a$} & 
\colhead{Subsystem A$^b$} & 
\colhead{Subsystem B$^c$} & 
\colhead{Subsystem C$^d$}}
\startdata
$\log \N{C^+}$ &$12.7\pm 0.11$&$13.4\pm 0.05$&$14.0\pm 0.05$&$13.3\pm 0.06$\\
$\log \N{N^{+4}}$ &$<13.9$&$<13.3$&$<13.8$&$<11.9$\\
$\log \N{O^{+5}}$ &$<14.9$&$<14.4$&$<14.8$&$<14.0$\\
\enddata
\tablenotetext{a}{Integrated AODM column densities for the interval containing \fgqso: $-200 \mkms < \delta v < +50 \mkms$ relative to $z_{fg}=2.4360$.}
\tablenotetext{b}{Integrated AODM column densities for Subsystem~A: $+50 \mkms < \delta v < +200 \mkms$ relative to $z_{fg}=2.4360$.}
\tablenotetext{b}{Integrated AODM column densities for Subsystem~B: $+200 \mkms < \delta v < +600 \mkms$ relative to $z_{fg}=2.4360$.}
\tablenotetext{d}{Integrated AODM column densities for Subsystem~C: $+600 \mkms < \delta v < +780 \mkms$ relative to $z_{fg}=2.4360$.}
\end{deluxetable}

\subsection{Constraints from High-Ion Observations}
\label{sec:high}

The current paradigm of baryons within galactic halos envisions a hot,
diffuse medium that has been heated by shocks during virialization
and/or feedback from supernovae within the galaxy.  Embedded within
this diffuse medium may be cooler, dense clouds that may be
photoionized by the local or background UV radiation field.  In the
Galaxy, this simple picture is supported by the observations of the
high velocity clouds \citep[cool, dense clumps, e.g.][]{wv97,pds+02}
and the widespread detection of \ion{O}{6} absorption which traces a
hotter, diffuse medium \citep{sws+03}.  This model is also supported
by quasar absorption line surveys, i.e.\ the association of
\ion{Mg}{2}, \ion{C}{4}, and \ion{O}{6} gas with galactic halos
\citep[e.g.][]{s93,clw01,fpl+07}.  On theoretical grounds, it has been
argued that a hot diffuse component pressure confines a population of
cold clouds, which are unlikely to be massive enough to be
self-gravitating \citep{mm96,mb04}.

With these scenarios in mind, we are motivated 
to search for absorption from a diffuse, hot component that could be 
associated with the halo of \fgqso.  The sightline to \bgqso\
intersects the presumed galactic halo of \fgqso\ at an impact
parameter $\rperp = 108$\,kpc. As we will see in \S~\ref{sec:NFW}, the
virial radius of the dark matter halo expected to host \fgqso\ is
$r_{\rm vir} \simeq 250\,{\rm kpc}$, so that our background sightline
easily resolves the virial radius.  In $\S$~\ref{sec:ion}, we discussed the
detection and analysis of low and intermediate-ion transitions
observed in \slls.  We also commented on the general absence of strong
absorption by high-ions like O$^{+5}$, C$^{+3}$, and N$^{+4}$.  These
ions trace gas either with a large ionization parameter ($\log U >
-2$) or with a high temperature $(T > 10^5$\,K).  Their observed
column densities, therefore, constrain the nature of a diffuse and/or
hotter phase within the halo of \fgqso.

In Figure~\ref{fig:highion}, we present the apparent column densities,
$N_a^i \equiv 10^{14.5761} \ln(1/I^i) / (f\lambda)$, of the
\ion{C}{4}, \ion{N}{5} and \ion{O}{6} profiles (smoothed by 5 pixels)
for the velocity interval spanning the quasar redshift and the three
subsystems of \slls.  The C$^{+3}$ and O$^{+5}$ ions exhibit multiple
transitions which allow one to identify line-blending by visual
inspection, i.e.\ regions
where the two profiles diverge significantly.  The \ion{C}{4}~1548 and
1550 profiles track each other closely for $\delta v > +200 \mkms$
indicating a positive detection without blending at these velocities.
At $\delta v < +200 \mkms$, the profiles diverge because the
\ion{C}{4}~1548 profile of component C blends into the \ion{C}{4}~1550
profile of component A (see also Fig~\ref{fig:metals}).  Therefore, the
\ion{C}{4}~1548 profile sets an upper limit to the column density of
C$^{+3}$ at these velocities.  At $\delta v \le 0\mkms$, there is no
statistically significant detection of \ion{C}{4}.

The \ion{N}{5} and \ion{O}{6} doublets lie within the \lya\ forest and
therefore are more likely to suffer from line blending.  Examining the
\ion{O}{6} transitions, the two profiles diverge at nearly all
velocities.  This indicates that the line-profiles are
severely affected by blends, most likely \lya\ and \lyb\ absorption
from gas at lower redshift.  The only interval where the \ion{O}{6}
profiles roughly coincide is at $400 \mkms < \delta v < 550 \mkms$.  One may
optimistically interpret the absorption as a positive detection of
O$^{+5}$ ions, but the more conservative approach is to report even
this absorption as an upper limit.  Turning to \ion{N}{5}, we present
only the \ion{N}{5}~1242 transition because the \ion{N}{5}~1238
profile is heavily blended with a strong \lya\ system at $z=2.502$.
We also expect that the broad absorption at $\delta v < 200 \mkms$
is associated with coincident \lya\ absorption; meanwhile, the optical depth
in subsystem~C ($\delta v < 600 \mkms$) is not statistically
significant.  The only plausible \ion{N}{5}~absorption, therefore, is
within subsystem~B where one notes the optical depth profile of
\ion{N}{5}~1242 resembles that of \ion{C}{4} in the interval
$300 \mkms < \delta v < 500 \mkms$.  
We caution, however,
that the \ion{O}{6} profiles do not track \ion{N}{5} in this velocity
interval.   This draws into question a positive detection of
\ion{N}{5} absorption because the O$^{+5}$ and N$^{+4}$ ions trace
gas with similar physical properties and one generally
expects coincident \ion{N}{5} and \ion{O}{6} absorption.  
It is possible, therefore, that the
apparent \ion{N}{5} profile instead tracks an unidentified blend and
we report the measured optical depth as an upper limit to
$\N{N^{+4}}$ in subsystem~B.

Table~\ref{tab:high} summarizes the column densities of the
high-ions in subsystems A, B, and C and in the velocity interval
containing $z = z_{\rm fg}$.  For C$^{+3}$ and N$^{+4}$,
the column densities are lower than 10$^{14} \cm{-2}$ and fall below 
$10^{13} \cm{-2}$ in several regions.  
For O$^{+5}$, the upper limits are a factor of ten higher owing to
noisier data and a higher frequency of line-blending.
All of these values lie below the $\gtrsim 10^{14} \cm{-2}$ 
column densities measured for the low-ions
of CNO in \slls\ (Table~\ref{tab:colm}).  This has significant
implications for the physical conditions of diffuse gas associated
with the galactic halo of \fgqso.

\begin{figure}
\begin{center}
\includegraphics[width=6.0in]{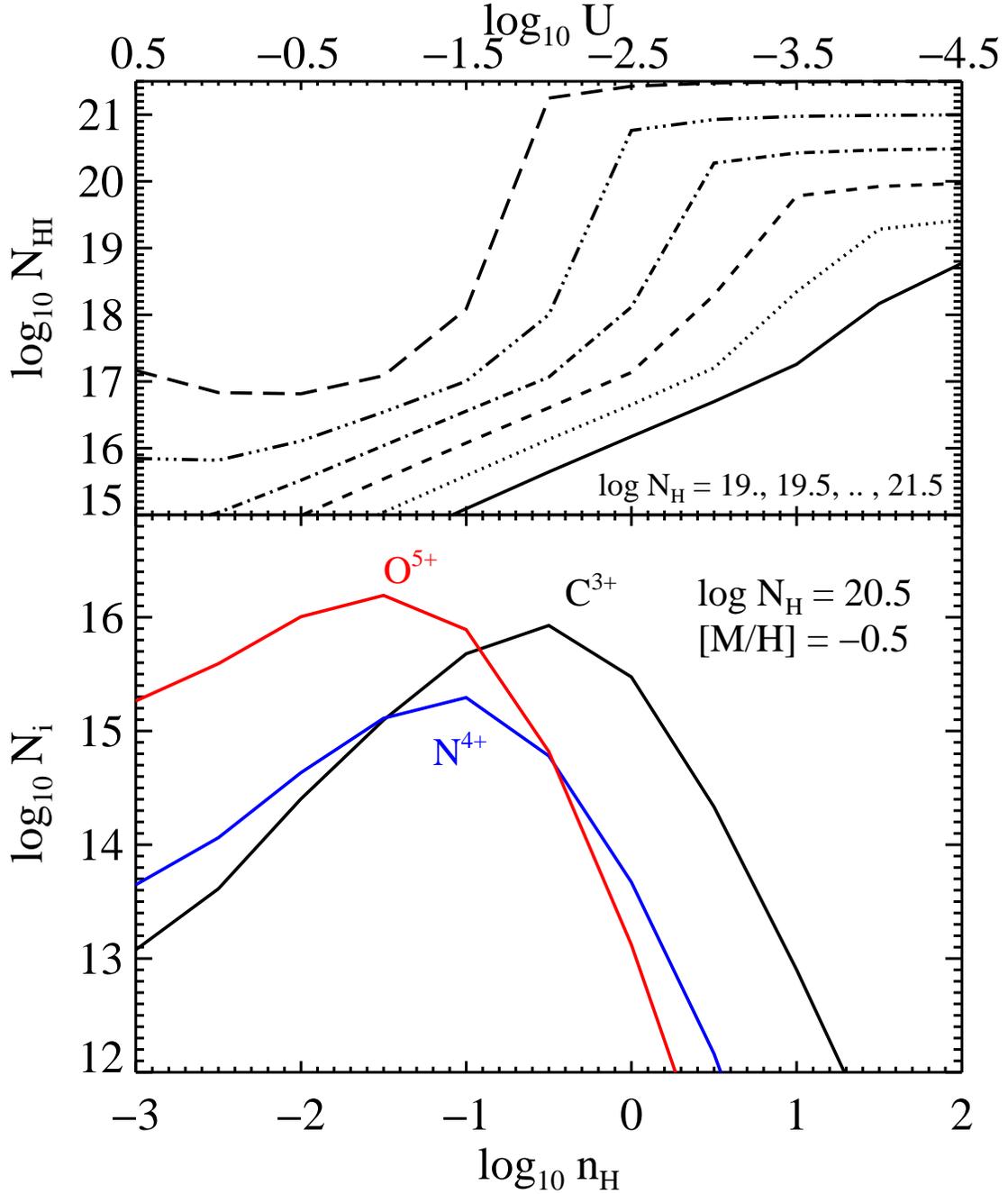}
\end{center}
\caption{ Upper panel: Predicted \ion{H}{1} column densities for a
  plane-parallel slab lying at a distance of 100\,kpc from \fgqso\ for
  a range of total H column ($\log N_H = 19, 19.5, ..., 21$) 
  and as a function of volume density
  $n_{\rm H}$. As $n_{\rm H}$ decreases, the effective ionization
  parameter increases and the gas eventually becomes optically thin
  (i.e.\ $\mnhi < 10^{17.2} \cm{-2}$).  Lower panel: Predicted column
  densities of several metal ions for a plane parallel slab with
  $N_{\rm H} = 10^{20.5} \cm{-2}$ illuminated by \fgqso\ at a distance
  of 100\,kpc.  As the assumed volume density of the gas is lowered,
  the gas is more photoionized and CNO are pushed into the high-ion
  states of C$^{+3}$, N$^{+4}$, and O$^{+5}$.  The predicted column
  densities for a metallicity of 1/3 solar are high and easily
  detectable even in a low-resolution spectrum (e.g.\ SDSS).  These
  predicted column densities are to be compared against
   the observed values
  (Figure~\ref{fig:highion}, Table~\ref{tab:high}).  }
\label{fig:evap}
\end{figure}

\begin{figure}
\begin{center}
\includegraphics[height=6.8in,angle=90]{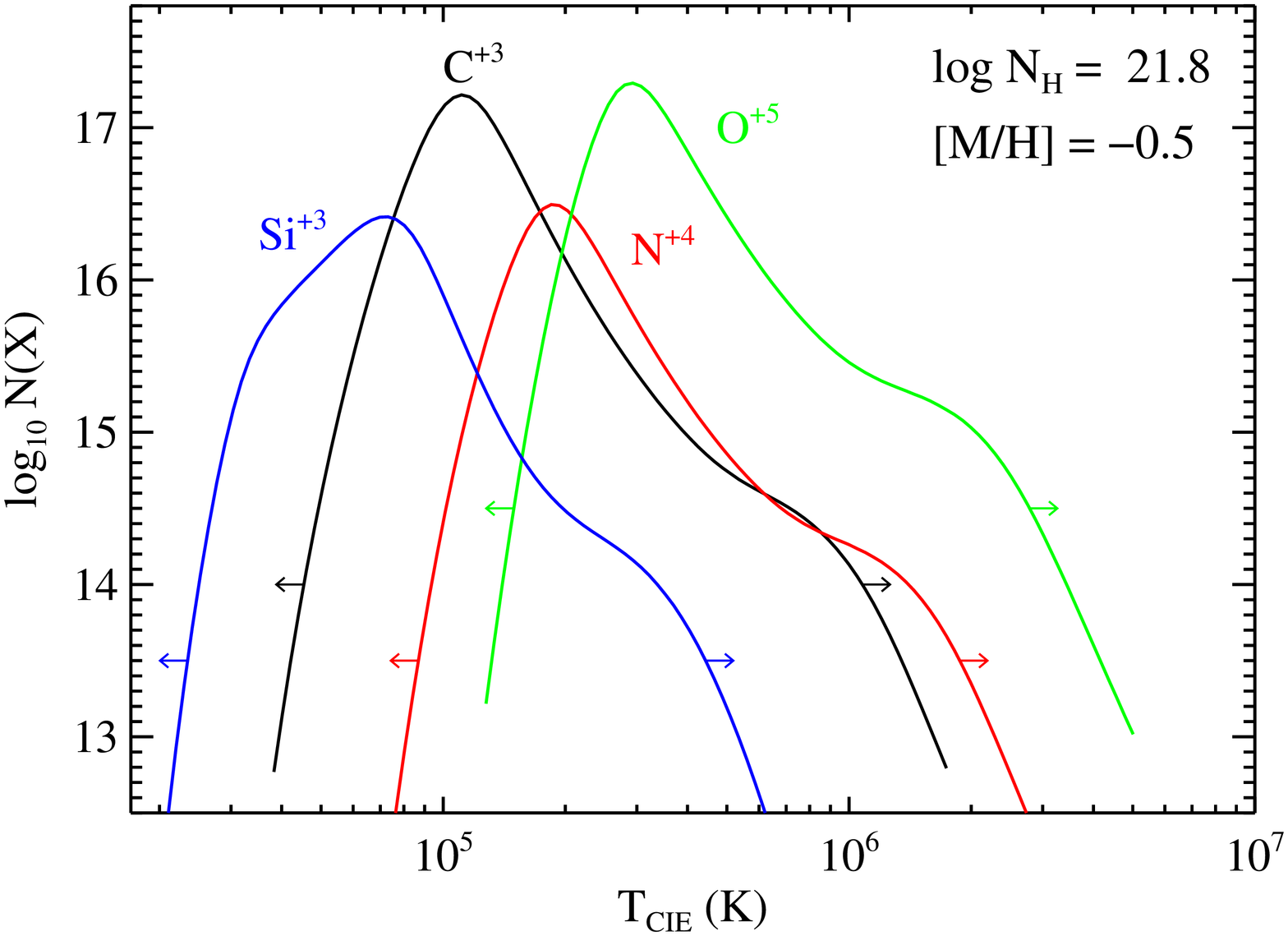}
\end{center}
\caption{
Predicted column densities for high-ion states of CNO and 
Si assuming collisional ionization equilibrium (CIE) for a gas with
total H column density $N_{\rm H} = 10^{21.8} \cm{-2}$ and
1/3 solar metallicity.  This hydrogen column density corresponds to the 
estimated baryonic surface density at a 100\,kpc
impact parameter from a dark matter halo with $M=10^{13} \msol$
and an NFW density profile.  
The arrows on the curves indicate constraints to the temperature
based on upper limits to the observed ionic column densities.
One notes that the data rule out 
$3 \sci{4} \, \rm K < T_{CIE} < 2 \sci{6} \, \rm K$.
}
\label{fig:cie}
\end{figure}

A direct conclusion from these observations is that the sightline does
not intersect a large column density of optically thin, photoionized
material.  In the lower panel of Figure~\ref{fig:evap}, we present the
predicted column densities of C$^{+3}$, N$^{+4}$, and O$^{+5}$ ions
for a wide range of physical density assuming an absorber with total
hydrogen column density $N_{\rm H} = 10^{20.5} \cm{-2}$ and metallicity
[M/H]~$=-0.5$\,dex.  For these calculations, we assume the gas is illuminated
by the \fgqso\ at a distance of $r = \rperp$.
Focusing on the parameter
space where the gas is optically thin ($n_{\rm H} < 0.3 \cm{-3}$), one predicts
column densities exceeding $10^{14.5} \cm{-2}$ for all of the ions.
Even if we assume a 10$\times$ lower metallicity, we can rule out such
a phase of gas along the sightline.

The absence of photoionized high-ions like O$^{+5}$ is not
particularly surprising, however, because the ambient medium in the
outer galactic halo surrounding a high-$z$ quasar almost certainly is
at a temperature so high that collisional ionization dominates over
photoionization.  This material may be observed independently of the
gas contributing to \slls.  A galaxy with mass $M > 10^{12} \msol$,
for example, has an implied virial temperature in excess of
$10^{6}$\,K and the dark matter halo expected to host \fgqso would
have a virial temperature $T_{\rm vir}\sim 10^{7}\,{\rm K}$ (see
$\S$~\ref{sec:NFW}).  In Figure~\ref{fig:cie} we present predicted
column densities for C$^{+3}$, N$^{+4}$, and O$^{+5}$ ions assuming
collisional ionization equilibrium \citep[CIE;][]{sd93}, $N_H =
10^{21.8} \cm{-2}$ and [M/H]~$= -0.5$\,dex.  This H column density is
the predicted value at $\rperp = 108\,$kpc in a $M = 10^{13.3} \msol$
halo with a Navorro-Frank-White (NFW) profile 
(concentration $c = 4$) and assuming a gas
fraction equal $f_g = 0.17$ equal to the cosmic baryon fraction
\citep[][see $\S$~\ref {sec:NFW}]{wmap05}.  The figure demonstrates that
the gas must have $T> 10^{6}$\,K to be consistent with the observed
upper limits to $\N{O^{+5}}$.  The dependence of $\N{O^{+5}}$ on $T$
is so strong that even if we assume considerably lower $N_H$ and [M/H]
values for the diffuse component, the gas must have $T > 10^6$\,K.  A
cooler phase is only allowed if $\log N_H + {\rm [M/H]} < 17.7$.

The principal conclusion of this section is that a diffuse medium in
the halo of \newline \fgqso, must have a temperature exceeding
10$^6$\,K.  This is consistent with the virial temperature
$T_{\rm vir}\sim 10^{7}\,{\rm K}$ for the gas in the $M \gtrsim
10^{13}\msol$ dark matter halo that is expected to host the f/g quasar
(see $\S$~\ref {sec:NFW}).  Note however that the absence of strong
high-ion transitions associated with \fgqso\ contrasts with similar
observations (using a b/g quasar sightline) of $z\sim 2.3$
star-forming galaxies where strong \ion{O}{6} and \ion{N}{5}
absorption is detected at similar impact parameters
\citep{simcoe06}.  In addition, the damped \lya\ systems (DLAs)
also frequently exhibit 
\ion{O}{6} absorption which reflects a highly ionized, diffuse
medium in the halos of these galaxies \citep{fpl+07}.  
The average $\N{O^{+5}}$ for the DLAs exceeds $10^{14} \cm{-2}$.  The DLA
sightlines presumably intersect galactic halos at much smaller impact
parameters than $\rperp = 108$\,kpc.  Nevertheless, the high-ion
material in DLAs is undoubtedly associated with the galactic halo
\citep{wp00a,mps+03,pcw+08} and may extend to many tens kpc.  The
fundamental difference is that DLA sightlines will rarely penetrate
the galaxies hosting high $z$ quasars and likely trace lower mass
halos with significantly lower virial temperatures.

\subsection{Is \fgqso\ Shining on \slls?}
\label{sec:shining}

Of considerable interest to studies of quasars and the buildup of
supermassive black holes in the Universe
\citep[e.g.][]{yu02,hoprich07,shankar07} is whether their optical-UV
emission is isotropic and or intermittent.  Evidence that quasars emit
anisotropically or intermittently comes from the null detections of
the transverse proximity effect in the Ly$\alpha$ forests of projected
quasar pairs \citep[][but see Jakobsen et al.
2003]{Crotts89,DB91,FS95,LS01,Schirber04,Croft04}, the anisotropic
clustering of optically thick absorbers near quasars
\citep{hp07,phh08}, and the absence of fluorescence detections in
optically thick gas proximate to quasars \citep[Hennawi \& Prochaska
2008, in prep.; but see][]{ask+06}. Together, these studies provide a
statistical case that quasars emit anisotropically or intermittently.

For an individual quasar, one can use a background sightline to
observe nearby gas in absorption, and test whether it is illuminated
by studying its ionization state. The general approach is to (1)
derive the ionization parameter $U$ of the gas near the quasar using
ionic metal line transitions (2) estimate the volume density of the
gas, and (3) compare the inferred intensity of the impingent radiation
field with that estimated for the nearby quasar.  \cite{wfw+07} and
\cite{gsp08} have conducted this test for highly ionized gas near
several $z \sim 2$ quasars.  The \cite{wfw+07} study was deemed
inconclusive.  \cite{gsp08}, meanwhile, identified several absorbers
proximate to quasars with high ionization parameters and estimated an
ionizing radiation field that exceeds the EUVB.  Instead, the
estimated fluxes roughly match the values calculated for the nearby
quasars under the assumption that they emit isotropically.  These
authors, however, did not place empirical constraints on the gas
density, did not consider the effects of collisional ionization, and
did not explore the effects of non-solar relative abundances.  In
these respects, we consider their results to be suggestive but
inconclusive.

We can similarly test for illumination of \slls\ by 
\newline \fgqso.
In $\S$~\ref{sec:ion}, we compared the observed
ratios of N, Si, Al, Fe, and C atoms and ions against photoionization
models assuming a single phase, plane-parallel slab of
gas illuminated by a quasar spectrum ($f_\nu \propto \nu^{-1.57}$).
For subsystems A+C of \newline \slls, the majority of the observational
constraints imply $\log U < -3.25$.  The only
exception is the large N$^+$/N$^0$ ratio which indicates
$\log U \approx -3$ for subsystem~A,  and $\log U > -4$ for subsystem~C.
Under the constraints of a single-phase model and a plane-parallel
geometry, one cannot identify an equilibrium photoionization
model that reproduces all of the observed ionic ratios.
By the same token, the observations are inconsistent with any
collisional ionization equilibrium model.   We
do not know if these inconsistencies highlight errors
in the atomic data (e.g.\ dielectronic recombination rates),
the oversimplification of a single-phase, plane-parallel geometry,
an erroneous shape for the radiation field, and/or the presumption of
equilibrium models.  We will proceed under
the relatively conservative assumption that the ionization
parameter $\log U < -3$ and we will comment on implications
for significantly smaller values.  
We also remind the reader that the observed ionic ratios 
are roughly consistent with a $T \approx 20,000$K
collisional ionization equilibrium model.  Any contribution
of collisional ionization processes to the observations would
lower the estimated flux of ionizing radiation.

The other key physical constraint is the volume density of the gas
$n_{\rm H}$.  For a predominantly ionized gas, this can be estimated
from the electron density $n_{\rm e}$.  Our analysis of the
fine-structure levels of the C$^+$ and Si$^+$ ions indicate $n_{\rm
  e}$ less than approximately $5 \cm{-3}$ for each of the subsystems.
These upper limits could be improved by obtaining higher S/N
observations of the undetected \ion{Si}{2}$^*$~1264 transition at
$\lambda \approx 4350$\AA.  In any case, we will adopt this
conservative limit to the electron density and assume $n_{\rm H} \approx
n_{\rm e}$ which is appropriate for a predominantly ionized medium.

Adopting $\log U < -3$ and $n_{\rm H} < 5 \cm{-3}$, we infer an upper limit
to the ionizing photon flux, 

\begin{equation}
  \Phi_{\rm obs} = U n_{\rm H} c \; < \; 1.5\sci{8} \;\;
{\rm photons \; s^{-1} \; cm^{-2}} \;\;\; .
\label{eqn:phiobs}
\end{equation}
We may compare this value with the ionizing flux of \fgqso\
under the assumptions that the gas lies at a distance equal
to the impact parameter and that the quasar emits isotropically.
Combining the SDSS
optical photometry of \fgqso\ with an assumed power-law spectral shape 
($f_\nu \propto \nu^{-1.57}$), we derive an AB magnitude $m_{912} = 21.42$
at 1\,Ryd.   At $z_{\rm fg} = 2.436$,
this gives a specific luminosity 
$L_{912} = 1.16\sci{30} {\rm erg \; s^{-1} \; cm^{-2} \; Hz^{-1}}$
and we estimate an ionizing photon flux at $r = \rperp = 108$\,kpc of

\begin{equation}
\Phi_{\rm QSO} = 9\sci{7} \; {\rm photons \; s^{-1} \; cm^{-2}} \;\;\; ,
\label{eqn:phiqso}
\end{equation}
assuming the quasar emits isotropically.
Therefore, the conservative upper limit to $\Phi_{\rm obs}$ roughly matches
the predicted photon flux from \fgqso\ at the observed impact
parameter.  In this conservative observational
limit, we do not have strong evidence for anisotropic
emission.

If we adopt less conservative but more realistic constraints on the
ionization parameter and volume density ($\log U < -4; n_{\rm H} <
1\cm{-3}$), we set an upper limit to the observed photon flux
$\Phi'_{\rm obs} < 3\sci{6} \;\; {\rm photons \; s^{-1} \; cm^{-2}}$.
To maintain the null hypothesis that \slls\ is illuminated, we would
require the gas be located at $r > 500$\,kpc.  This large distance is
disfavored by other arguments.  The extreme velocity field, high
metallicity, and quasar-like relative abundances suggest the gas is
local to the quasar host galaxy.  At $r=500$\,kpc one could associate
\slls\ with a protocluster containing \fgqso, but not its galactic
halo.  Furthermore, the strong clustering of optically thick absorbers
around quasars at $z\sim 2.5$ implies a high probability that $r \approx
\rperp$ \citep{hp07}.  In Figure~\ref{fig:cluster}, we show the
probability of the absorber lying at a distance less than $r$ from the f/g 
quasar, given the knowledge that it is within a velocity interval $v_{\rm
  vmax} = \pm 700 \mkms$ and has an impact parameter $\rperp =
108$\,kpc.  The upper x-axis of the figure shows the corresponding cumulative
probability of the ionizing flux $\phi_{\rm QSO}$, assuming that \fgqso emits
isotropically.  We calculate a 70\% probability that \slls\ is located
within 500\,kpc of the quasar.

We independently test whether the transverse direction is illuminated
by searching for \lya\ fluorescence from the optically thick gas near
the b/g quasar sightline. By coadding a series of longslit exposures
of \sdssj\ using Gemini/GMOS, amounting to a 7200s total integration,
we have not detected any extended emission \citep[Hennawi \& Prochaska
2008, in prep.; but see][]{ask+06} at the f/g quasar redshift near the
impact parameter of the b/g quasar sightline. The gas clouds
comprising \slls\ have estimated sizes of $\sim 10-100\,{\rm pc}$,
making them so small that it would be impossible (with current
sensitivity) to detect \lya\ emission from the \emph{actual} clouds we
see in absorption in \slls.  However, the high covering factor of
optically thick gas near foreground quasars with $\rperp \lesssim
100\,{\rm kpc}$ characterized statistically by \citet{hpb+06},
together with the detection of three such optically thick subsystems
in \slls, argue that the covering factor of optically thick gas is
indeed roughly unity at $\rperp$ near \fgqso. If this gas is
illuminated it should be emitting detectable \lya\ photons. The
fluorescent surface brightness of clouds illuminated by \fgqso\ at a
distance of $\rperp = 108\,{\rm kpc}$ would be \newline ${\rm SB}_{\rm
  Ly\alpha} < 4.8 \sci{-17} {\rm ergs \, s^{-1} \, \cm{-2} \,
  arcsec^{-2}}$. From a preliminary analysis, we conservatively
estimate that we could have detected extended emission a factor of ten
fainter than this expectation, implying that the distance to the
clouds would have to be $r > \sqrt{10}\,\rperp$ to render them
undetectable if the foreground quasar emits isotropically. Thus,
isotropic emission from \fgqso\ is ruled out unless all of the
optically thick gas lies beyond $r > 340\,{\rm kpc}$.

\subsection{Why Aren't Similar Absorbers Observed along the Line of
  Sight to Quasars?}
\label{sec:nal}

\cite{hp07} have demonstrated that high-$z$ quasars exhibit a high
covering fraction to optically thick absorbers at impact parameters
$\rperp \lesssim 200$\,kpc, and argued that the incidence of such
transverse absorption exceeds the incidence along the line-of-sight to
bright high-$z$ quasars.  \cite{phh08} similarly concluded that there
are fewer damped \lya\ systems near $z>2$ quasars then expected from
clustering arguments and \cite{wkw+08} have recently reported a
similar deficit of strong \ion{Mg}{2} absorbers along the
line-of-sight compared to the transverse direction.  Together these
observations argue that quasars emit ionizing radiation
anisotropically or intermittently. Hence a paucity of optically thick
absorbers are detected along the line-of-sight because this gas, which
is by construction illuminated, is photoionized by the quasar.  One
test of this hypothesis is to search for an enhancement of highly
ionized material along the sightline, e.g.\ \ion{Si}{4}, \ion{C}{4},
\ion{N}{5}, \ion{O}{6}. Indeed, \cite{wkw+08} report a higher
incidence of \ion{C}{4} absorption along the same sightlines where the
lower ionization potential transition, \ion{Mg}{2}, is deficient.
There is also a high incidence of highly ionized absorbers within
several thousand \kms\ of the quasar redshift commonly termed narrow
associated absorption line systems \citep[NAAL;][]{dcr+04}.  We now
explore the physical conditions required for \fgqso\ to
photoionize/photoevaporate \slls\ and also the expected properties of
\slls\ if it were illuminated by the quasar, and thus exposed to a
more intense radiation field.  These calculations are similar to those
performed by \cite{cmb+08} and we refer the reader to their work for
additional analysis.

In Figure~\ref{fig:evap} we present a series of calculations from the
Cloudy software package for constant density gas slabs placed 100\,kpc
from \fgqso\ along the sightline to Earth.  The quasar's specific
luminosity was assumed to be a power-law ($L_\nu \propto \nu^{-1.57}$)
and was normalized to SDSS photometry: $L_\nu = L_{912}
(\nu/\nu_{912})^{-1.57}$ with $L_{912} = 1.16 \sci{30} {\rm erg \,
  s^{-1} \, Hz^{-1}}$.  In all of the calculations we assume a gas
metallicity [M/H]~$=-0.5$\,dex with solar relative abundances.  In the
upper panel of the Figure, we show the \ion{H}{1} column densities for
slabs with a series of total H column densities\footnote{These
  calculations are nearly self-similar with respect to the ionization
  parameter $U$, i.e.\ we derive similar results if we modify the
  distance to the slab, the quasar luminosity and $n_{\rm H}$ provided
  that $U$ remains constant.}  $\log N_H = 19, 19.5, ..., 21$ as a
function of the gas density $n_{\rm H}$.  For slabs with $N_H =
10^{19} \cm{-2}$, the gas has $\mnhi < 10^{17.2}$ and is hence
optically thin to ionizing photons, provided $n_{\rm H} < 10 \cm{-3}$; while 
slabs with $N_H = 10^{21} \cm{-2}$ are optically thin only if $n_{\rm
  H} < 0.1 \cm{-3}$.

In $\S$~\ref{sec:ion}, we estimated $N_H \approx 10^{20} \cm{-2}$ and
$n_{\rm H} \approx 1 \cm{-3}$ for subsystems A and C based on the
relative column densities of several ion pairs, our estimate of the
\ion{H}{1} column density (uncertain by $\pm 0.4$\,dex), 
and an analysis of \ion{C}{2} and
\ion{Si}{2} fine-structure transitions.  According to the calculations
presented in Figure~\ref{fig:evap}, this gas would have $\mnhi \approx
10^{17} \cm{-2}$ if it were at 100\,kpc from \fgqso\ along our
sightline.  If the volume density is just a little lower (permitted by
the fine-structure analysis), the gas would be optically thin.  We
conclude that subsystems~A and C of \slls\ would be photoionized by
\fgqso\ if it were placed along our sightline.  Subsystem B, however,
exhibits a larger \nhi\ value and has a larger implied
$N_H \approx 10^{20.3} \cm{-2}$ value.  Assuming $n_{\rm H} = 1
\cm{-3}$, this system would have
$\mnhi \approx 10^{17.3} \cm{-2}$ if illuminated by the \fgqso\
at a distance of 100\,kpc. To be optically thin and illuminated at
this distance requires $n_H < 1 \cm{-3}$.
It is important to note that \fgqso\ is
relatively faint compared to the brighter quasars studied to establish
the low incidence of optically thick absorbers near quasars
\citep[e.g.][]{reb06,phh08,wkw+08}. These quasars are a factor of
$\sim 5-10$ brighter and the resulting ionization parameters are higher
by the same factor.  Thus clouds with properties similar to \slls\
could still appear optically thin if illuminated, resulting in no
conflict with previous observations. Similarly, it would be very
fruitful to characterize the incidence of proximate optically thick
absorbers as a function of quasar luminosity. 

If an absorber like \slls\ is photoionized by a nearby quasar, the
system may still give rise to several discernible features.  If the
\ion{H}{1} column density exceeds $10^{14} \cm{-2}$, it would exhibit
a strong \lya\ absorption feature.  This would be difficult to
distinguish from the ambient \lya\ forest, however.  If this gas has
even a modest metallicity, it may show unique metal-line signatures.
The lower panel of Figure~\ref{fig:evap} presents the predicted column
densities of several high-ions assuming $N_H = 10^{20.5} \cm{-2}$,
[M/H]~$=-0.5$\,dex, and that the cloud is at a distance of 100\,kpc
from \fgqso\ along our sightline.  Once the cloud becomes optically
thin (at $n_{\rm H} \lesssim 0.5 \cm{-3}$), large column densities of
C$^{+3}$, N$^{+4}$, and O$^{+5}$ ions are expected.  The alkali
doublets associated with these ions saturate at column densities of
$\approx 10^{14} \cm{-2}$ and, therefore, would yield absorption lines
that could be detected at even the modest resolution of the SDSS
spectroscopic survey.  These column densities scale directly with the
assumed metallicity and are also sensitive to the $N_{\rm H}$ value.
Combining the high incidence of optically thick gas observed in the
transverse direction with the high ionization parameter along the
sightline, and using our knowledge of the physical properties of the
(unilluminated) absorbing clouds from \slls, we conclude that that a
large fraction of bright quasars should exhibit strong \ion{C}{4},
\ion{N}{5}, and \ion{O}{6} absorption at $z \approx z_{\rm em}$ along
their sightlines.

Absorption systems with strong high-ion absorption near quasars with
$z_{\rm abs} \approx z_{\rm em}$ or small velocity intervals ($\Delta
v < 1000 \mkms$) have been detected in quasar spectra, and are
referred to as narrow associated absorption line (NAAL) systems
\citep[e.g.][]{wbp93,sp00}.  The most comprehensive survey of these
NAALs using high resolution spectra was performed by
\cite{dcr+04} who analyzed the UVES key project of $z \approx 2$
quasars.  Of the 22 quasars in the sample, \cite{dcr+04} report that
16 exhibit significant \ion{C}{4} absorption within $\pm 5000 \mkms$
of $z_{\rm em}$ for a total of 34 \ion{C}{4} systems, 15 of which also
exhibit \ion{N}{5} absorption.  \cite{dcr+04} studied the properties
of a subset which allowed for detailed photoionization modeling.  The
characteristics of this subset are similar in metallicity, density,
and relative abundances to \slls\ (see also $\S$~\ref{sec:chem}).
Furthermore, the authors used observed limits on the volume density of
the gas (from fine-structure analysis) and estimates of partial
coverage to infer distances of 10 to 200\,kpc from the quasar.  We
conclude, therefore, that if \slls\ experienced an $\approx 10$ times
higher ionization parameter, it would show characteristics common to
NAAL systems.  Indeed, \fgqso\ is approximately 2.5\,mag (10$\times$)
fainter than the quasars of the \cite{dcr+04} sample.  We speculate
that an absorption-line survey of intrinsically faint quasars will
show a high incidence of absorbers with properties similar to \slls\
and a correspondingly lower incidence of NAAL systems.  One could
perform such a survey with current and future echellette spectrometers
on 10m-class telescopes.

\section{New Clues about Massive Galaxy Formation}
\label{sec:model}

In this section we examine inferences one might draw from our 
measurements of \slls\ in the context of galaxy formation 
models and quasar feedback.
With only a single sightline, it is too early to draw firm conclusions
on these processes.  In part, the following section sets the framework
for future discussion.
We begin with a review of the expected properties of the dark matter
halo for a $z \sim 2.5$ quasar.

\subsection{Expected Properties of \fgqso's Dark Halo}
\label{sec:NFW}

\citet{pmn04} measured the clustering of quasars in the redshift range
$0.8 < z < 2.1$, which allowed them to estimate that quasars are
hosted by dark matter halos with mass $M \sim 10^{13}~\msol$. As they
found strong evolution of the clustering with redshift, which
continues to even higher redshift ($z > 3.0$; see \citet{shen07}), we
henceforth assume that \fgqso\ resides in a dark matter halo
$M=10^{13.3}~\msol$, but we caution that this mass is uncertain by a factor
of $\sim 0.5$\,dex, due to both measurement error and the intrinsic
distribution of quasar host halo masses.  

For a dark matter halo of this mass with an NFW profile \citep{nfw97}, 
the virial radius is 
\be 
r_{\rm
  vir} = 251\left(\frac{M}{10^{13.3}~\msol}\right)^{1/3}~{\rm kpc},
\ee 
Thus, if \fgqso\ resides at the center of its host dark matter halo,
the \newline \bgqso\ sightline probes it at $\lesssim 50\%$ of the virial
radius. If we assume a  concentration parameter $c=4$ (characteristic of high 
redshift halos), the peak circular velocity of the dark matter halo is
\be
v_{\rm circ} =
604\left(\frac{M}{10^{13.3}~\msol}\right)^{1/3}\,\mkms  \label{eqn:vcirc},  
\ee
which implies a dynamical time at the impact parameter $\rperp$
of the background sightline of
\be
t_{\rm dyn} \sim \rperp/v_{circ} 
= 1.7\sci{8}\left(\frac{M}{10^{13.3}~\msol}\right)^{-1/3}\,{\rm years}\label{eqn:tdyn}. 
\ee

The circular velocity we deduce is comparable to the
extreme kinematics observed in the subsystems of \slls. It is thus
conceivable that the motions in \slls\ represent gravitational infall
or virial motion of `cold gas clouds' orbiting in the dark matter halo
potential, just as galaxies would. To make this more precise we note
that the maximum extent spanned by the three components of \slls\ is
$\Delta v \approx 650~\mkms$. For a Gaussian distribution, we find
that the maximum extent measured from three samples is related to the
dispersion by $\Delta v = 1.69\sigma$, so we estimate that the dark
matter halo hosting \fgqso\ and \slls\ has a line-of-sight velocity
dispersion $\sigma \simeq 380~\mkms$. \citet{tormen97} found that the
maximum circular velocity of an NFW halo is a factor of $\approx 1.4$
times larger than the maximum of the one dimensional velocity
dispersion, so our kinematics suggest $v_{\rm circ} = 540~\mkms$. The
velocity width of \slls\ is consistent with the characteristic
velocity for the dark matter halos hosting $z \sim 2.5$ quasars
(eqn.~\ref{eqn:vcirc}) , thus gravitational dynamics in the halo
surrounding \fgqso\ could explain the observed kinematics.  

One challenge to this scenario is that the velocities of the clouds
comprising \newline \slls\ are systematically offset to positive values
relative to \newline \fgqso.  Such an offset is characteristic of organized
motions (e.g.\ rotation, infall along a filament) rather than a random
velocity field and we question whether a highly organized velocity
field with large velocity width can occur in the outer halo.  It is
worth noting, however, that the offset resembles a similar trend
observed for optically thick absorbers (\ion{Mg}{2} systems) at $z
\sim 1$.  For these absorbers, \cite{sks+02} and \cite{kcs+07} report
a systematic velocity offset between the gas and the systemic velocity
of its host galaxy.  The origin of this asymmetry is not
understood. Another explanation for the observed kinematics is that
\slls\ represents cold material swept up in large scale outflow, which
we consider in more detail in \S~\ref{sec:outflow}.


In the standard structure formation picture, gas falling into a massive
dark matter halo will be shock heated to near the virial temperature of the 
dark halo, converting its gravitational kinetic energy into thermal energy. 
The virial temperature is defined as 
\be
T_{\rm vir} = \frac{\mu m_{\rm p} v_{\rm circ}}{2k_{\rm B}} = 1.3\sci{7}\left(\frac{M}{10^{13.3}~\msol}\right)^{2/3}\,{\rm K}\label{eqn:Tvir}, 
\ee
where $\mu$ is the mean atomic weight, which we take as $\mu=0.59$ for a fully
ionized primordial gas. 
Gas at these high temperatures will be collisionally ionized and 
should result in negligible \ion{H}{1} absorption. 

As we detect a large column of enriched and 
cold photoionzed gas at $T\simeq 20,000\,$K in the background sightline, it 
is of interest to calculate the time that it would take gas shock heated to 
the virial temperature to cool at the distance of our impact parameter. 
The cooling time of a gas is conventionally taken to be the ratio of 
its specific thermal energy to the volumetric cooling rate, or 
$t_{\rm cool} = [3(n_e+n)kT]/[2n_e n \Lambda]$
where $\Lambda$ is the cooling function
\citep[e.g.][]{sd93} and
$n$ is the total ion density. 
We can estimate the ion density of the quasar host
halo assuming that the total mass (dark matter + gas) distribution
follows an NFW density profile $\rho(r)$, with the gas density being a
fraction $f_{\rm g}$ of the total density $\rho_{\rm g}(r) = f_{\rm
  g}\rho(r)$. We have then $n =\rho_{\rm g}\slash \mu m_{\rm p}$, or
\be
n(r=108\,{\rm kpc}) \approx 7.5\sci{-3}\left(\frac{f_{\rm g}}{0.17}\right)
\left(\frac{M}{10^{13.3}~\msol}\right)^{0.73}\,{\rm cm}^{-3}\label{eqn:nnfw}, 
\ee
where we took $f_{\rm g}= \Omega_{\rm b}\slash\Omega_{m} = 0.17$ to be the 
cosmic baryon fraction from \citet{wmap05}. 
We finally find for the cooling time at $r = 108$\,kpc,
$t_{\rm cool} \approx 3.2\sci{9}
(M/10^{13.3}~\msol)^{-0.22}{\rm years}$.
Note that at the high virial temperatures $T\sim 10^7\,$K 
characteristic of our dark matter halo, cooling is dominated
by thermal bremsstrahlung radiation and hence does not depend strongly
on metallicity.
According to the current structure formation paradigm, the bulk of the
gas interior to the virial radius should have been shock heated to
$T_{\rm vir}$, and the cooling time of this hot phase at $R\simeq
100\,$kpc is comparable to the age of the Universe at $z=2.4360$,
$t_{\rm H}=2.6\sci{9}\,$years. Thus the impact
parameter $\rperp = 108\,{\rm kpc}$ for \bgqso\ probes \fgqso\ at
approximately the cooling radius (where $t_{\rm cool} = t_{\rm H}$ ,
which is $r_{\rm cool} = 99\,$kpc). Furthermore, because the dynamical
time (eqn.~\ref{eqn:tdyn}) is a factor of $\sim 20$ shorter than the
cooling time, the dark matter halo we consider satisfies the condition
for stable shock formation \citep{db06} and is in the pressure
supported ``hot halo'' regime.

\subsection{The Distribution of Cold Gas}
\label{sec:cold}

From the foregoing discussion, we expect the bulk of the gas near
\fgqso\ to have been shock heated to the virial temperature of its
hot-halo (see eqn.~\ref{eqn:Tvir}).  
However, our analysis of \slls\ indicates a
column density $\mnhi = 10^{19.65} \cm{-2}$, and hence a substantial
amount of ``cold'' gas at $\rperp\simeq 108\,$kpc. 
Similarly, a high
covering factor of high column density ($\mnhi > 10^{19}~{\rm
  cm^{-2}}$) cold gas was seen statistically by \citet{hpb+06} and
\citet{hp07}. In what follows we combine the \citet{hp07} measurement
of the clustering of cold gas around quasars, with our measurements
of the properties of \slls\ (see Table~\ref{tab:summ}), to estimate the 
covering factor, density, and volume filling factor of cold gas
in the halos surrounding high $z$ massive galaxies.


At a transverse separation $\rperp$ from a foreground quasar, the covering
factor of absorbers can be written \citep{hp07}
\be 
C(\rperp) = \left\langle\frac{dN}{dD}\right\rangle \left[1 + \chi_{\perp}(\rperp,\Delta D)\right] \Delta D, \label{eqn:cover}
\ee 
where we search over a radial (comoving) distance interval $\Delta D$
corresponding to a redshift interval $\Delta z$ in the background
quasar spectrum. Here $\left\langle\frac{dN}{dD}\right\rangle = n_{\rm
  abs}A$ is the incidence of absorption line systems per unit comoving
distance, which is simply the product of the comoving number density
of the clouds which give rise to absorption line systems and their
absorption cross section $A$ (in comoving units).  The transverse
correlation function $\chi_{\rm \perp}(R,\Delta D)$ accounts for the
enhancement due to clustering and can be written as an integral of
the 3-d quasar-absorber correlation function, $\xi_{\rm QA}(r)$, over
the search window \citep[see][for details]{hp07}.
Assuming a power law shape for the quasar-absorber cross-correlation
function, \citet{hp07} measured $r_0 = 9.2^{+1.5}_{-1.7}$\,\hMpc\ for
$\gamma=1.6$, or $r_0 = 5.8^{+1.0}_{-0.6}$\,\hMpc\ for
$\gamma=2$. These fits imply covering factors of $C(\rperp=108\,{\rm kpc})
= 0.29\pm 0.08$ and $0.35^{+0.13}_{-0.07}$, respectively. For this 
calculation we chose a projection distance $\Delta D = 19.2$\,Mpc along the 
line of sight, which corresponds to a velocity interval 
$\pm 700\mkms$, motivated by the $650\mkms$ extent of the 
absorption components in \slls. 

For the simplified case of identical clouds of mass $M_{\rm c}$, the mass
density of cold gas (in proper units) is 
\be
\rho_{\rm cold}(r) = \frac{1}{a^3}n_{\rm abs}\left[1 + \xi_{\rm QA}(r)\right]M_{\rm c} = \frac{1}{a}\left\langle\frac{dN}{dD}\right\rangle \left[1 + \xi_{\rm QA}(r)\right]N_{\rm H}\frac{1}{X} \label{eqn:rho}
\ee
where $a$ is the scale factor, $M_{\rm c}$ is the mass of the clouds,
$X=0.76$ is the hydrogen mass fraction, and $\mu_{\rm c}=0.61$ is the
mean atomic weight of the cold gas\footnote{We assume one electron per
  Helium atom in the cold phase gas.}. 
Plugging in numbers using the parameters of \slls\, we find
\be
\rho_{\rm cold}(r=108\,{\rm kpc})\approx 3.0\sci{-6}\,
\left[\frac{\left(108\,{\rm kpc}\slash r_0\right)^{-\gamma}}{472}\right]
\left(\frac{N_{\rm H}}{10^{20.6}\,{\rm cm^{-2}}}\right)
\,\msol\,{\rm pc^{-3}}.\label{eqn:rho_c}
\ee
We can compare this to our expectation for the baryon density in an NFW halo 
at this distance (see eqn.~\ref{eqn:nnfw}), and thus compute the cold gas 
fraction 
\be
\frac{\rho_{\rm cold}}{\rho_{\rm g}}(r=108\,{\rm kpc})\approx 0.03
\left(\frac{f_{\rm g}}{0.17}\right)^{-1}
\left[\frac{\left(108\,{\rm kpc}\slash r_0\right)^{-\gamma}}{472}\right]
\left(\frac{N_{\rm H}}{10^{20.6}\,{\rm cm^{-2}}}\right)
\left(\frac{M}{10^{13.3}\,\msol}\right)^{-0.73} \; . \label{eqn:frac}
\ee

Unfortunately, this estimate suffers from a large uncertainty.
Besides the $\sim 0.5$\,dex in the dark halo mass, the quasar-absorber
correlation function has significant errors and an uncertain
slope. Also we assumed a single cloud population with a total hydrogen
column density $\mnh=10^{20.6}\,{\rm cm^{-2}}$ equal to the total in
\slls. This value may not be representive of the cloud
population near quasars as a whole, although the neutral column of \slls\
$\mnhi = 10^{19.65} \cm{-2}$ is lower than the average $\langle
\mnhi\rangle \approx 10^{20.1} \cm{-2}$ of the four absorbers with $R
\lesssim 200$\,kpc in the \citet{hp07} clustering analysis.  Finally, it 
is important to note that we have assumed a maximal value for the gas fraction, 
$f_{\rm g} = \Omega_{\rm b}\slash \Omega_{\rm m} = 0.17$, equal to the
universal baryon fraction. While in principle the gas fraction can be this 
large, it is probably smaller because of gas loss from outflows. In this
regard the cold gas fraction in eqn.~(\ref{eqn:rho_c}) could
easily be a factor $\sim 2$ higher. 

Despite the uncertainties, eqn.~(\ref{eqn:rho_c}) illustrates how
measurements of the covering factor and column density distribution of
optically thick absorbers combined with photoionization modeling, can be used 
to measure the cold gas density \emph{as a function of radius} from massive
dark halos. This new observable\footnote{We note that previous workers have 
  studied the distribution of \ion{H}{1} near star-forming 
  galaxies \citep{adel03,ass+04,simcoe06}, but these are significantly
  less massive galaxies $M_{\rm halo}\sim 10^{11.5}$ and none estimated a cold 
  gas density.} thus provides a direct test of ideas which 
are currently popular in structure formation models, such as the cooling 
radius, cold-accretion, the formation of pressure supported hot-halos, and
multiphase media (see below). Furthermore, 
although the current estimate crudely assumed that all gas clouds are the same
as those in \slls, a larger statistical 
analysis could use the distribution of cloud column densities 
measured from high-resolution spectra and photoionization modelling.

Next we use our measurement of the electron density (see
\S~\ref{sec:ion} and Table~\ref{tab:summ}) and hence size 
of the absorbing clouds to determine their volume filling factor. 
The filling factor can be  written 
\be
C_{\rm v}(r)=\frac{1}{a^3}n_{\rm abs}\left[1 + \xi_{\rm QA}(r)\right]V_{\rm c} 
= \frac{1}{a}\left\langle\frac{dN}{dD}\right\rangle\left[1 + \xi_{\rm QA}(r)\right] \frac{N_{\rm H}}{n_{\rm H}}
\ee
where we used the fact that the ratio of the cloud volume to cloud area is the
absorption path length $ V_{\rm c}\slash (a^2 A) = \ell = N_{\rm H}\slash n_{\rm H}$. Plugging in numbers we find
\be
C_{\rm v}(r=108\,{\rm kpc})\approx 2.2\sci{-5}
\left[\frac{\left(108\,{\rm kpc}\slash r_0\right)^{-\gamma}}{472}\right]
\left(\frac{\mnhi}{10^{20}\,{\rm cm^{-2}}}\right)
\left(\frac{n_{\rm e}}{1.7\,{\rm cm^{-3}}}\right)^{-1},\label{eqn:vfill}
\ee
where we have used the \ion{H}{1} column and electron density
measurements for component C (see \S~\ref{sec:ion}). The mass of the
clouds 
$M_{\rm c}\sim \frac{4\pi}{3}n_{\rm c}\mu_{\rm c} m_{\rm p}(\ell\slash 2)^3$, 
where $n_{\rm c}$ is the cold gas ion density and $R_{\rm c}$ is the cloud 
radius. For a hard sphere the cloud radius is related to the average absorption 
path by $R_{\rm c} = \frac{3}{4} \ell$. Again, plugging in numbers for 
component C, we
find
\be
M_{\rm c} = 720\left(\frac{N_{\rm H}}{10^{20}\,{\rm cm^{-2}}}\right)^{3}
\left(\frac{n_{\rm e}}{1.7\,{\rm cm^{-3}}}\right)^{-2}\msol, 
\ee
however we caution that this estimate is highly uncertain because of its
dependence on large powers of quantities with significant errors.

Our conclusion from eqn.~(\ref{eqn:frac}) that a few percent of the
total baryon supply is in a ``cold'' gas phase at $T\simeq 20,000\,$K,
implies that \fgqso\ has intercepted a multi-phase medium. This is of
particular interest in light of considerable theoretical work which
indicates that the hot gas associated with massive galaxies should be
thermally unstable and prone to fragmentation instabilities
\citep[][but see \citet{malagoli90}]{field65,fr85,ml90,mm96,mb04}. The
expected result of these instabilities is a fragmented distribution of
cooled material at $T\sim 10^4\,$K, in pressure equilibrium with a hot
gas background \citep{mm96,mb04,db06}.  These instabilities have not
yet been seen in hydrodynamical simulations of galaxy formation
\citep{katz92,tw95,syw01,yss+02,hcf+03}, which lack the mass
resolution necessary to resolve them \citep{kmw+06}.  Some have
suggested that the formation of this multiphase medium can
dramatically influence galaxy formation, resulting in a maximum galaxy
mass in massive halos \citep{mb04}, or providing an important source
of feedback energy \citep{db08}.

To this end we compute the pressure $P_{c} = n_{\rm c}T_{\rm c}$ of the 
cold clouds that we detect in absorption as \slls\ to be
\be
P_{\rm c}=7.1\sci{4}\left(\frac{n_{\rm e}}{1.7\,{\rm cm}^{-3}}\right)
\left(\frac{T}{20,000~{\rm K}}\right)\,{\rm K~cm^{-3}}. 
\ee
This value can be compared to our estimate of the pressure in the NFW
halo at $r=108\,$kpc thought to be hosting \fgqso, by 
combining eqns.~(\ref{eqn:Tvir}) and (\ref{eqn:nnfw}),  
\be
P_{\rm h}\approx 9.8\sci{4}\left(\frac{f_{\rm g}}{0.17}\right)
\left(\frac{M}{10^{13.3}~\msol}\right)^{1.4}{\rm K~cm^{-3}} \label{eqn:Ph}.
\ee
It is compelling that these pressures are comparable. Our determination 
of the pressure of the absorbing gas in \slls\ indicates that the gas clouds
could very well be pressure confined by the hot $T\sim 10^{7}{\rm K}$ 
gas which is expected to permeate the dark matter halo of \fgqso.

\subsection{Could the Absorbing Clouds be in Galaxies Clustered around
  the Quasar?}
\label{sec:sfgal}

We can calculate the number density of the absorbing clouds 
from eqn.~(\ref{eqn:cover}) if we have knowledge of the absorption cross 
section $A$, 
\be
n(r) = n_{\rm abs}\left[1 + \xi_{\rm QA}(r)\right] = 
\frac{C(\rperp)}{A \Delta D}
\frac{1 + \xi_{\rm QA}(r)}{1 + \chi_{\perp}(\rperp,\Delta D)}. \label{eqn:nabs}
\ee
We use our estimate for the radius of the absorber from component C,
$r_{\rm abs}=15\,$pc (see \S~\ref{sec:ion}), to determine $A=\pi
r_{\rm abs}^2$. Combined with $C(\rperp=108\,{\rm kpc}) = 0.35$ from
\citet{hp07}, eqn.~(\ref{eqn:nabs}) gives for the number density
\be
n(r = 108\,{\rm kpc}) = 3.7\sci{7}\,{\rm Mpc}^{-3}\left(\frac{C}{0.35}\right)
\left(\frac{r_{\rm abs}}{15\,{\rm pc}}\right)^{-2}
\left(\frac{1+\xi_{\rm QA}}{473}\right)
\left(\frac{1 + \chi_\perp}{28}\right)^{-1}\label{eqn:ncloud}
\ee

This extremely high number density of clouds leads us to consider an
alternative scenario where the small clouds we observe with
characteristic size $r_{\rm abs}=15\,$pc are organized in larger
galactic structures (with characteristic size $r_{\rm gal}$) which are
clustered around the quasar. This scenario would still give rise to
the small absorption path length of $r_{\rm abs}\sim 15\,$pc, but
since the clouds are associated with galaxies the macroscopic
absorption cross-section would be $A \pi r_{\rm gal}^2$, and
hence the implied number density of galactic hosts $n_{\rm abs}$ would
be significantly smaller than determined from
eqn.~(\ref{eqn:ncloud}). This scenario could also help explain why the
metallicities deduced for \slls\ are so high in spite of the large
impact parameter $\rperp = 108\,$kpc from the quasar. If the absorbers
are associated with nearby galaxies, then the characteristic size of
these enriched regions would be $r_{\rm gal}$, obviating the need for
the quasar to have transported the metals to large distances, perhaps
via a large scale outflow (see \S~\ref{sec:outflow}).

First, we assume the absorbers are as numerous as the faintest most
abundant star-forming galaxies which have been studied to date and
determine the corresponding size of the enriched regions.
\citet{rsp+07} calculated the luminosity function for a combined
sample of brighter spectroscopic LBGs and the fainter photometric LBGs
from the Hubble Deep Field North studied by \citet{Steidel99}.  The
comoving number density of the faintest star-forming galaxies at
$z\sim 3$ is $n_{\rm LBG} = 6.9\sci{-3}\,{\rm Mpc}^{-3}$ for a flux
limit of $M_{\rm AB}$(1700\AA) $< -18.2$, corresponding to ${\cal R} <
26.9$ or $L > 0.09 L_\ast$. \citet{as05} measured the clustering of
LBGs around luminous quasars in the redshift range ($2 \lesssim z
\lesssim 3.5$), and found a best fit correlation length of
$r_0=4.7~$\hMpc\ for a slope of $\gamma=1.6$, which should be
considered an upper limit for the much fainter galaxies we consider, which 
are likely to cluster less strongly.  If we use the 
\citet{as05} correlation function to calculate
the factor of $\chi_{\perp}$ in eqn.~(\ref{eqn:cover}), then we find
that the size of the absorber galaxies has to be 
\be r_{\rm gal} =
91\,{\rm kpc}\left(\frac{C}{0.35}\right)^{1/2} \left(\frac{n_{\rm
      LBG}}{6.9\sci{-3}\,{\rm cm}^{-3}}\right)^{-1/2} \left(\frac{1 +
    \chi_\perp}{9}\right)^{-1/2}\label{eqn:rgal}.  
\ee 
Since this
value for $r_{\rm gal}$ is very similar to $R_{\rm perp}$ there is
little point in distinguishing between the star-forming galaxy halo
and the quasar halo, or stated in a different way, the average number
of LBGs in the cylindrical volume set by $R_{\rm perp}$ and $\Delta
D$ is, including the effects of clustering, about one ($N_{\rm LBG} =
0.8$). Thus associating the high-metallicity absorbing clouds with
faint LBGs does not help explain the high covering factor of enriched
material at large impact parameter. 

If the absorption is to be associated with galaxies near the quasar,
they must arise from a population much more abundant than $L \sim 0.1
L_\ast$ galaxies. These dwarf systems will be significantly fainter
and smaller. If we arbitrarily choose a value of $r_{\rm abs}=20\,$kpc
for their absorption cross-section radius, then eqn.~(\ref{eqn:cover})
implies a comoving number density of $n_{\rm dwarf} = 0.14\,{\rm
  Mpc}^{-3}$. Hence, if galaxies clustered around the quasar have
$r_{\rm abs}=20\,$kpc then they must be twenty times more abundant
than the faintest photometric LBGs studied to date.  
Extrapolating the \citet{rsp+07} luminosity function fit to achieve
this number density implies $L \sim 10^{-3} L_\ast$, or an apparent
magnitude limit of ${\cal R} < 31.7$.  The challenge to this scenario,
however, is achieving a nearly solar metallicity in these extremely
faint, high-$z$ dwarf-galaxies. Indeed the damped \lya\ systems are
thought to arise from a population of faint galaxies and are likely to
have an absorption size $r_{\rm abs}$ comparable what we assume, but
their metallicities are much lower (see Figure~\ref{fig:relabnd}) and
similarly low metallicities are seen locally in dwarf galaxies.

\subsection{Are the Kinematics Tracing an Outflow?}
\label{sec:outflow}  

How do we explain the presence of such a large gas mass of highly
enriched material with a broad-line-region-like enrichment pattern at
such a large distance from the quasar ($r\sim 100\,{\rm kpc})$?  It is
tempting to associate the absorbing material with a large scale
outflow or galactic wind from \fgqso\ and/or its host galaxy.  Indeed,
quasars exhibit fast outflows in the form of broad absorption line
systems on small scales (within 100pc of the engine) and on large
scales as radio jets extending to tens of kpc.  Furthermore, high-$z$
quasars are believed to reside in actively star-forming galaxies
perhaps with highly elevated star-formation rates.  Such galaxies may
drive fast outflows that, in principal, could extend to tens or even
hundreds of kpc.  Therefore, in terms of an outflow, there are two
extreme scenarios to consider: (a) a highly collimated outflow such as
a radio jet associated with the quasar and (b) a large-scale,
expanding `bubble' of cold swept up material driven by star-formation
feedback and/or the quasar accretion power.  The first ``jet''
scenario naturally yields an asymmetric velocity field and may
accelerate gas to very high speeds.  However, it has at least two
serious drawbacks.  First, the cross-section of a highly collimated
outflow is, by definition, small.  This is not a viable model for
reproducing the high incidence of optically thick absorbers near
quasars \citep{hpb+06,hp07}, although it might explain the single
example presented here.  Second, the large relative velocities among
the subsystems of \slls\ contradicts the notion of a collimated
flow --- in order for an outflow to remain collimated out to $\sim
100$\, kpc is must have a very small velocity shear.



For the case of a large-scale outflow driven by intense
star-formation or accretion onto the AGN, there are three issues to
consider: (i) is there a viable physical mechanism present to drive a
wind to the observed speed?; (ii) will the outflow propagate to a
distance exceeding the observed impact parameter, that is $\gtrsim
100$\,kpc?; and (iii) will the outflow yield an asymmetric velocity
field relative to $z_{\rm fg}$?  The last point requires an asymmetric
outflow which might be achieved by an irregular gas distribution in
the halo of \fgqso\ (e.g.\ variable inertia in different directions)
or if the gas has been asymmetrically photoionized by the quasar.
The required outflow speed, meanwhile, is a very challenging
constraint.  Allowing for viewing angle, portions of the outflow would
need speeds of greater than 1000\,\kms.  This surpasses
the speed of outflows generally observed in star-forming galaxies
\citep{pss+01,martin05}.  Finally, theoretical treatments of galactic
outflows suggest these will not extend to this large distance
in halos with masses characteristics of $z>2$ quasar host galaxies
\citep{fl01,ahs+01}, but of course this depends on the assumed
energetics of the flow.

In what follows we will use the observed properties of \slls\ to
estimate, in an order of magnitude sense, the energetics of the
presumed outflow, under the assumption that we have detected cold
swept up material.  We then consider if an outflow with these
energetics could be plausibly produced by a starburst or AGN.

The mass of a shell of material at radius $r_{\rm w}$, with 
column density $N_{\rm w}$, and covering solid angle $\Omega$ is 
$M_{\rm w} =m_p N_H \Omega r_w^2$. 
If this shell
is outflowing at velocity $v_{\rm w}$ in a wind, the mass 
outflow rate is ${\dot M_{\rm w}} \sim M_{\rm w} v_{\rm w}\slash r_{\rm w}$.
For our fiducial values and $\Omega=2\pi$, we derive 
$M_{\rm w} \sim 3 \sci{11} \msol$ and 
$\dot M_{\rm w} \sim 3000 \msol \rm yr^{-1}$
%
Note that we have taken the outflow velocity to be $v_{\rm w} = 1000\mkms$
because \newline \slls\ shows velocity separations as 
large as $\simeq 700\mkms$
from the f/g quasar, but this represents only the radial component of the
velocity vector. 
The corresponding energy, power, and momentum deposition rate of the wind, are 
$E_{\rm w} \sim 1\slash 2 M_{\rm w} v_w^2$, 
${\dot E_{\rm w}} \sim 1\slash 2 {\dot M_{\rm w}} v_w^2$, 
and ${\dot P_{\rm w}}\sim {\dot M_{\rm w}}v_{\rm w}$, respectively. With
our fiducial values, these are 
$E_{\rm w} \sim 3\sci{60} \, \rm erg$, 
${\dot E_{\rm w}} \sim 9\sci{44} \, \rm erg \, s^{-1}$,
and 
${\dot P_{\rm w}} \sim 2\sci{37} \, \rm g \, cm \, s^{-2}$.

These crude estimates can be compared with the
momentum and energy driven winds of a starburst and/or AGN. 
Following the formalism presented in \cite{murray05},
the SN deposition rate for
a starburst galaxy with star-formation
rate ${\dot M_{\ast}}$ is \newline
${\dot P_{\rm SN}} \sim 2\sci{33} {\dot M_{\ast}}/(\msol\,{\rm yr^{-1}}) \,
\rm g \, cm \, s^{-2}$.
A similar deposition rate is inferred from radiation pressure from a
assuming the gas is optically thick to absorption by dust grains.
We conclude that if the absorption in \slls\ is from
material swept up in a large scale wind, the star formation rate
required to power the momentum flow is extremely large ${\dot
  M_{\ast}}\sim 10^4\,\msol\,{\rm yr^{-1}}$. Furthermore, this
estimate shuld be considered a lower limit because it assumes
all momenta deposited into the galaxy are assumed
to add coherently which can significantly overestimate the deposition
\citep{socrates06}. 

The rest-frame near ultraviolet luminosity from such a high
star-formation rate can be estimated \citep{kennicutt98}, and it is
$L_{\rm UV}\sim 10^{47}{\rm erg \, s^{-1}}$ -- this would outshine the
quasar bolometric luminosity $L_{\rm QSO} \simeq 1.4\sci{46}{\rm erg
  \, s^{-1}}$ by an order of magnitude, which can easily be ruled out
since a starburst spectrum is not observed in \fgqso\ or detected in
the SDSS imaging. While the starburst could be significantly extincted
by dust (this is in fact an implicit assumption),
the fact that we observe the f/g quasar unextincted
along our line-of-sight implies that we should have been able to see
some evidence for a starburst ten times brighter. Finally, we note
that radiation pressure from the quasar radiation is also unable to
drive the presumed outflow\footnote{Here we are imagining that the
  quasar, viewed from another direction, emits a similar bolometric
  luminosity but is obscured by dust grains which absorb momentum.},
since ${\dot P_{\rm QSO}} = L_{\rm QSO}\slash c \sim 5\sci{35} {\rm
  g\,cm\,s^{-2}}$, which is a factor of forty lower than our estimate.

Next we consider the wind power, again following the formalism
in \citet{murray05}. 
For SNe, assuming a rate of one per 100 yr per $1\,\msol\,{\rm
  yr^{-1}}$,  
the SNe energy can only power the outflow for an extremely large star
formation rate: ${\dot M_{\ast}}\sim 3\sci{4}\,\msol\,{\rm yr^{-1}}$. 
We conclude that the kinematics of \slls\ are not driven by SN winds
and consider, instead, the energy from the AGN.
The absolute $B$-band magnitude\footnote{We compute the cross 
filter K-correction
  $K_{Bi}(z)$, between apparent magnitude $i$ and absolute magnitude
  $B$, which allows us to determine $M_B$ from the SDSS $i$-band
  photometry.} of \fgqso\ is $M_B = -24.8$, which
corresponds to a bolometric luminosity of $L_{\rm
  QSO}=1.4\sci{46}\,{\rm erg\,s^{-1}}$, where we used the
\citet{mclure04} fit to the \citet{elvis94} bolometric correction
to determine $L_{\rm QSO}$ from $M_B$. The ratio of the outflow power
to the accretion power is then 
\be 
{\dot E_{\rm w}}\slash L_{\rm QSO}
\sim 0.06 \left(\frac{\Omega}{2\pi}\right) \left(\frac{N_{\rm
      H}}{10^{20.6}\,{\rm cm^{-2}}}\right) \left(\frac{\rperp}{108\,{\rm
      kpc}}\right) \left(\frac{v_{\rm w}}{1000\mkms}\right)^3\,{\rm
  erg}.   \label{eqn:edotratio}
\ee 
If indeed the AGN is powering an outflow, than $\sim 6\%$ of the
accretion power has been coupled to the host via a large scale wind. A
similar comparison can be made between the energy of the outflow,
which is $E_{\rm w} \sim 4\sci{60}\,{\rm erg}$ 
and the rest-mass energy liberated to grow the supermassive black hole
$E_{\rm BH} = \epsilon_{\rm rad}M_{\rm BH}c^2$ . Assuming an Eddington
ratio of $f_{\rm Edd} \equiv L_{\rm QSO}\slash L_{\rm Edd} =0.1$,
consistent with recent estimates for a quasar at $z\sim
2.5$ near the luminosity of \fgqso\ \citep{juna06,shen08}, we deduce a 
black hole mass of
$M_{\rm BH} = 1.1\sci{9}\left(\frac{f_{\rm Edd}}{0.1}\right)^{-1}\,\msol$. 
Thus we can write
\be
E_{\rm w}\slash E_{\rm BH} \sim 0.01\left(\frac{f_{\rm Edd}}{0.1}\right)
\left(\frac{\epsilon_{\rm rad}}{0.1}\right)^{-1} \label{eqn:eratio}. 
\ee

To summarize, if the \ion{H}{1} and metal line absorption in \slls\ is
from material swept up in an large scale outflow, then the energetics
are extreme. Starburst feedback is highly unlikely unless we are
willing to consider unprecedented star-formation rates ${\dot
  M_{\ast}}\gtrsim 10^4\,\msol\,{\rm yr^{-1}}$. Radiation pressure
feedback from the quasar cannot drive the outflow even if all of the
bolometric luminosity were absorbed by dust grains. Although large,
the implied power of the presumed outflow is only a few percent of the
bolometric luminosity of the f/g quasar, or similarly, its energy is of order
a percent of the radiated rest-mass energy required to grow a $\sim
10^9\msol$ black hole.  It is intriguing that the we are led to deduce
a few percent coupling between the black hole accretion and the host
galaxy. This is very similar to the coupling factors used in
simulations of quasar feedback. 
For instance \citet{Springel05} must
inject a fraction $\eta = 0.05$ of the accreted energy into the host
galaxy to reproduce the observed $M_{\rm BH} - \sigma$ correlation,
where $\eta$ is defined by $E_{\rm feedback} = \eta \epsilon_{\rm rad}
M_{\rm BH} c^2$ \citep[see also][]{gds+04}. 
However, if such a large amount of energy is being
deposited in the host galaxy, radiative losses from the wind could be
significant, which we have not considered. Thus, in effect, our estimate
of the coupling factor is a lower limit. 


Finally, it is worth noting that high speed outflows of cold gas have
been identified in a small sample of post-starburst galaxies at $z
\sim 0.5$ \citep{tmd07}.  In several cases the wind speeds exceed
1000\kms\ and the authors also argue that the energetics may require
prior quasar activity. Similarly, the high speeds of narrow associated
absorption lines (NAALs) detected in quasars are also suggestive of
fast outflows \citep{dcr+04,wkw+08}.  However the mass and energetics of these
outflows, detected along the line-of-sight to background sources,
cannot be reliably estimated because of the highly uncertain column
density of the absorbers and the unknown distance to the absorbing
gas.






\section{Summary}
\label{sec:summ}

In this paper, we introduced a novel technique to study the physical
state of gas in the halos of luminous quasars, which has the
potential to provide powerful constraints on the physical processes
governing the formation of massive galaxies.  By mining the sky for
very rare close associations of quasars \citep{thesis,hso+06,hpb+06}, we
previously discovered \sdssj, a close projected quasar pair with
angular separation $\theta = 13.3''$ corresponding to $\rperp
=108$\,kpc at the redshift of the foreground quasar $z_{\rm fg}=
2.4360 \pm 0.0005$, precisely determined from Gemini/GNIRS near-IR
spectroscopy. The spectral and photometric properties of \fgqso\ make
it an unremarkable quasar at $z \sim 2.4$. It has an SDSS $i$-band
magnitude of $i=20.5$, from which we estimate a bolometric luminosity
of $L_{\rm QSO} \simeq 1.4\sci{46}{\rm erg\, s^{-1}}$, corresponding
to a supermassive black hole mass $M_{\rm BH} \simeq
1.1\sci{9}\left(\frac{f_{\rm Edd}}{0.1}\right)^{-1}\,\msol$, if the
black hole accretes at one tenth of the Eddington limit.  The
luminosity of \fgqso\ places it near the `knee' of the $z\sim 2.5$ quasar
luminosity function \citep{croom04,richards06}, and the clustering of such
quasars indicates they inhabit massive dark halos $M \sim
10^{13.3}\msol$ \citep{croom01,pmn04,croom05}, making them the progenitors
of massive red-and-dead galaxies observed today. Our impact parameter
$\rperp$ easily resolves the expected virial radius $r_{vir} =
250\,{\rm kpc} (M/10^{13.3} \msol)^{1/3}$, and pierces the halo of
\fgqso\ at about the cooling radius, where gas shock heated to the
virial temperature should take about a Hubble time to cool.

Rather, the only remarkable thing about \fgqso\ is that it forms a
close projection with a b/g quasar ($z_{\rm bg} = 2.53$) which is
bright enough ($r=19.0$) for high resolution spectroscopy. Our Keck
HIRES Echelle spectrum of \bgqso, the first ever to probe the halo gas
of a f/g quasar, resolves the velocity fields of the absorbing gas and
allows us to measure precise column densities for \ion{H}{1} and the
ionic transitions of metals like Si, C, N, O, and Fe. These
measurements allow us to place constraints on the physical state of
the gas near the f/g quasar, such as its kinematics, temperature,
ionization structure, chemical enrichment patterns, volume density,
the size of the absorbers, the intensity of the impingent radiation
field, as well as test for the presence of hot collisionally ionized
gas. We first summarize the results for our single sightline, and then
discuss their implications for massive galaxy formation and the quasar
phenomenon.

\subsection{Model Independent Constraints}
  
In our Keck/HIRES spectrum of \bgqso, we identify a super Lyman limit
system (SLLS) with a redshift $z=2.44$.  A Voigt profile fit to the
\lya\ and \lyb\ profiles imply a total \ion{H}{1} column density
$\mnhi = 10^{19.65 \pm 0.15} \cm{-2}$.  The \ion{H}{1} absorption
occurs redward of $z_{\rm fg}$, spanning from $\delta v = +50 \mkms$ to
$+780 \mkms$.  There is no \ion{H}{1} absorption detected ($\mnhi <
10^{13.5} \cm{-2}$) in the velocity interval $\delta v = -1500 \mkms$
to 0\kms.  We identify a series of metal lines coincident in velocity
with the \ion{H}{1} absorption (distributed in three primary
components) that includes transitions from a range of ionization states of C, O,
Fe, Si, Al, and N.  We observe an integrated
O$^0$/H$^0$ ratio of $\log[\N{O^0}/\N{H^0}] = -3.9$\,dex which indicates a
highly enriched gas.  Ignoring ionization corrections to this ratio,
which should be negligible, the average gas metallicity is [O/H]~$=
-0.5$\,dex.  Both the extreme kinematics and high metallicity of this
system are highly uncommon for intervening SLLSs.

The Doppler parameters of the low and intermediate ions are small $(b
\lesssim 5\mkms)$ indicating a gas temperature $T \simeq 20,000$K.
At all velocities we observe an ionic ratio N$^+$/N$^0 > 1$ which
implies the gas is predominantly ionized.  We observe weak, (if any)
\ion{C}{4}, \ion{N}{5}, and \ion{O}{6} absorption indicating a modest
(or negligible) quantity of metal-enriched gas with $T \approx 10^5$
to 10$^6$K in the halo surrounding \fgqso.


\subsection{Model Dependent Constraints}

Under the assumption that the gas is photoionized by a hard radiation
field (e.g.\ the nearby quasar or the EUVB), we estimate an ionization
parameter $\log U = -3.0 \pm 0.3$ for the gas.  This implies an
ionization fraction $x = 0.96$ for the two lower column density
components, and $x=0.2$ for the component with the largest column. The
total implied hydrogen column density of the system is $N_H =
10^{20.6} \cm{-2}$.  Applying ionization corrections to the observed
N$^0$/O$^0$ ratio, we infer solar to super-solar N/O abundances.  At
high $z$, such large N/O values are only found in gas associated with
quasar environments.  At $\delta v \approx +700 \mkms$, we detect
absorption from the excited fine-structure level of C$^+$,
\ion{C}{2}*~1335.  Under the assumptions that electron collisions
dominate the C$^+$ excitation and that $\N{C^+} = \N{Si^+} + \log(\rm
C/Si)_\odot$, we estimate an electron density $n_{\rm e} = 0.2$ to $10
\cm{-3}$.  By comparing this volume density to the ionization
corrected column density, we determine characteristic sizes $\ell \sim
10$ to 100\,pc for the absorbing `clouds'.

The properties of the SLLS toward \bgqso\ -- $z \approx z_{\rm fg}$,
extreme kinematics, high metallicity, and a solar N/O ratio -- argue
that this gas is located within the halo of \fgqso.  There is an
additional statistical argument: the strong clustering of optically
thick absorbers around $z \sim 2$ quasars \citep{hp07} implies the
SLLS is likely to lie at a distance near the observed transverse
distance (see Figure~\ref{fig:cluster}).  We summarize in the next
subsection several interpretations drawn from associating \slls\ with
the galactic halo of \fgqso.

We tested the hypothesis that the \slls\ is illuminated by ionizing
radiation from f/g quasar by comparing the radiation field intensity
inferred from analysis of the SLLS against the value expected from
\fgqso.  Adopting the most conservative parameters for the SLLS ($\log
U = -3$ and $n_{\rm H} = 5 \cm{-3}$), illumination of the SLLS cannot
be ruled out if it is located at $r = \rperp = 108$\,kpc from \fgqso.
Adopting more realistic values ($\log U = -4$ and $n_{\rm H} < 1
\cm{-3}$), we conclude that \slls\ is not illuminated by \fgqso,
unless the SLLS is located at an unlikely distance of greater than
500\,kpc.  We further demonstrated that if a system like \slls\ were
located along the sightline to a bright quasar at $r < 100$\,kpc, and
hence illuminated, then it would exhibit properties characteristic of
the narrow associated absorption line (NAAL) systems.  This would
include strong absorption by high-ion transitions of \ion{O}{6},
\ion{N}{5}, and \ion{C}{4}, optically thin \ion{H}{1} absorption, and
solar chemical abundances.  The absence of strong high-ion absorption
in this SLLS further suggests it is not illuminated by \fgqso\ and it
also indicates that there is not a large reservoir of warm ($T\approx
10^5$ to 10$^6$K) gas in the halo surrounding the quasar.  We suggest
that there may still be a diffuse shock heated medium but that it has
a high temperature ($T>10^6$K), characteristic of the virial
temperature of the massive dark matter halo $M \gtrsim 10^{13}\msol$
expected to be hosting the f/g quasar. If it exists, this material
undetectable in absorption with the transitions accessible with our
b/g quasar spectrum.

\subsection{Interpretation and Outlook}

The covering factor of cold $T\sim 10^4{\rm K}$, neutral, optically
thick gas is nearly unity for transverse sightlines to $z \sim 2$
quasars with $\rperp \lesssim 100\,{\rm kpc}$ \citep{hpb+06}, although
this high column density absorbing gas is rarely observed along the
line-of-sight to individual quasars \citep{reb06,phh08}. The
explanation for this anisotropic absorption is that the transverse
direction is much less likely to be illuminated by the f/g quasar
ionizing flux, either because of obscuration or intermittent emission
\citep{hpb+06,hp07,phh08}. Besides this anisotropic absorption
pattern, several independent lines of evidence corroborate this
picture for the case of \sdssj. First, we modeled the ionization state
of \slls\, which indicates that the gas is unlikely illuminated by
\fgqso\ (see \S~\ref{sec:shining}). Second, we failed to detect
high-ion transitions like \ion{N}{5} or \ion{O}{6} that should have
been seen if the gas were highly ionized by a hard radiation field
(see \S~\ref{sec:high}). Third, we showed that if gas clouds similar
to those in \slls\ were at a similar distance along the line-of-sight
(and hence illuminated), they would explain the population of much
more highly ionized associated NAALs, commonly detected in quasar
spectra (see \S~\ref{sec:nal}).  Fourth, fluorescent \lya\ emission is
not observed near $\rperp$ in \sdssj\ \citep{qpq4}; whereas the clouds
responsible for the high covering factor of optically thick gas
\citep{hpb+06} near quasars (three such clouds were detected in
\slls), would be emitting detectable \lya\ photons if
illuminated \citep{qpq4}. The upshot of this anisotropic illumination
picture is that the b/g quasar sightline probes the physical state of
gas near \fgqso\ which is \emph{shadowed and hence unaltered by the
  intense ionizing radiation emitted by the f/g quasar}.

By combining the statistical covering factor measured by \citet{hp07}
with the total $N_{\rm H}$ column determined for \slls, we argued in
\S~\ref{sec:cold} that the amount of cold gas at $r \sim 100$\,kpc is
significant, amounting to $\sim 3\% (\frac{f_{\rm g}}{\Omega_{\rm
    b}\slash \Omega_{\rm m}})^{-1}$ of the total expected gas density
of the f/g quasar's dark matter halo, if the gas fraction $f_{\rm g}$ is
equal to the cosmic baryon fraction $\Omega_{\rm b}\slash \Omega_{\rm
  m}$.  Similarly, if one assumes the material is distributed in a
thin shell of radius $\rperp$ which subtends $\Omega = 2\pi$, the
implied gas mass is $M \sim 3\sci{11}\msol$. Although we have deduced
much about the physical state of this cold gas (see
Table~\ref{tab:summ}), its origin is still unclear.  The biggest clue
could lie in its extreme enrichment patterns. The nearly solar
metallicity of \slls\ and its roughly solar N/O relative abundance
make it anomalous relative to the population of intervening SLLS and
DLAs (see Figure~\ref{fig:relabnd}).  Indeed at $z \sim 2.5$ such high
metallicities have only been observed on kpc scales in starburst
galaxies \citep{pss+01,sas+03} and solar N/O has only been observed in
the broad-line regions of quasars \citep{dhs03,agk+07} 
or in the NAALs \citep{dcr+04}.

How do we explain the presence of such a large cold gas mass of highly
enriched material with a broad-line-region-like enrichment pattern at
such a large distance from the quasar $r\sim 100\,{\rm kpc}$?  It is
tempting (and fashionable) to associate the absorbing material with a
large scale outflow or galactic wind.  If we are observing dense
material swept up by an outflow then the energetics are extreme ${\dot
  E_{\rm w}} \sim 10^{45}\,{\rm erg\,s^{-1}}$. A starburst cannot
drive this wind unless we are willing to consider unprecedented
star-formation rates ${\dot M_{\ast}}\gtrsim 10^4\,\msol\,{\rm
  yr^{-1}}$. Radiation pressure feedback from the quasar cannot drive
an outflow with this power even if its bolometric luminosity were
completely absorbed by dust grains (along a different direction than
our line-of-sight).  However, the power of the flow is only a few
percent of \fgqso's estimated bolometric luminosity, or similarly, its
energy is of order a percent of the radiated rest-mass energy required to
grow a $\sim 10^9\msol$ black hole. So the feedback hypothesis
suggests a coupling between black hole accretion and outflow which is
comparable to the value used by simulators to reproduce the
supermassive black hole $M_{\rm BH}-\sigma$ correlation
\citep[e.g.][]{Springel05}. Furthermore, radiative losses from such an
energetic flow could be significant, so our estimate of the coupling
factor must be considered a lower limit.  If we are in fact observing
feedback in \slls, then the kinematics, radial extent, and high
metallicity of the emergent outflow bear intriguing similarities to
the giant \lya\ nebulae observed in high-z radio galaxies.  However,
it is difficult to explain why such an energetic outflow would result
in so little material at intermediate temperature $T \sim 10^{4.5} -
10^{6}{\rm K}$, which we would have easily detected (but did not, see
\S~\ref{sec:high}).

But we may not be observing an outflow at all. The observed kinematics
in \newline \slls, although extreme relative to the intervening SLLS
population, are consistent with the expected gravitational motions if
the f/g quasar is indeed hosted by a massive dark matter halo
$M\gtrsim 10^{13}\,\msol$, as indicated by the strong clustering of
$z\sim 2.5$ quasars \citep{croom01,pmn04,croom05}. In this alternative
scenario, the absorption is being caused by cold clouds, with sizes
$r_{\rm abs} = 10-100\,{\rm pc}$, unit covering factor and implied
volume filling factor $C_{\rm v} \sim 10^{-5}-10^{-4}$, which are
either infalling or virialized in the deep potential well of the
massive dark matter halo.  We estimated the pressure of these clouds
to be $P \sim 10^{5}\,{\rm K\,cm^{-3}}$, which intriguingly matches
the pressure of the $T_{\rm vir} \sim 10^{7}\,{\rm K}$ shock-heated
gas expected to permeate the massive dark matter halo hosting the f/g
quasar. This pressure equilibrium is reminiscent of a large class of
galaxy formation models that postulate cold $T \sim 10^{4}{\rm K}$
clouds pressure confined by a hot shock heated virialized medium
\citep{mm96,mb04,db06}, and these scenarios might explain the lack of
significant intermediate temperature $T\sim 10^{5-6}{\rm K}$ gas near
the f/g quasar.  However, if we are indeed detecting pressure confined
cold clouds undergoing gravitational motions, why should these clouds
have such a high metallicity? This question is all the more puzzling
considering that the expected cooling time of the tenuous virialized
hot gas would be comparable to the Hubble time at $r\sim 100\,{\rm
  kpc}$. While it may be more plausible to associate the cold gas and
metals with star-forming galaxies clustered around the quasar, they
would need to be extremely abundant $n \sim 0.1\,{\rm Mpc}^{-3}$ to
produce the near unit covering factor of cold gas and metals.  These
faint $L \sim 10^{-3} L_\ast$ dwarf galaxies would need to enrich
spheres of $r_{\rm abs}\sim 20\,$kpc to solar metallicity, which seems
implausible in light of the low metallicities observed in most DLAs
(see Figure~\ref{fig:relabnd}) and the similarly low values seen
locally in dwarf galaxies.

\begin{figure}
\begin{center}
\includegraphics[width=6.0in]{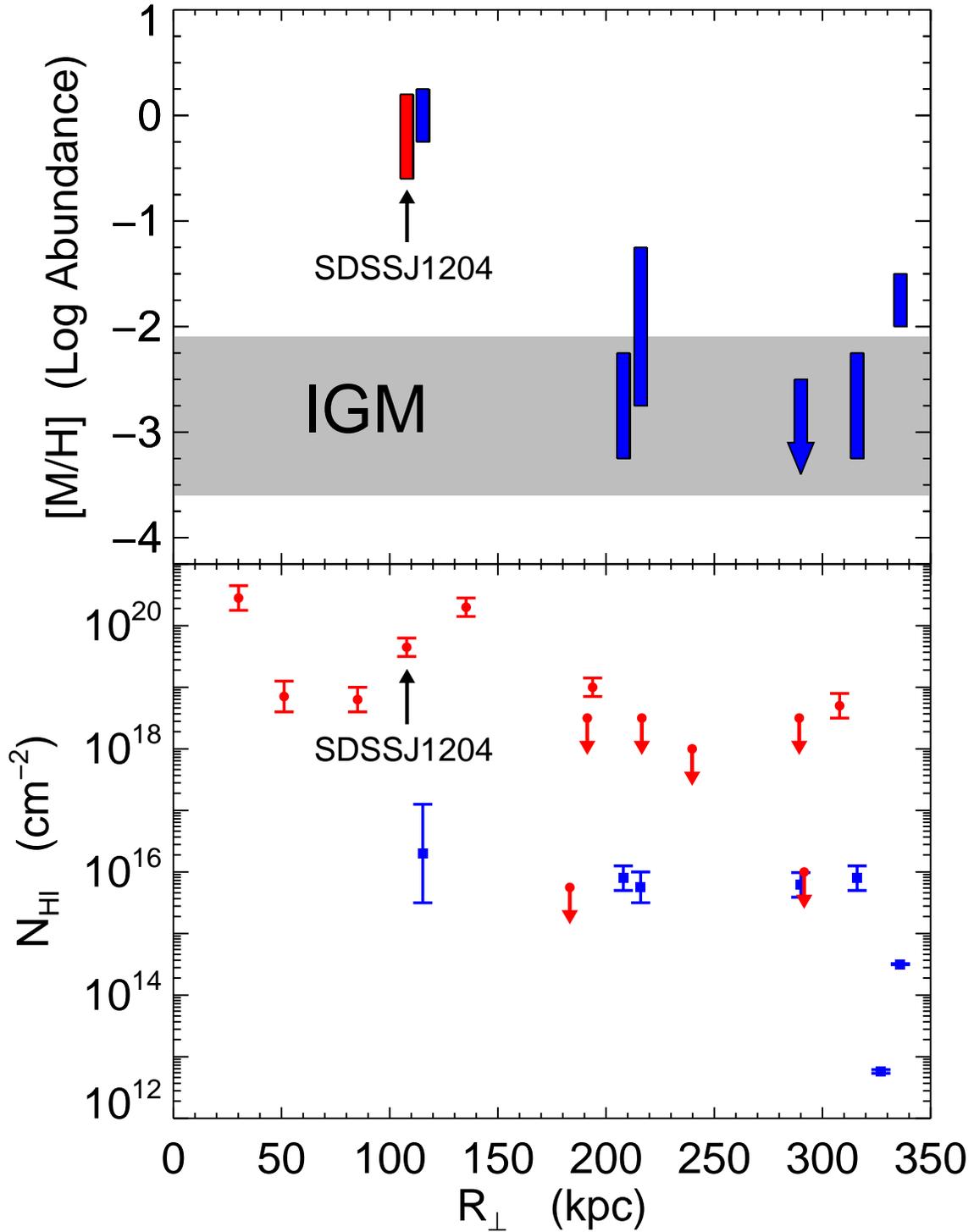}
\end{center}
\caption{Comparison of absorption line properties of $z\sim 2.5$ f/g
  quasars to $z\sim 2.5$ star forming galaxies. The lower panel plots
  the \ion{H}{1} column density versus impact parameter for absorbers,
  detected in a b/g quasar spectrum, coincident with the f/g quasar
  (or galaxy) redshift. The (red) circles are thirteen projected
  quasar pairs uncovered in \citet{hpb+06} with $\rperp < 350\,{\rm
    kpc}$, and (blue) squares are seven f/g $z\sim 2.5$ star-forming
  galaxies studied in absorption against a bright b/g quasar by
  \citet{simcoe06}. Downward pointing arrows are upper limits on \nhi\
  for the f/g quasars, which reflects our inability to measure
  reliable column densities below $\mnhi \lesssim 18.5$ from the
  moderate resolution (FWHM $\sim 125 \mkms$) b/g spectra used in
  \citet{hpb+06}. Because the \citet{simcoe06} study used a high
  resolution b/g spectrum they do not suffer from this limitation. The
  gray shaded region indicates the range of abundances observed in the
  low density IGM.  }
\label{fig:simcoe}
\end{figure}

With just a single sightline, we cannot distinguish between a quasar
powered outflow or cold clouds undergoing gravitational
motions. However, similar observations of a statistical sample of
projected quasar pairs stand to teach us a tremendous amount about the
physics of massive galaxy formation.  This is nicely illustrated by
the preliminary comparison of the absorption line properties of $z
\sim 2.2$ quasars to $z\sim 2.3$ star-forming galaxies presented in
Fig~\ref{fig:simcoe}. The primary motivation for this comparison is
that, across the two populations, we are afforded a mass baseline of
more than an order of magnitude.  The clustering of star-forming
galaxies at $z\sim 2$ indicates that they inhabit dark matter halos
with $M \lesssim 10^{12}~\msol$, making them the likely progenitors of
$\sim L_{\ast}$ galaxies that inhabit the ``blue-cloud'' of the color
magnitude diagram today \citep{cst+07}; whereas, the stronger
clustering of quasars at $z\sim 2$ implies larger halo masses $M
\gtrsim 10^{13}~\msol$ \citep{croom01,pmn04,croom05}, making them
progenitors of massive red-and-dead galaxies on the red sequence.
Hence in comparing quasars to star-forming galaxies at $z\sim 2$, \emph{we are
  effectively comparing the progenitors of quenched galaxies to those
  of un-quenched galaxies.}

The lower panel of Figure~\ref{fig:simcoe} plots the \ion{H}{1} column
density of absorbers at the f/g quasar (or galaxy) redshift versus the
impact parameter to the b/g quasar sightline. The (red) circles are the
thirteen f/g quasars ($\langle z_{\rm fg}\rangle = 2.2$) with a
background quasar within $\rperp < 350\,{\rm kpc}$ uncovered by
\citet{hpb+06}, and (blue) squares are seven f/g star-forming galaxies
($\langle z_{\rm fg}\rangle = 2.3$) studied in absorption against a
bright b/g quasar by \citet{simcoe06}. The vertical rectangles in the
upper panel illustrate the range of metallicities encountered in the
individual components of each system at each impact parameter. The only
f/g quasar metallicity measured thus far is for \sdssj\ (this work),
whereas \citet{simcoe06} measured metallicities near six f/g galaxies.

Two notable features of Fig~\ref{fig:simcoe} warrant further
discussion. First, at $\rperp \simeq 100\,{\rm kpc}$, one galaxy and
one quasar metallicity have been measured, and both indicate
abundances near solar. \citet{simcoe06} attributed the high
metallicity of their smallest impact parameter system (MD103) to
galaxy formation feedback, and we similarly cited the high metallicity
of \sdssj\ as the most compelling argument for an outflow powered by the
quasar. By mapping out the run of abundance with impact parameter in
this plot, one could characterize the size of the enriched regions
around protogalaxies with a significant mass lever-arm. If feedback is
responsible for the metals at large impact parameters, measuring the
abundance-impact parameter relation would constrain the energetics and
transport processes characterizing the relevant feedback
mechanism. Furthermore, this measurement would be fundamental to any
discussion of the enrichment history of the IGM and the ICM.  

The second noteworthy feature of Figure~\ref{fig:simcoe} is that on
small scales the f/g quasars appear to have \nhi\ column densities
significantly larger than the f/g galaxies.  This comparison is
complicated by the fact that the dark matter halos that host the
quasars are expected to be larger and more massive than those hosting
the galaxies. For reference, \citet{cst+07} estimated that the average
halo mass of $z\sim 2$ star-forming galaxies to be $M = 10^{12}\msol$,
which implies a virial radius $r_{\rm vir} = 89\,{\rm kpc}$; whereas,
at $z\sim 2.4$ we are using $M = 10^{13.3}\msol$ implying $r_{\rm vir}
= 250\,{\rm kpc}$ \citep{pmn04}.  At an impact parameter of $\rperp =
100\,{\rm kpc}$, the ratio of the total hydrogen column densities of
these dark matter halos is $N^{\rm QSO}_{\rm H}\slash N^{\rm gal}_{\rm
  H} = 5$, whereas the quasars in Fig~\ref{fig:simcoe} have \ion{H}{1}
column densities a factor $\gtrsim 100$ times larger than the galaxies
for $\rperp \lesssim 200\,{\rm kpc}$.  Although we caution that the
statistics in the lower panel of Figure~\ref{fig:simcoe} are still
very poor, it is nevertheless interesting to speculate about the
implications of this tentative result.  Why would the quasars have a
much larger reservoir of cold gas at $\rperp \sim 100\,{\rm kpc}$, of
order a few percent of the total expected gas density (see
\S~\ref{sec:cold}), than star-forming galaxies? Can this difference be
attributed to a distinct feedback mechanism operating in quasars that
does not occur in star-forming galaxies?  If instead the absorbers at
$\rperp \sim 100\,{\rm kpc}$ arise from cold gas accretion rather than
outflows, then the larger relative density of cold gas around quasars
is particularly puzzling. Cosmological hydrodynamical simulations
predict the opposite trend with dark halo mass: cold accretion
accounts for a smaller fraction of the total accreted gas in higher
mass halos where shocks become stable throughout the halo and the cold
filamentary mode of accretion disappears
\citep{kkw+05,db06,ocvirk08}. Whether the cold gas around quasars is
ejected by feedback or accreted in gravitational collapse, what is the
ultimate fate of this material? Does it eventually fall back onto the
quasar host galaxy or do the gas and metals blend into the IGM. Do we
expect to observe a similar cold halo of gas around nearby massive
elliptical galaxies, since their progenitors are $z\sim 2$ quasars like
\fgqso?

Although preliminary, the putative trends in Figure~\ref{fig:simcoe}
provoke fundamental questions about feedback, quenching and the
physics of massive galaxy formation.  Yet this discussion represents
just two measurements (metallicity and \nhi) gleaned from our analysis
of the spectrum of \bgqso. Other properties of the gas such as its
kinematics, temperature, ionization structure, relative abundance
patterns, the presence or absence of a hot collisionally ionized
phase, and the volume density and size of the clouds, have a similar
potential to constrain the physics of massive galaxy formation if they
can be mapped out for statistical samples. Such a statistical analysis
is well within reach. To date, our pair confirmation program
\citep{thesis,hso+06,hpb+06} has confirmed about 90 pairs of quasars
with impact parameter $R < 300~{\rm kpc}$ and $z_{\rm fg} > 1.6$, and
about a comparable number still remain to be discovered in the
existing SDSS photometric data.  Of the known pairs, only about five
are bright enough $r\lesssim 19$ for high resolution (FHWM$\sim
10\mkms$) echelle spectroscopy like the HIRES spectrum used to study
\slls. However, higher throughput but lower resolution echellette
spectrographs, such as the Echellette Spectrograph and Imager
\citep[ESI,][]{ESI} on Keck II, the Magellan Echellette (MagE) on the
Magellan Clay Telescope, or the planned X-Shooter spectrograph
\citep{xshooter} for the VLT, can deliver spectra with FHWM$\sim
30-50\mkms$ and the required signal to noise ratio ($\sim 10$ per
resolution element), in 1-2 hours for $r \lesssim 21.5$, which would
make \emph{all} of the quasar pairs known observable.  Although these
instruments have lower spectral resolution, they roughly resolve the
\ion{H}{1} Lyman series affording precise estimates of \nhi.
Furthermore, such data would yield metallicity and relative abundance
estimates to a precision of 0.3\,dex, and would allow for the
construction of photoionization models with an accuracy comparable to
that achieved in this work.

The most important aspect of the approach taken to understanding
galaxy formation in this study and \citet{simcoe06} (see also
Adelberger et al.\ 2003), is that they provide the first observational
constraints on the physical state of the \emph{gas} on scales
$\gtrsim$ kpc in high-$z$ protogalaxies. Although the ideas behind
quenching and feedback were introduced to explain the observed
properties of local galaxy stellar populations, such as the bimodality
in the color magnitude diagram
\citep{Strateva01,baldry04,bell04,blanton05,faber07} and the sharp
cutoff in the luminosity function \citep{bbf+03}, these are
fundamentally gas dynamics problems.  Many of the relevant
hydrodynamical and physical processes can be (or are nearly) resolved
by simulation grids with current technology. Conversely, predicting
how this gas physics manifests itself in the stellar populations of
local or high-$z$ galaxies requires that the uncertain ``sub-grid''
physics of star formation be inserted by hand, or with the aid of
semi-analytical recipes.  Thus, observational constraints on the gas
provide especially fruitful comparisons to theory. If the
techniques presented here for a single object can be expanded to
statistical samples, that will constitute a significant step on the
road toward understanding massive galaxy formation.

\acknowledgments We acknowledge helpful discussions with T.J. Cox,
P. Hopkins, P. Madau, B. Mathews, E. Ramirez-Ruiz, R. Simcoe, and
A. Wolfe.  We are grateful to R. Simcoe for kindly providing the
galaxy data in Figure~\ref{fig:simcoe} in electronic form.  For part
of this work JFH was supported by NASA through Hubble Fellowship grant
\# 01172.01-A awarded by the Space Telescope Science Institute, which
is operated by the Association of Universities for Research in
Astronomy, Inc., for NASA, under contract NAS 5-26555.  JXP
acknowledges funding through an NSF CAREER grant (AST-0548180) and NSF
grant (AST-0709235).  JFH is currently supported by the National
Science Foundation through the Astronomy and Astrophysics Postdoctoral
Fellowship program (AST-0702879).

The conclusions of this work are partly based on observations obtained
at the Gemini Observatory through Gemini Program ID
GS-2006A-Q-3. Gemini Observatory is operated by the Association of
Universities for Research in Astronomy (AURA) under a cooperative
agreement with the NSF on behalf of the Gemini partnership: the
National Science Foundation (United States), the Science and
Technology Facilities Council (United Kingdom), the National Research
Council (Canada), CONICYT (Chile), the Australian Research Council
(Australia), CNPq (Brazil) and CONICET (Argentina).

The conclusions of this work are based on data collected from
observatories at the summit of Mauna Kea. The authors wish to
recognize and acknowledge the very significant cultural role and
reverence that the summit of Mauna Kea has always had within the
indigenous Hawaiian community.  We are most fortunate to have the
opportunity to conduct observations from this mountain.




\end{document}